\documentclass{aa}  
\usepackage{natbib}
\bibpunct{(}{)}{;}{a}{}{,} 
\usepackage{orcidlink}
\usepackage{hyperref}
\usepackage{xcolor}

\usepackage{ulem}
\usepackage[document]{}
\usepackage{graphicx}
\usepackage{txfonts}
\begin{document}


    \title{ALMA-IMF XIX: C$^{18}$O ($J$ = 2$-$1): Measurements of turbulence in 15 massive protoclusters}{}
    
     
    \author{A. Koley$^{1}$\orcidlink{0000-0003-2713-0211}, A. M. Stutz$^{1,~2}$\orcidlink{0000-0003-2300-8200}, F. Louvet$^{3}$\orcidlink{0000-0003-3814-4424}, F. Motte$^{3}$\orcidlink{0000-0003-1649-8002}, A. Ginsburg$^{4}$\orcidlink{0000-0001-6431-9633}, R. Galv$\acute{\text{a}}$n-Madrid$^{5}$\orcidlink{0000-0003-1480-4643}, R. H. $\acute{\text{A}}$lvarez-Guti$\acute{\text{e}}$rrez$^{1}$\orcidlink{0000-0002-9386-8612}, P. Sanhueza$^{6,~7}$\orcidlink{0000-0002-7125-7685}, T. Baug$^{8}$\orcidlink{0000-0003-0295-6586}, N. Sandoval-Garrido$^{1}$\orcidlink{0000-0001-9600-2796}, J. Salinas$^{1}$\orcidlink{https://orcid.org/0009-0009-4976-4320}, G. Busquet$^{9,~10,~11}$\orcidlink{0000-0002-2189-6278}, J. Braine$^{12}$\orcidlink{0000-0003-1740-1284}, H.-L. Liu$^{13}$\orcidlink{0000-0003-3343-9645}, T. Csengeri$^{14}$\orcidlink{0000-0002-6018-1371},  A. Gusdorf$^{15,~16}$\orcidlink{0000-0002-0354-1684}, M. Fern$\acute{\text{a}}$ndez-L$\acute{\text{o}}$pez$^{17}$\orcidlink{0000-0001-5811-0454}, N. Cunningham$^{18}$\orcidlink{0000-0003-3152-8564}, L. Bronfman$^{19}$\orcidlink{0000-0002-9574-8454}, M. Bonfand$^{20}$\orcidlink{0000-0001-6551-6444}\\
    \vspace{5mm}
    (Affiliations can be found after the references)}
    \authorrunning{Koley et al.}{}
          \institute{}

   \date{Received xxxxxxx; accepted xxxxxx}{}


 
  \abstract
   {\color{black}{ALMA-IMF is a large program of the Atacama Large Millimeter/submillimeter Array (ALMA) that aims to determine the origin of the core mass function (CMF) of 15 massive Galactic protoclusters ($\sim$ 1.0 $-$ 25.0 $\times$ 10$^{3}~$\(\textup{M}_\odot\) within $\sim$ 2.5 $\times$ 2.5 pc$^{2}$) located towards the Galactic plane. In addition, the objective of the program is to obtain a thorough understanding of their physical and kinematic properties. Here we study the turbulence in these protoclusters with C$^{18}$O (2$-$1) emission line using the sonic Mach number analysis ($M$$_{\text{s}}$) and the size-linewidth relation. The probability distribution functions (PDFs) for $M$$_{\text{s}}$ show a similar pattern, exhibiting no clear trend associated with evolutionary stage, peaking in the range between 4 and 7, and  then extending to $\sim$ 25. Such values of $M$$_{\text{s}}$ indicate that the turbulence in the density regime traced by the C$^{18}$O line inside the protoclusters is supersonic in nature. In addition, we compare the non-thermal velocity dispersions ($\sigma_{\text{nth, C$^{18}$O}}$) obtained from the C$^{18}$O (2$-$1) line with the non-thermal line widths ($\sigma_{\text{nth, DCN}}$) of the cores obtained from the DCN (3$-$2) line. We observe that, on average, the non-thermal linewidth in cores is half that of the gas surrounding them. This suggests that turbulence diminishes at smaller scales or dissipates at the periphery of the cores. Furthermore, we examine the size-linewidth relation for the structures we extracted from the position-position-velocity C$^{18}$O (2$-$1) line emission cube with dendrogram algorithm. The power-law index ($p$) obtained from the size-linewidth relation is between 0.41 and 0.64, steeper than the Kolmogorov law of turbulence, as expected for compressible media. In conclusion, this work is one of the first to carry out such a statistical study of turbulence for embedded massive protoclusters. } }

   {}
   {}
   {}
   {}


   \maketitle
%

\newpage

\section{Introduction}\label{section_0}

Turbulence plays a crucial role in shaping the interstellar medium (ISM), which is essential for understanding the evolution of molecular clouds \citep{1987ApJ...319..730S,2004ApJ...615L..45H,2019MNRAS.483..593K,2023MNRAS.525..962P,2023PASA...40...46K,2024ApJ...966...51G}. The interplay between turbulence, magnetic fields, and gravity leads to the fragmentation of dense clumps, resulting in the formation of cores and eventually in the formation of new stars \citep{2016ApJ...819..139B,2020ApJ...905..158W,2016A&A...590A...2S,2018MNRAS.473.4890S,2021ApJ...912..159P,2021MNRAS.501.4825K,2022MNRAS.516L..48K,2024MNRAS.528.1460R,2024arXiv241208790S,2024MNRAS.529.2220R}. Various studies have indicated that in molecular clouds turbulence is supersonic in nature. {\color{black}On a large scale, supersonic turbulence can prevent global collapse and supports the structure of molecular clouds \citep{2004RvMP...76..125M,2015MNRAS.450.4035F}. On a small scale, turbulence also affects the fragmentation of dense cores, and stars begin to form when turbulence dissipates \citep{1983ApJ...270..105M,1998ApJ...504..223G,1998ApJ...494..587N,2023PASA...40...53K}. In the interstellar medium, turbulence plays a key role in determining the star formation rate (SFR) \citep{2012ApJ...761..156F,2015MNRAS.450.4035F, 2017A&A...599A..99O, 2018PhT....71f..38F}. It is thus important to examine the nature of turbulence for a better understanding of the star formation process.\\

In molecular clouds, turbulence is measured by a variety of methods. These include analysis of the sonic Mach number ($M_{\text{s}}$), examination of the size-linewidth relationship, principal component analysis (PCA) and power spectrum analysis \citep{2004ApJ...615L..45H,2017A&A...599A..99O,2023MNRAS.525..962P}. For example, using the examination of $M_{\text{s}}$, it was shown in multiple studies that in molecular clouds turbulence is supersonic in nature \citep{2017A&A...599A..99O,2023A&A...674A..46W}. In addition, using the size-linewidth relation, previous studies have shown that in molecular clouds the power-law index ($p$) is close to 0.50. For example, \cite{1987ApJ...319..730S} examined the size-linewidth relation for 273 molecular clouds using the CO molecular line and noted that $p$ is 0.50. Similarly, using PCA \cite{2004ApJ...615L..45H} demonstrated that the velocity structure function varies as a function of spatial scale with a value of $p$ of 0.49. In addition to these observational studies, several numerical studies have also shown that the power-law index of velocity dispersion in the molecular cloud is steeper than the standard Kolmogorov law of turbulence ($p$ = 0.33) and close to 0.50 \citep{1941DoSSR..30..301K,1983ApJ...272L..45F,2007ApJ...666L..69K}.\\

In this context, the ALMA-IMF is an Atacama Large Millimeter/submillimeter Array (ALMA) Large Program designed to observe 15 massive protoclusters in our Galaxy. Here, the distance of the 15 protoclusters ranges between 2.0 kpc and 5.5 kpc with a mean value of $\sim$ 3.9 kpc. This suitable distance range allows us to trace the area of more than 1 pc$^{2}$ towards the highest column density area obtained from the ATLASGAL survey \citep{2017A&A...601A..60C}. In addition, the unprecedented resolution of ALMA allows us to achieve and investigate $\sim$ 0.01 pc core-scale structures in these regions. The gas masses of the ALMA-IMF protocluster sample are $\sim$ 1.0 $-$ 25.0 $\times$ 10$^{3}~$\(\textup{M}_\odot\). The majority of the protoclusters are located within $\pm$ 0.5$^{\circ}$ of the Galactic plane, where most of the massive gas clouds reside. \cite{2022A&A...662A...8M} and \cite{2024ApJS..274...15G} classified the evolutionary stage of these 15 protoclusters based on the flux ratio between the 1.3 and 3.0 mm continuums and the H41$\alpha$ emission. In their study, the authors found that during the evolution of the protoclusters, flux ratio between 1.3 mm and 3.0 mm continuums decreases, and H41$\alpha$ emission increases at the same time. Based on these constraints, they classified six protoclusters as young, five protoclusters as intermediate, and the remaining four protoclusters as evolved. We provide a general overview of the 15 protoclusters in Table \ref{tab:table1}.\\

This large program utilizes data from ALMA 12m and 7m arrays as well as the total power (TP) in two bands: Band 3 (3.0 mm) and Band 6 (1.3 mm) \citep{2022A&A...662A...8M,2022A&A...662A...9G,2023A&A...678A.194C}.  The observations include several spectral lines and continuum in two bands. Two continuum images at 1.3 and 3.0 mm mainly tracing thermal dust emission are used to investigate the key properties of the cores (which are on the verge of star formation) such as their luminosity, temperature, and mass \citep{2018NatAs...2..478M,2022A&A...664A..26P,2024arXiv240109203A,2024A&A...690A..33L}. Likewise, the spectral energy distribution (SED) of thermal dust emission is also used to determine the dust temperature within the protoclusters \citep{2022A&A...665A.140B,2024arXiv240215023B,2024A&A...687A.217D,2024arXiv241202011M}. Furthermore, the DCN (3$-$2) spectral line provides insight into core kinematics \citep{2023A&A...678A.194C}. Another prominent spectral line, N$_{2}$H$^{+}$(1-0), allows us to study the kinematics of the dense cold gas inside the protoclusters \citep{2024A&A...689A..74A,2024arXiv241009843S}. In addition, the CO (2$-$1), SiO (5$-$4) lines allow us to characterize the outflow in the protoclusters \citep{2020A&A...636A..38N,2023A&A...674A..75N,2024ApJ...960...48T,2024A&A...686A.122A}.\\

In this work, we study the turbulence in 15 massive protoclusters using C$^{18}$O (2$-$1) line (Band 6 in ALMA) using sonic Mach number ($M_{\text{s}}$) analysis and size-linewidth relation. The rest frequency and the critical density ($n_{\text{cr}}$) of this line are 219.56035800 GHz {\color{blue}\footnote{{\color{blue}https://splatalogue.online}}} and 9.33 $\times$ 10$^{3}$ cm$^{-3}$ (at 20 K), respectively \citep{2016A&A...585A..44M}. This $n_{\text{cr}}$ is relatively low compared to the other spectral species in the ALMA-IMF survey, such as DCN (3$-$2), N$_{2}$D$^{+}$(3$-$2), for which $n_{\text{cr}}$ varies between $\sim$ 10$^{6}$ and 10$^{7}$ cm$^{-3}$ \citep{2022A&A...662A...8M,2023A&A...678A.194C}. Consequently, C$^{18}$O (2$-$1) line traces extended gas compared to the lines mentioned above \citep{1997ApJ...482..245U,2015PASP..127..266M,2022ApJ...936...80S,2023A&A...678A.194C}. Compared to other isotopes such as $^{12}$CO and $^{13}$CO, C$^{18}$O's optical depth ($\tau_{\nu}$)  is relatively low \citep{2016A&A...591A.104H,2019MNRAS.487.1259L,2022ApJ...936...80S}. Several previous studies have shown that $\tau_{\nu}$ of C$^{18}$O (2$-1$) line is generally lower than 1 even in high mass star-forming regions. In contrast, it can be one order of magnitude higher for $^{12}$CO and $^{13}$CO lines \citep{2000ApJ...536..393H,2015ApJS..216...18N,2019MNRAS.490.4489S}. Through C$^{18}$O line, various phenomena, such as cloud-cloud collisions, large-scale velocity oscillations, rotation, infall, and even outflow can be traced \citep{2019MNRAS.487.1259L,2021ApJ...908...86A,2022MNRAS.513.2942D,2023ApJ...957...61H}. In addition to these various phenomena, it is also possible to measure several other effects, such as the nature of turbulence in star-forming regions \citep{2017A&A...599A..99O,2018A&A...617A..14P,2021AJ....161..229K,2021MNRAS.500.1721M,2023JApA...44...34M}.\\


Our paper is organized as follows. In \S~\ref{section_1}, we present the data. In \S~\ref{section_2}, we show the average C$^{18}$O (2$-$1) spectra for 15 protoclusters. We present moment maps for 15 protoclusters in \S~\ref{section_3}.  In \S~\ref{section_4}, we examine the turbulence in the protoclusters using sonic Mach number ($M_{\text{s}}$) analysis and size-linewidth relation.  We discuss our main results in \S~\ref{s:discussion}. Finally, we draw our main conclusions in \S~\ref{s:concl}.


\begin{flushleft}
	
	\begin{table*}[ht]
		\caption{Overview of 15 ALMA-IMF protoclusters.}
		
		\begin{tabular}{  c c c c c c c c c c c }
			\hline

			Protocluster & \hspace{-3.190mm} R.A.  & \hspace{-3.190mm} Dec.  & \hspace{-3.190mm} Long.  & \hspace{-3.190mm} Lat. & \hspace{-3.190mm} $V$$_{\text{sys}}$$^{\texttt{a}}$ & \hspace{-3.190mm}\textit{d}$^{\texttt{b}}$ &   \hspace{-3.190mm} Mass$^{\texttt{c}}$ & \hspace{-3.190mm} Evolutionary & \hspace{-3.190mm} Imaged area & \\ [0.5 ex]
			
			(name) & \hspace{-3.190mm} [ICRS] & \hspace{-3.190mm}[ICRS] & \hspace{-3.190mm} [deg.] & \hspace{-3.190mm} [deg.] & \hspace{-3.190mm} [km s$^{-1}$] & \hspace{-3.190mm} [kpc]& \hspace{-3.190mm} [ $\times $10$^{2}$ \(\textup{M}_\odot\)]  & \hspace{-3.190mm} stage [Y / I / E]$^{\texttt{d}}$  & \hspace{0.0001mm} [pc $\times$ pc] & \\ [0.5 ex]
			
			\hline
			
			W43-MM1 & \hspace{-3.190mm} 18:47:47.00  & \hspace{-3.190mm} -01:54:26.0   & \hspace{-3.190mm} \hspace{-1.8mm}30.82   & \hspace{-3.190mm} -0.057  & \hspace{-3.190mm}\hspace{-2.7mm} +96.7 & \hspace{-3.190mm} 5.5$\pm$0.4  & \hspace{-3.890mm} 170$\pm$60 & \hspace{-3.190mm} Y & \hspace{-3.090mm}  3.1 $\times$ 2.3 & \\ [0.5 ex]
			
			W43-MM2 & \hspace{-3.190mm} 18:47:36.61  & \hspace{-3.190mm} -02:00:51.1   & \hspace{-3.190mm} \hspace{-1.8mm}30.70   & \hspace{-3.190mm} -0.067  & \hspace{-3.190mm}\hspace{-2.7mm} +92.0 & \hspace{-3.1290mm} 5.5$\pm$0.4  & \hspace{-3.190mm} 150$\pm$60  & \hspace{-3.190mm} Y & \hspace{-3.190mm} 2.6 $\times$ 2.4   & \\ [0.5 ex]
		
		\hspace{-3.2mm}G327.29  & \hspace{-3.190mm} 15:53:08.13  & \hspace{-3.190mm} -54:37:08.6   & \hspace{-3.190mm} 327.29   & \hspace{-3.190mm} -0.579   & \hspace{-3.190mm}\hspace{-2.7mm} -45.5 & \hspace{-3.190mm} 2.5$\pm$0.5  & \hspace{-3.890mm}80$\pm$40  & \hspace{-3.190mm} Y & \hspace{-3.190mm}  1.3 $\times$ 1.3  & \\ [0.5 ex]
		
		\hspace{-3.2mm}G328.25  & \hspace{-3.190mm} 15:57:59.68  & \hspace{-3.190mm} -53:58:00.2   & \hspace{-3.190mm}  328.25  & \hspace{-3.190mm}  -0.532  & \hspace{-3.190mm}\hspace{-2.7mm} -44.3 & \hspace{-3.190mm} 2.5$\pm$0.5  & \hspace{-6.1990mm} 10$\pm$5  & \hspace{-3.190mm} Y & \hspace{-3.190mm}  1.4 $\times$ 1.4 & \\ [0.5 ex]
		
		\hspace{-3.2mm}G337.92 & \hspace{-3.190mm} 16:41:10.62   & \hspace{-3.190mm} -47:08:02.9   & \hspace{-3.190mm}  337.92   & \hspace{-3.190mm} -0.477  & \hspace{-3.190mm}\hspace{-2.7mm} -39.8 & \hspace{-3.190mm} 2.7$\pm$0.7  &\hspace{-4.590mm}  30$\pm$20   & \hspace{-3.190mm} Y & \hspace{-3.190mm}  1.2 $\times$ 1.1 & \\ [0.5 ex]
		
		\hspace{-3.2mm}G338.93 & \hspace{-3.190mm} 16:40:34.42  & \hspace{-3.190mm} -45:41:40.6    & \hspace{-3.190mm} 338.93    & \hspace{-3.190mm} +0.554   & \hspace{-3.190mm}\hspace{-2.7mm} -63.1 & \hspace{-3.190mm} 3.9$\pm$1.0  & \hspace{-3.890mm}  40$\pm$30    & \hspace{-3.190mm} Y & \hspace{-3.190mm}  1.6 $\times$ 1.6 & \\ [0.5 ex]
		
		\hline
		
		W43-MM3  & \hspace{-3.190mm} 18:47:41.46 & \hspace{-3.190mm} -02:00:27.6   & \hspace{-3.190mm}  \hspace{-1.8mm}30.72      & \hspace{-3.190mm}  -0.082   & \hspace{-3.190mm}\hspace{-2.7mm} +93.2 & \hspace{-3.190mm} 5.5$\pm$0.4  & \hspace{-3.890mm} 80$\pm$30  & \hspace{-3.190mm} I & \hspace{-3.190mm} 2.7 $\times$ 2.4  & \\ [0.5 ex]
		
		\hspace{-5.5mm}W51-E & \hspace{-3.190mm} 19:23:44.18  & \hspace{-3.190mm} +14:30:29.5   & \hspace{-3.190mm} \hspace{-1.8mm}49.49   & \hspace{-3.190mm} -0.389   & \hspace{-3.190mm}\hspace{-2.7mm} +57.8 & \hspace{-3.190mm} 5.4$\pm$0.3  & \hspace{-2.490mm} 240$\pm$70  & \hspace{-3.190mm} I & \hspace{-3.190mm} 2.6 $\times$ 2.4   & \\ [0.5 ex]
		
		\hspace{-3.2mm}G008.67  & \hspace{-3.190mm} 18:06:21.12   & \hspace{-3.190mm} -21:37:16.7  & \hspace{-3.190mm} \hspace{-3.6mm}8.68   & \hspace{-3.190mm} -0.361  & \hspace{-6.0mm} +36.2 & \hspace{-3.190mm} 3.4$\pm$0.3 & \hspace{-4.190mm} 30$\pm$10  & \hspace{-3.190mm} I & \hspace{-3.190mm}  2.2 $\times$ 1.4  & \\ [0.5 ex]
		
		\hspace{-3.2mm}G351.77  & \hspace{-3.190mm} 17:26:42.62  & \hspace{-3.190mm} -36:09:20.5  & \hspace{-3.190mm} 351.77    & \hspace{-3.190mm}  -0.537 & \hspace{-3.190mm}\hspace{-4.7mm} -3.2 & \hspace{-3.190mm} 2.0$\pm$0.7 & \hspace{-4.190mm} 20$\pm$10  & \hspace{-3.190mm} I & \hspace{-3.190mm}  1.3 $\times$ 1.3 & \\ [0.5 ex]
		
		\hspace{-3.2mm}G353.41 & \hspace{-3.190mm} 17:30:26.28    & \hspace{-3.190mm} -34:41:49.7 & \hspace{-3.190mm}  353.40    & \hspace{-3.190mm}  -0.361  & \hspace{-3.190mm} \hspace{-2.7mm}-16.6 & \hspace{-3.190mm} 2.0$\pm$0.7 & \hspace{-4.190mm} 20$\pm$10 & \hspace{-3.190mm} I & \hspace{-3.190mm}  1.3 $\times$ 1.3 & \\ [0.5 ex]
		
		\hline
		
		W51-IRS2 & \hspace{-3.190mm} 19:23:39.81  & \hspace{-3.190mm} +14:31:03.5 & \hspace{-3.190mm} \hspace{-2.30mm} 49.49   & \hspace{-3.190mm} -0.369   & \hspace{-3.190mm}\hspace{-2.7mm} +59.3 & \hspace{-3.190mm} 5.4$\pm$0.3 & \hspace{-2.090mm} 250$\pm$80 & \hspace{-3.190mm} E & \hspace{-3.190mm}  2.6 $\times$ 2.4 & \\ [0.5 ex]
		
		\hspace{-3.2mm}G010.62 & \hspace{-3.190mm} 18:10:28.84   & \hspace{-3.190mm} -19:55:48.3  & \hspace{-3.190mm} \hspace{-2.30mm} 10.62   & \hspace{-3.190mm} -0.384  & \hspace{-3.190mm} \hspace{-4.2mm}-2.7 & \hspace{-3.190mm} \hspace{+1.4mm}4.95$\pm$0.5 & \hspace{-3.190mm} 110$\pm$40 & \hspace{-3.190mm} E & \hspace{-3.190mm} 2.3 $\times$ 2.2   & \\ [0.5 ex]
		
		\hspace{-3.2mm}G012.80  & \hspace{-3.190mm} 18:14:13.37   & \hspace{-3.190mm} -17:55:45.2  & \hspace{-3.190mm} \hspace{-2.30mm} 12.80    & \hspace{-3.190mm} -0.199  & \hspace{-3.190mm}\hspace{-2.7mm} +35.5 & \hspace{-3.190mm} 2.4$\pm$0.2 & \hspace{-3.190mm} 120$\pm$40 & \hspace{-3.190mm} E & \hspace{-3.190mm}  1.5 $\times$ 1.5  & \\ [0.5 ex]
		
		\hspace{-3.2mm}G333.60   & \hspace{-3.190mm} 16:22:09.36  & \hspace{-3.190mm} -50:05:58.9  & \hspace{-3.190mm}  333.60    & \hspace{-3.190mm} -0.212   & \hspace{-3.190mm} \hspace{-2.7mm}-47.4 & \hspace{-3.190mm} 4.2$\pm$0.7 & \hspace{-2.790mm} 130$\pm$60 & \hspace{-3.190mm} E & \hspace{-3.190mm}  2.9 $\times$ 2.9 & \\ [0.5 ex]
		
		\hline
		
		\end{tabular}
		
		\vspace{3mm}
		\label{tab:table1}

\textbf{Notes.} Col. 1: Protocluster names. Col. 2, Col. 3, Col. 4, and Col. 5 represent the Right Ascension (R.A.), Declination (Dec.), Longitude, and Latitude of the centre of the mosaic observation that have been used in the ALMA-IMF survey. Col. 6: Systemic velocity ($V _{\text{sys}}$) of the protoclusters. Col. 7: Distance ($d$) of the protoclusters. Col. 8: Mass of the protoclusters within the area of 1.3 mm continuum. Col. 9: Evolutionary stage of the protoclusters and Col. 10: Imaged area in the 1.3 mm band.\\
$^{\texttt{a}}$$V$$_{\text{sys}}$ is measured based on the C$^{18}$O ($J$= 2$-$1) line, which is discussed in Section \ref{section_2}.\\
\hspace{-17mm}$^{\texttt{b}}$ Distance of the protoclusters taken from \cite{2022A&A...662A...8M}.\\
\hspace{-4mm}\hspace{-36mm}$^{\texttt{c}}$ {\color{black}Cloud mass computed from the point process mapping (PPMAP) analysis of the dust continuum emission \citep{2024A&A...687A.217D}}.\\
\hspace{-4.3mm}\hspace{-22mm}$^{\texttt{d}}$ Y $\rightarrow$ young, I $\rightarrow$ intermediate, and E $\rightarrow$ evolved phase of the protocluster, as defined by \cite{2022A&A...662A...8M}.

	\end{table*}

\end{flushleft}

\begin{table*}
	\caption{Summary of the observational parameters of C$^{18}$O (2$-$1) spectral line.}
	\begin{tabular}{  c c c c c c }
		\hline
        
		Protocluster & \hspace{0.001mm} Spectral resolution & \hspace{0.001mm} Beam major ($\theta_{major}$)  & \hspace{0.001mm} Beam minor ($\theta_{minor}$)  & \hspace{0.001mm} Position angle (PA) & \hspace{0.001mm} RMS noise ($\sigma_{\text{rms}}$)\hspace{0.001mm} \\ [0.5 ex]
		(name) & \hspace{0.001mm} (km s$^{-1}$) & \hspace{0.001mm} ($\arcsec$)  & \hspace{0.001mm} ($\arcsec$)  & \hspace{0.001mm} ($^\circ$) & \hspace{0.001mm} (mJy beam$^{-1}$)\hspace{0.001mm} \\ [0.5 ex]
		\hline
		
		W43-MM1 & \hspace{1.8mm} 0.17$^{a}$ & \hspace{0.001mm} 0.65  & \hspace{0.001mm} 0.49  & \hspace{0.001mm} -80.63 & \hspace{0.001mm} 5.0 \hspace{0.001mm} \\ [0.5 ex]
		
		W43-MM2 & \hspace{0.001mm} 0.33 & \hspace{0.001mm} 0.79  & \hspace{0.001mm} 0.63  & \hspace{0.001mm} -83.67 & \hspace{0.001mm} 4.1 \hspace{0.001mm} \\ [0.5 ex]
		
		\hspace{-3.2mm}G327.29 & \hspace{0.001mm} 0.33  & \hspace{0.001mm} 0.86  & \hspace{0.001mm} 0.79  & \hspace{0.001mm} -55.97 & \hspace{0.001mm} 18.6 \hspace{0.001mm} \\ [0.5 ex]
		
		\hspace{-3.2mm}G328.25 & \hspace{0.001mm} 0.33 & \hspace{0.001mm} 1.23  & \hspace{0.001mm} 1.21  & \hspace{0.001mm} -37.49 & \hspace{0.001mm} 29.8 \hspace{0.001mm} \\ [0.5 ex]
		
		\hspace{-3.2mm}G337.92 & \hspace{0.001mm} 0.33 & \hspace{0.001mm} 0.70  & \hspace{0.001mm} 0.55  & \hspace{0.001mm} -49.02 & \hspace{0.001mm} 9.6 \hspace{0.001mm} \\ [0.5 ex]
		
		\hspace{-3.2mm}G338.93 & \hspace{0.001mm} 0.33 & \hspace{0.001mm} 0.77  & \hspace{0.001mm} 0.70  & \hspace{0.001mm} 76.36 & \hspace{0.001mm} 8.0 \hspace{0.001mm} \\ [0.5 ex]
		
		W43-MM3 & \hspace{0.001mm} 0.33 & \hspace{0.001mm} 0.76  & \hspace{0.001mm} 0.67  & \hspace{0.001mm} 84.49 & \hspace{0.001mm} 4.8 \hspace{0.001mm} \\ [0.5 ex]
		
		\hspace{-5.5mm}W51-E & \hspace{0.001mm} 0.33 & \hspace{0.001mm} 0.42  & \hspace{0.001mm} 0.33  & \hspace{0.001mm} 29.81 & \hspace{0.001mm} 6.3 \hspace{0.001mm} \\ [0.5 ex]
		
		\hspace{-3.2mm}G008.67 & \hspace{0.001mm} 0.33 & \hspace{0.001mm} 0.97  & \hspace{0.001mm} 0.78  & \hspace{0.001mm} -82.98 & \hspace{0.001mm} 15.7 \hspace{0.001mm} \\ [0.5 ex]
		
		\hspace{-3.2mm}G351.77 & \hspace{0.001mm} 0.33 & \hspace{0.001mm} 1.14  & \hspace{0.001mm} 0.89  & \hspace{0.001mm} 88.68 & \hspace{0.001mm} 20.0 \hspace{0.001mm} \\ [0.5 ex]
		
		\hspace{-3.2mm}G353.41  & \hspace{0.001mm} 0.33 & \hspace{0.001mm} 1.19  & \hspace{0.001mm} 0.91  & \hspace{0.001mm} 85.55 & \hspace{0.001mm} 21.0 \hspace{0.001mm} \\ [0.5 ex]
		
		\hspace{-1.6mm}W51-IRS2 & \hspace{0.001mm} 0.33 & \hspace{0.001mm} 0.76  & \hspace{0.001mm} 0.68  & \hspace{0.001mm} -26.43 & \hspace{0.001mm} 4.5 \hspace{0.001mm} \\ [0.5 ex]

		\hspace{-3.2mm}G010.62 & \hspace{0.001mm} 0.33 & \hspace{0.001mm} 0.65  & \hspace{0.001mm} 0.49  & \hspace{0.001mm} -67.97 & \hspace{0.001mm} 12.8 \hspace{0.001mm} \\ [0.5 ex]

		\hspace{-3.2mm}G012.80 & \hspace{0.001mm} 0.33 & \hspace{0.001mm} 1.33  & \hspace{0.001mm} 0.90  & \hspace{0.001mm} 77.41 & \hspace{0.001mm} 19.0 \hspace{0.001mm} \\ [0.5 ex]

		\hspace{-3.2mm}G333.60  & \hspace{0.001mm} 0.33 & \hspace{0.001mm} 0.76  & \hspace{0.001mm} 0.69  & \hspace{0.001mm} -23.64 & \hspace{0.001mm} 8.1 \hspace{0.001mm} \\ [0.5 ex]

		\hline
	\end{tabular}
	
	\vspace{2mm}
	\label {tab:table2}
\textbf{Notes.} Col. 1: Source name of the protoclusters. Col. 2: Spectral resolution of the observations. Col. 3: Size of the major axis of
the clean beams. Col. 4: Size of the minor axis of the clean beams. Col. 5: Position angle (PA) of the clean beams. Col. 6: Achieved RMS noises of the C$^{18}$O (2$-$1) feathered image cubes.\\
$^{a}$ Only for W43-MM1 region, the spectral resolution is 0.17 km s$^{-1}$, because this data was taken from the pilot project of \cite{2018NatAs...2..478M}.
\end{table*}



\begin{figure*}
	\centering 
	\includegraphics[width=7.7in,height=5.4in,angle=0]{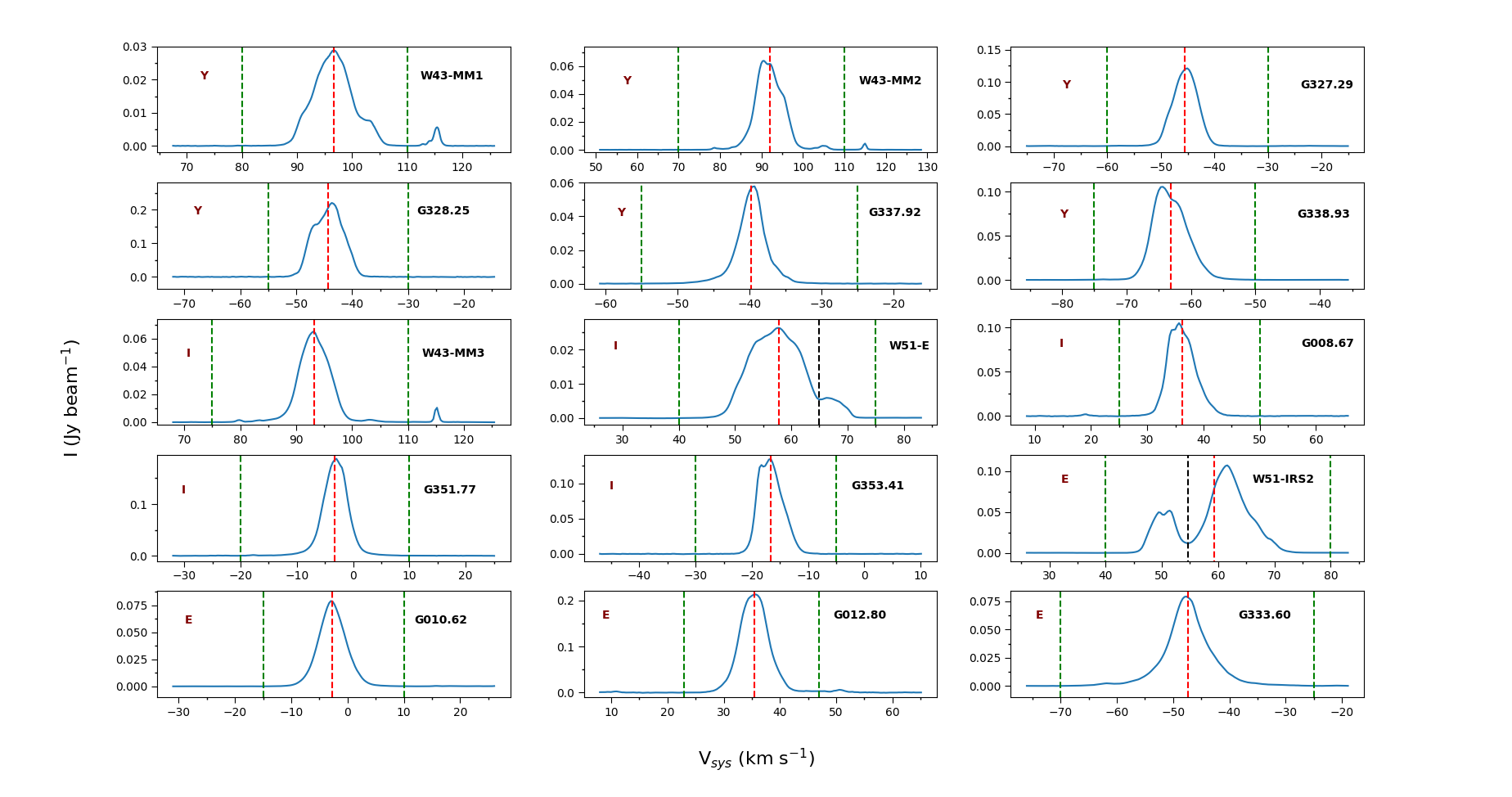}
	\caption{Average C$^{18}$O ($J$=2$-$1) spectrum for 15 different protoclusters. Two vertical green dashed lines indicate the velocity cut-off to measure the systemic velocity ($V_{\text{sys}}$). The red dashed line represents the measured $V_{\text{sys}}$ based on the C$^{18}$O ($J$=2$-$1) line. In the W51-E and W51-IRS2 regions, the black dashed line indicates the minima between the separation of the two broad clouds. Symbols Y, I and E in the figures denote young, intermediate and evolved protoclusters respectively (see Table~\ref{tab:table1}).}
		\label{fig:fig1}
	\end{figure*}

\begin{figure*}
	\centering 
	\includegraphics[width=2.5in,height=1.8in,angle=0]{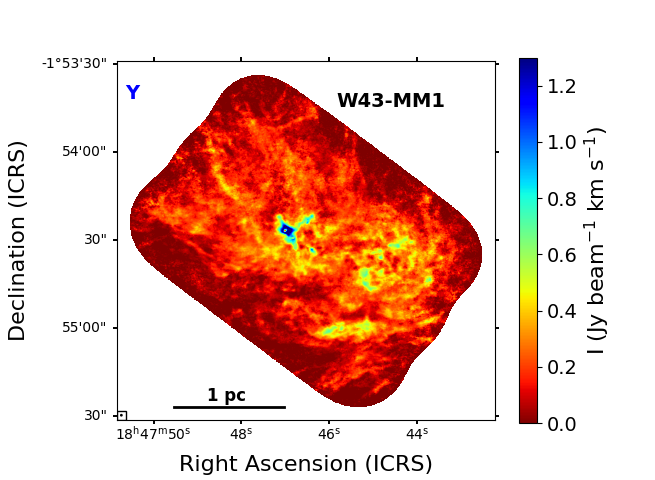}\includegraphics[width=2.5in,height=1.8in,angle=0]{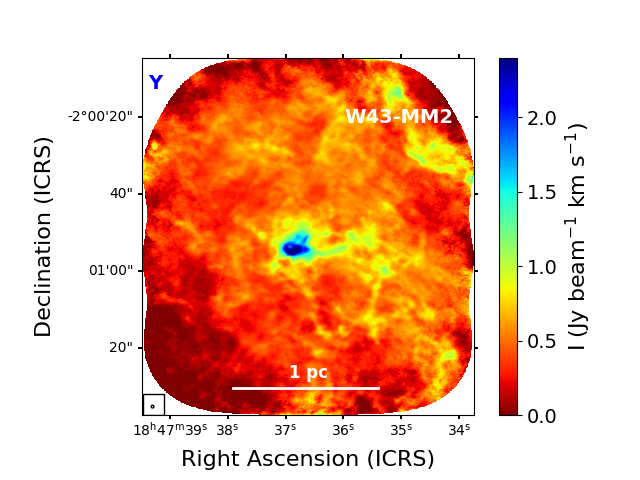}\includegraphics[width=2.5in,height=1.8in,angle=0]{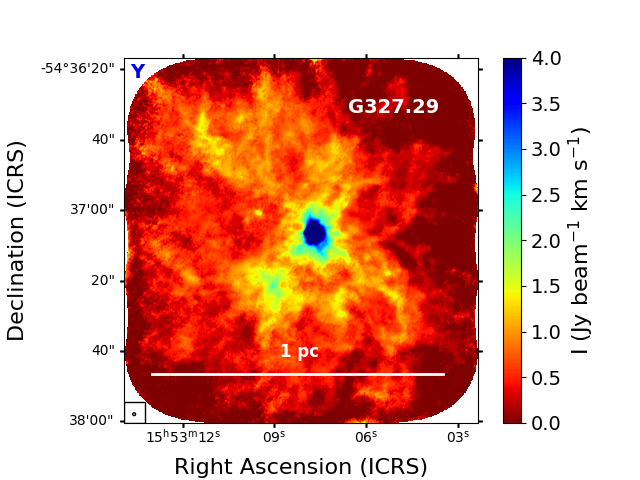}\\ \includegraphics[width=2.5in,height=1.8in,angle=0]{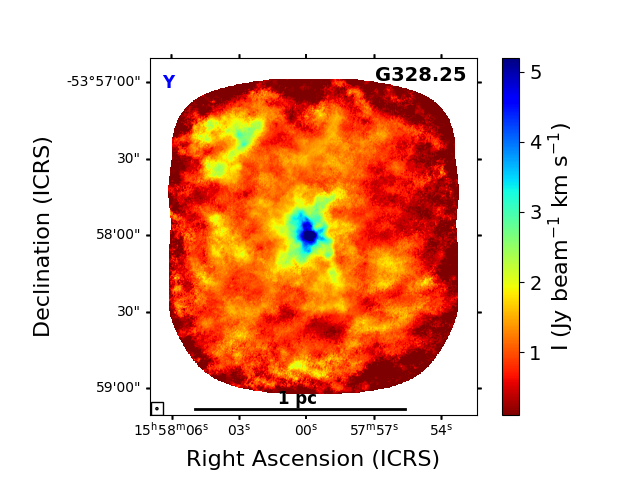}\includegraphics[width=2.5in,height=1.8in,angle=0]{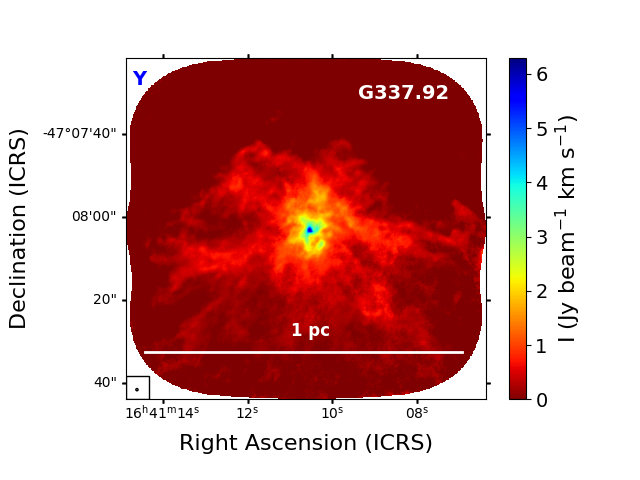}\includegraphics[width=2.5in,height=1.8in,angle=0]{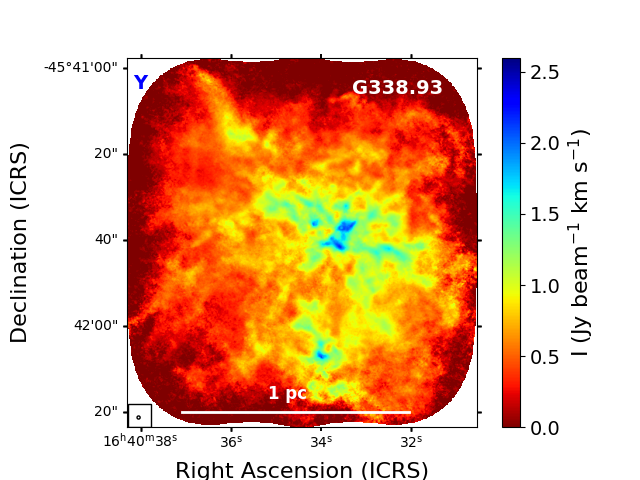}\\
	\includegraphics[width=2.5in,height=1.8in,angle=0]{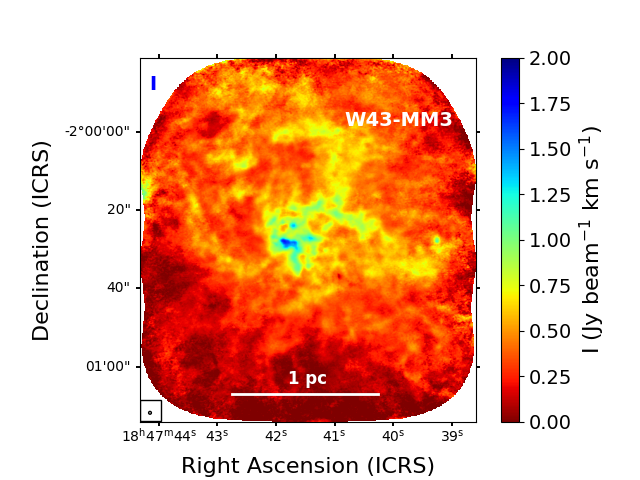}\includegraphics[width=2.5in,height=1.8in,angle=0]{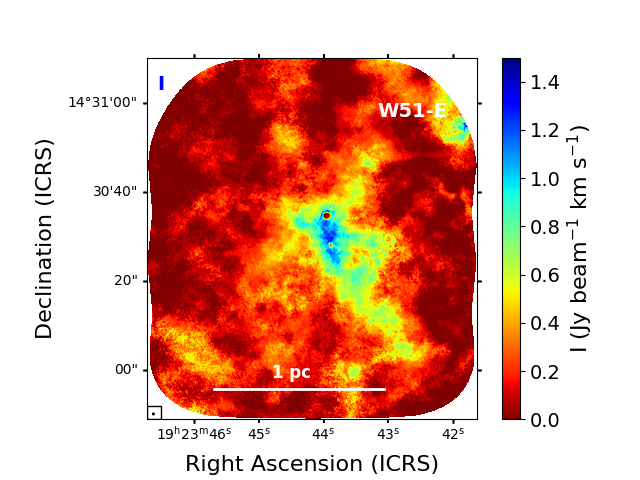}\includegraphics[width=2.5in,height=1.8in,angle=0]{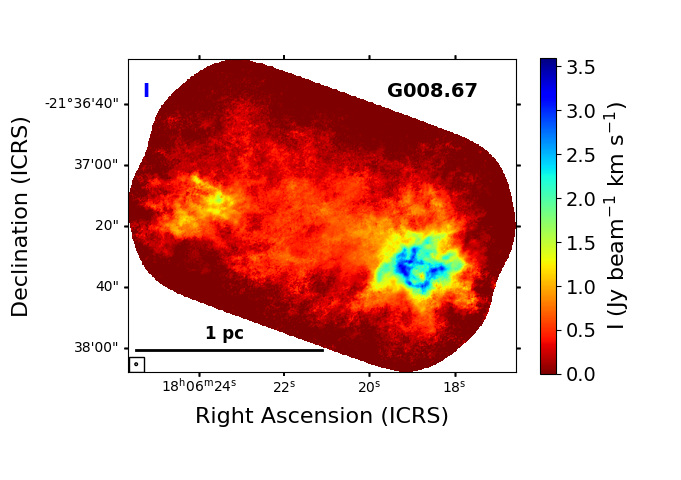}\\ 
	\includegraphics[width=2.5in,height=1.8in,angle=0]{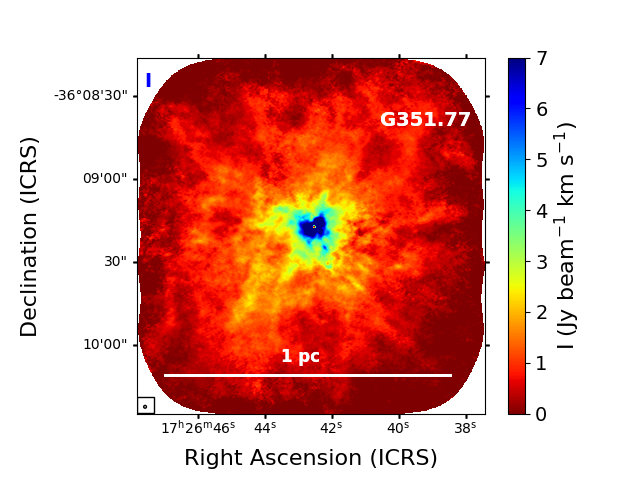}\includegraphics[width=2.5in,height=1.8in,angle=0]{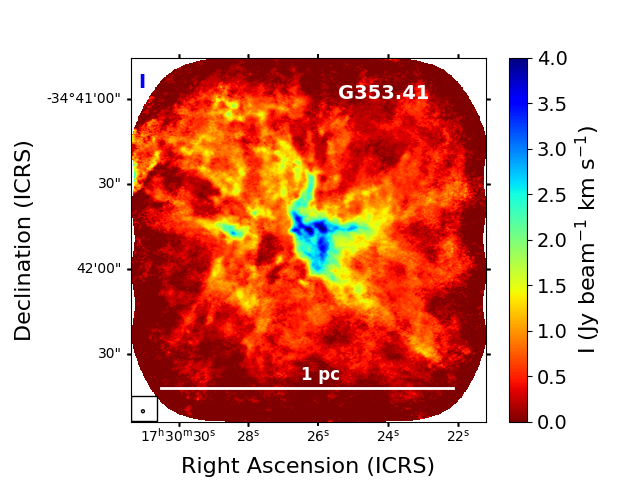}\includegraphics[width=2.5in,height=1.8in,angle=0]{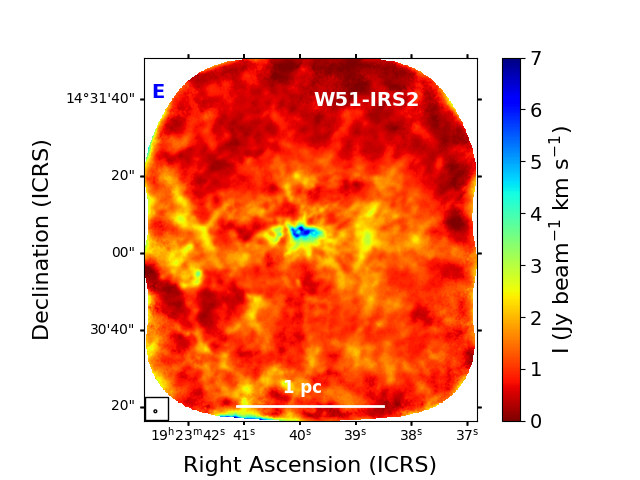}\\
	\includegraphics[width=2.5in,height=1.8in,angle=0]{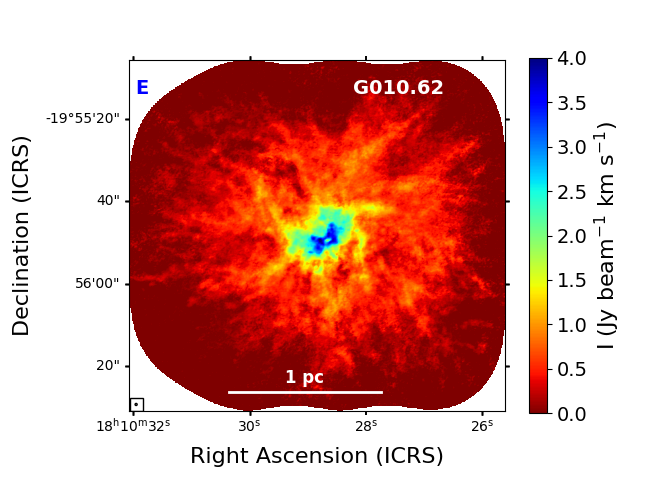}\includegraphics[width=2.5in,height=1.8in,angle=0]{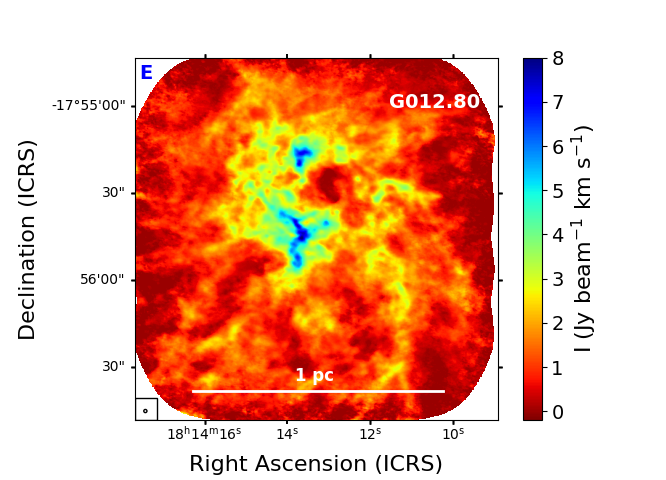}\includegraphics[width=2.5in,height=1.8in,angle=0]{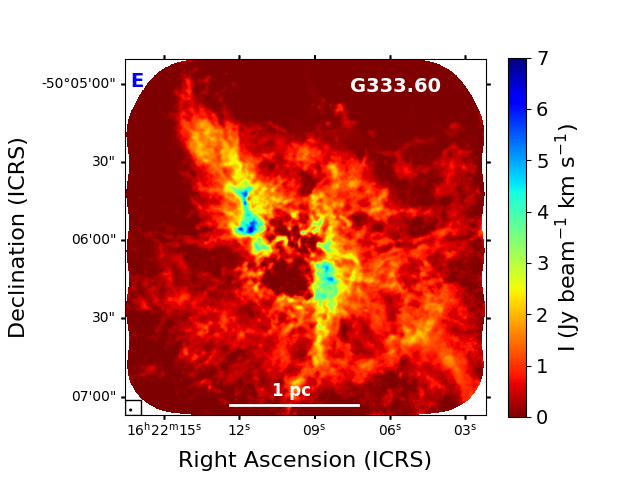}\\ 
	\caption{Integrated intensity (moment 0) maps of the C$^{18}$O ($J$=2$-$1) lines for 15 protoclusters. Symbols Y, I and E in the figures indicate young, intermediate and evolved protoclusters respectively (see Section \ref{section_0}).}
	\label{fig:fig2}
\end{figure*}

\begin{figure*}
	\centering 
	\includegraphics[width=2.5in,height=1.8in,angle=0]{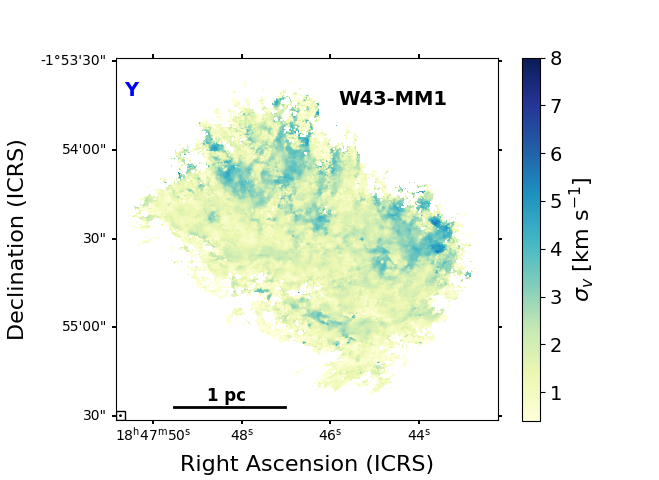}\includegraphics[width=2.5in,height=1.8in,angle=0]{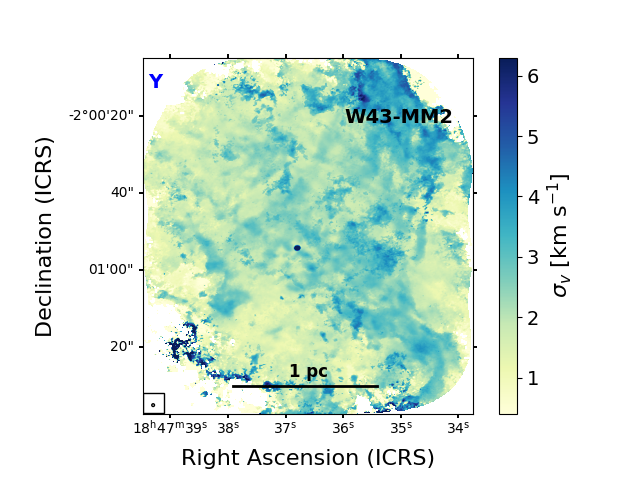}\includegraphics[width=2.5in,height=1.8in,angle=0]{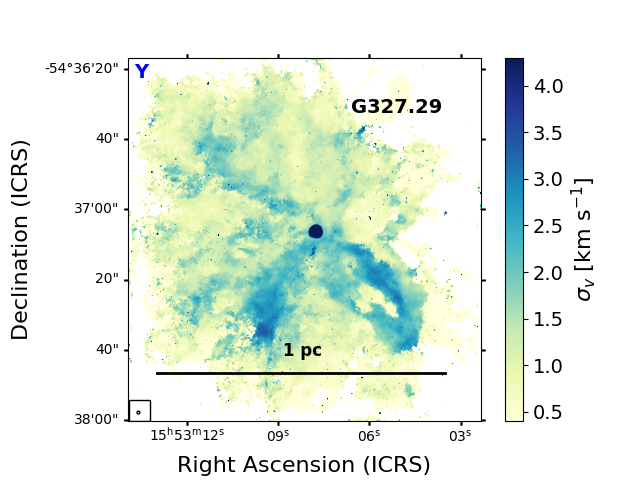}\\\includegraphics[width=2.5in,height=1.8in,angle=0]{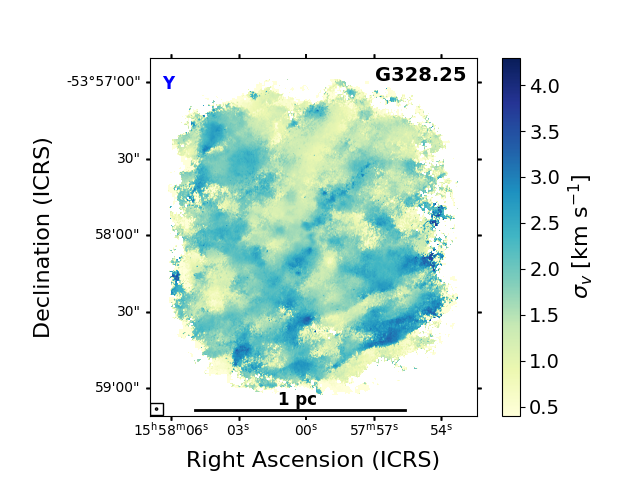}\includegraphics[width=2.5in,height=1.8in,angle=0]{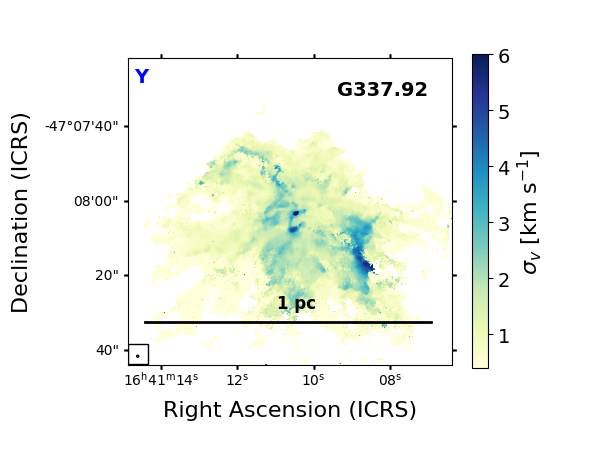}\includegraphics[width=2.5in,height=1.8in,angle=0]{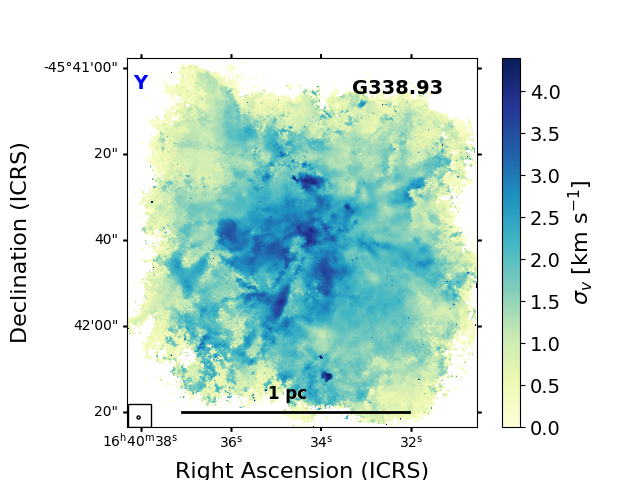}\\\includegraphics[width=2.5in,height=1.8in,angle=0]{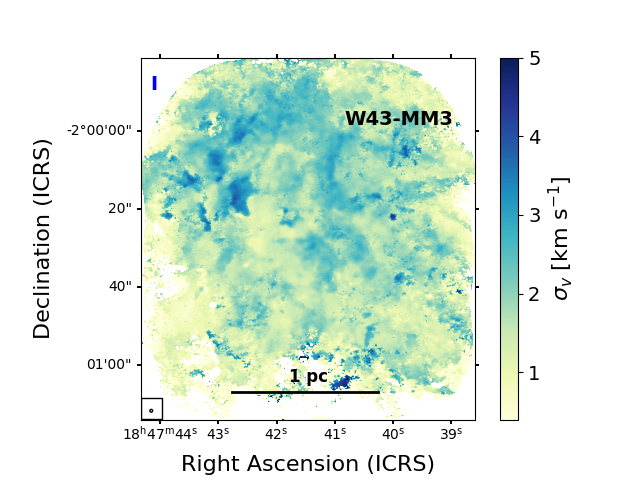}\includegraphics[width=2.5in,height=1.8in,angle=0]{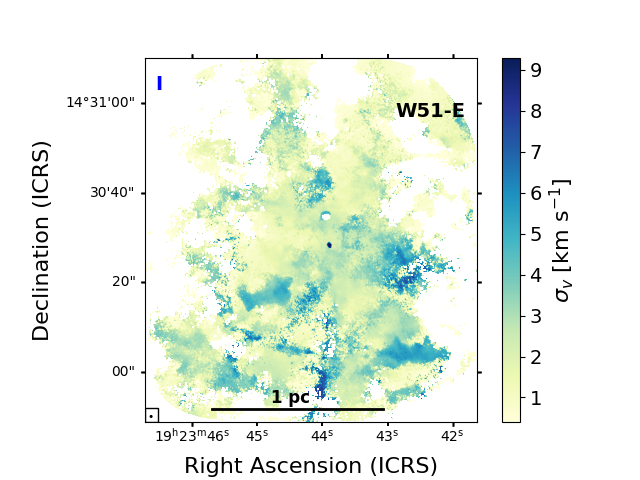}\includegraphics[width=2.5in,height=1.8in,angle=0]{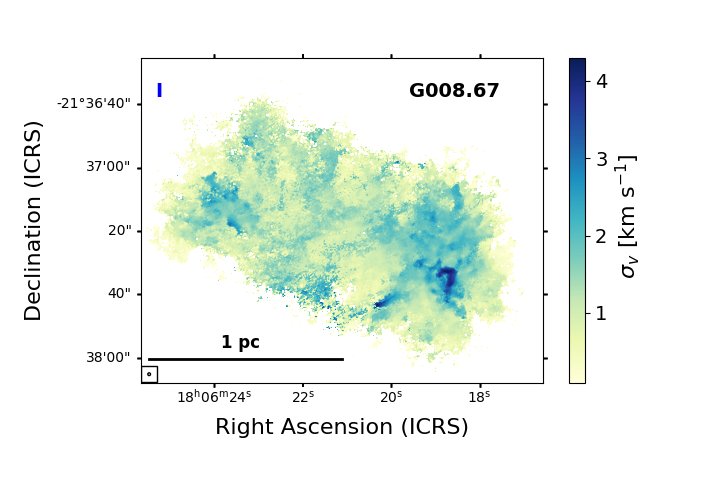}\\
	\includegraphics[width=2.5in,height=1.8in,angle=0]{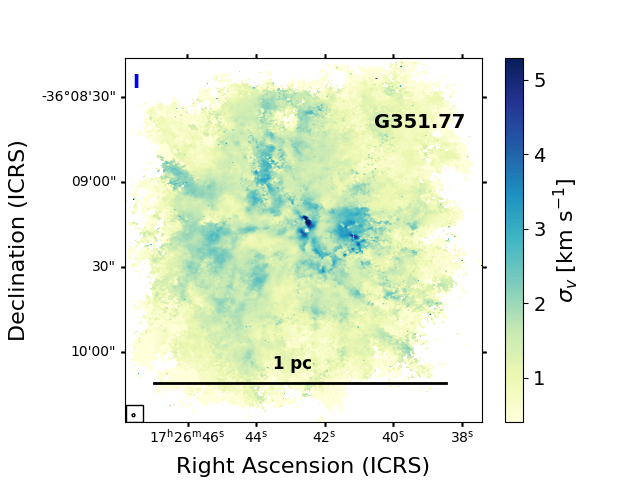}\includegraphics[width=2.5in,height=1.8in,angle=0]{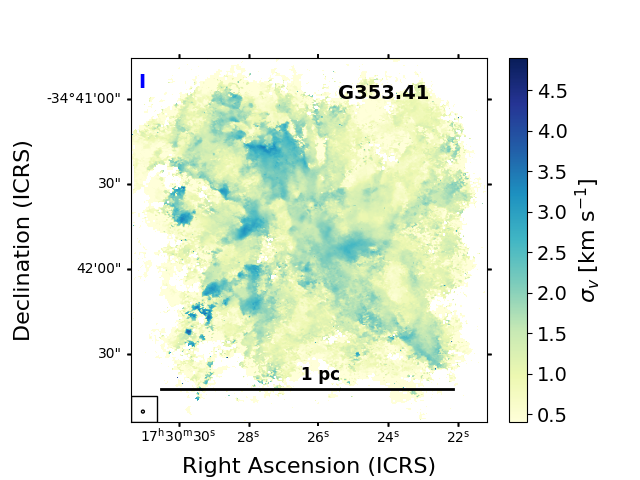}\includegraphics[width=2.5in,height=1.8in,angle=0]{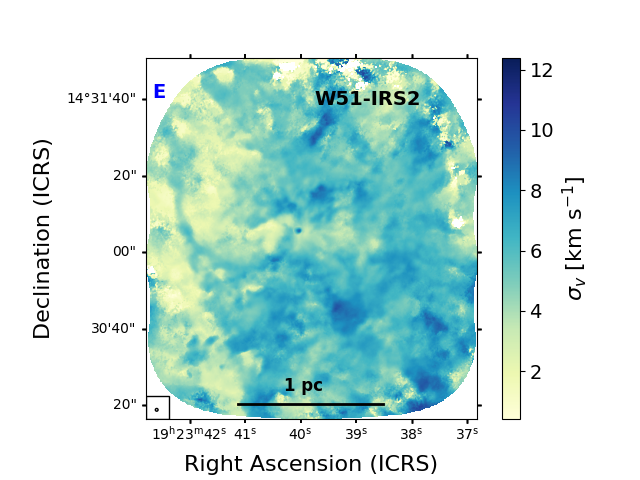}\\
	\includegraphics[width=2.5in,height=1.8in,angle=0]{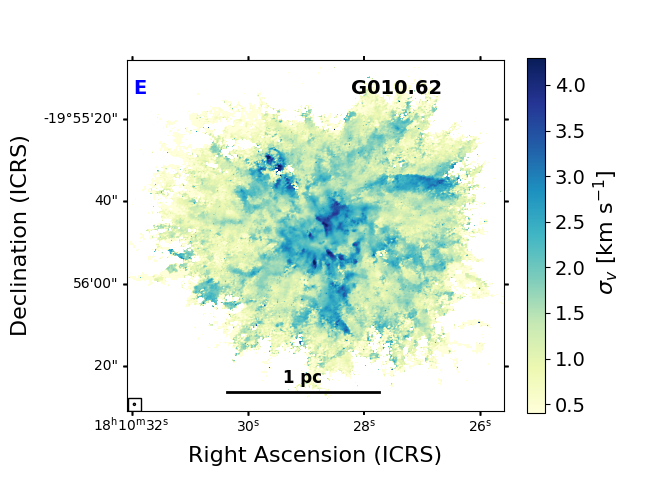}\includegraphics[width=2.5in,height=1.8in,angle=0]{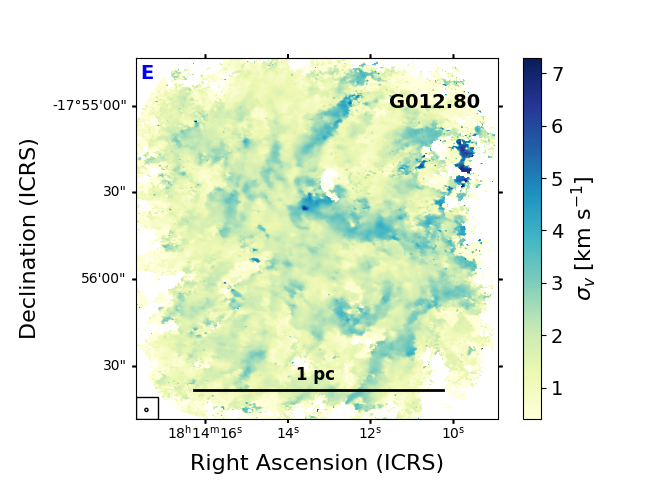}\includegraphics[width=2.5in,height=1.8in,angle=0]{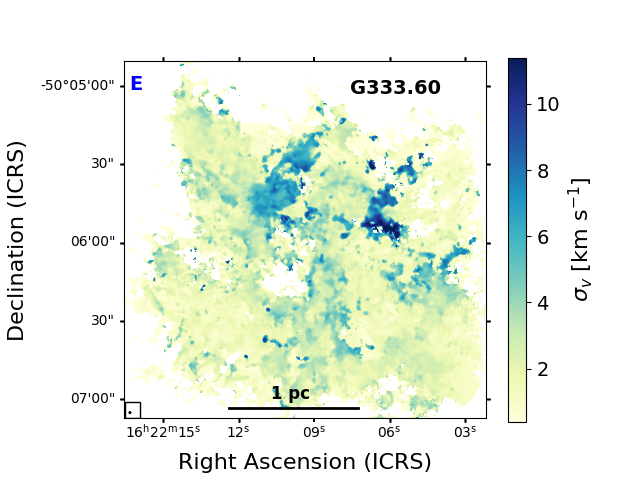}
	\caption{Integrated intensity weighted velocity dispersion (moment 2) maps of the C$^{18}$O ($J$=2$-$1) lines for 15 protoclusters. Symbols Y, I and E in the figures indicate young, intermediate and evolved protoclusters respectively (see Section \ref{section_0}).}
	\label{fig:fig3}
\end{figure*}

\vspace{-2mm}

\section{Observation and data analysis}\label{section_1}

The complete continuum and spectral line setup of the ALMA-IMF Large Program are described in \cite{2022A&A...662A...8M}. In this study, we analyze the C$^{18}$O (2$-$1) spectral line which is in Band 6 in the ALMA-IMF large program. The primary goal of this analysis is to examine the turbulence in the protoclusters using C$^{18}$O (2$-$1) line. The kinematic properties of the protoclusters will be discussed in an upcoming paper (Koley et al., in prep). To study the turbulence in the protoclusters, in addition to C$^{18}$O (2$-$1) line, we require dust temperature ($T_{\text{d}}$) maps in the protoclusters, which are obtained from the work of \cite{2024A&A...687A.217D}. We also take the molecular hydrogen column maps of the protoclusters from the same work \citep{2024A&A...687A.217D} to examine the effect of $\tau_{\nu}$ of the C$^{18}$O (2$-$1) line in the protoclusters.\\


\subsection{C$^{18}$O (J = 2$-$1) data}
Detailed information about how the data were restored from the measurement sets and what further reprocessing was performed to obtain the final data products is provided in the  work of \citet{2023A&A...678A.194C}. To produce calibrated and imaged continuum and line cubes for the full spectral line windows, \citet{2022A&A...662A...9G} developed the ALMA-IMF data pipeline. This pipeline has been written in the Common Astronomy Software Applications (CASA) environment {\color{blue}\footnote{\color{blue}https://casa.nrao.edu}}. Custom Python scripts can be found on the ALMA-IMF GitHub repository {\color{blue}\footnote{{\color{blue}https://github.com/ALMA-IMF/reduction}}}.  For imaging the line cubes, we use two script files: \texttt{line$\_$imaging.py} and \texttt{imaging$\_$parameters.py}, where the necessary parameters for the cleaning task \texttt{tclean} are contained. First, we image 7M+12M calibrated measurement (.ms) files and then combine them with the total power (TP) image cube using the CASA task \texttt{feathering}. We iteratively adjust the cleaning parameters to optimize the final imaging results. We use multiscale cleaning using the  \texttt{deconvolver = $'
$multiscale$'$} to capture all the structures from small to diffuse large scales and use the primary beam limited mask through  the parameters \texttt{usemask} and \texttt{pbmask}. We note that, regarding the multiscales, we use four to five scales in geometric progression. After increasing the number of scales by more than five,  we do not observe any flux improvement in the final image cube. In some cases, it has also been observed that processes diverge when too many scales are added during the cleaning process. The first scale is set at 0, the second scale is approximately equal to the beam size (in terms of the pixel unit), and three further scales are then added in geometric progression. The clean beams of C$^{18}$O ($J$=2$-$1) lines  of these 15 protoclusters and the RMS noises achieved from the feathered cubes are listed in Table \ref{tab:table2}. In addition to these, we adjust various parameters, including \texttt{threshold}, and \texttt{cyclefactor}.  For threshold we use 3$\sigma$ value and for \texttt{cyclefactor} we use 3 and sometimes if cleaning artifacts arise we use 4. Additionally, we set the pixel size between 1/3 and 1/5 of the minor axis of the clean beam of the deconvolved image cube, which is most effective for cleaning. {\color{black}After producing the 7m+12m data cube, we first smooth the image cube with a common beam across all channels using the CASA task \texttt{imsmooth} where we set \texttt{kernel=$'$commonbeam$'$}. Subsequently, we subtract the continuum from the image cube using the CASA task \texttt{imcontsub} with \texttt{fitorder}=0. To subtract the continuum, we use the line-free channels on both sides of the spectral line.\\

After obtaining the 7M+12M continuum subtracted image cubes, we combine them with the total power (TP) data cubes using the CASA task \texttt{feathering}. In task \texttt{feathering}, we insert the high resolution 7M + 12M data in the parameter \texttt{highres} and insert the low-resolution total power (TP) data in the parameter \texttt{lowres}. For the others, we retain the default parameters.} In one region (G012.80), negative bowls are still present after feathering the 7M+12M and TP data. In such a case, we modify the pbmask after checking the negative bowls in the feathered image cube and the 1.3 mm continuum. As there is no continuum towards this portion of the image cube, we confirm that this negative feature is primarily the result of the cleaning artifacts. Following masking of the area (for the few channels where negative features are present), user mask is used to clean the 7M + 12M image cube. In the end, while comparing the final image cube with the earlier one, we notice a slight improvement in terms of the negative bowls. Consequently, for this G012.80 region, we use the final feathered image cube obtained from the manual cleaning. \\

\vspace{-2mm}
\subsection{Hydrogen column density [$N$($\text{H}$$_{2})$] and dust temperature ($T_{\text{d}}$) maps}
In addition to  C$^{18}$O ($J$ = 2$-$1) line, molecular hydrogen column density [$N$(H$_{2})$] and dust temperature ($T_{\text{d}}$) maps have been taken from the work of  \cite{2024A&A...687A.217D}. These maps were obtained from point process mapping (PPMAP) analysis, which is based on Bayesian statistics. Using the prior information on the opacity index ($\kappa_{\nu}$), dust temperature ($T_{\text{d}}$), etc., and using the resulting spectral energy distribution (SED), the average properties along the line of sight are obtained \citep{2006AJ....132.1789M,2015MNRAS.454.4282M,2017MNRAS.471.2730M}. In their analysis, \cite{2024A&A...687A.217D} used the 1.3 mm continuum, SOFIA / HAWC + (53 $\mu$ m, 89 $\mu$ m, and 214 $\mu$ m), APEX / SABOCA (350 $\mu$ m), and APEX / LABOCA (870 $\mu$ m) data sets. {\color{black}The resulting image cubes for the molecular hydrogen column density map [$N$($\text{H}$$_{2}$)] and dust temperature ($T_{\text{d}}$) maps are at angular resolution of 2.5$''$.} For details, we refer to the work of \cite{2024A&A...687A.217D}

\section{Average spectra of C$^{18}$O ($J$ = 2$-$1)}\label{section_2}
In Fig. \ref{fig:fig1}, we show the average spectra (over the entire field of view) of C$^{18}$O ($J$=2$-$1) lines for 15 protoclusters. From these spectra, we characterize the systemic velocity ($V_{\text{sys}}$) for these regions, which are weighted by intensity (see Table \ref{tab:table1}). 
In Fig. \ref{fig:fig1}, we indicate the velocity cuts with green dashed lines for each spectrum. We consider the velocity range up to these velocity cuts on both sides of the spectrum to calculate the systemic velocity ($V_{\text{\text{sys}}}$). We indicate the measured intensity-weighted systemic velocity ($V_{\text{\text{sys}}}$ = $\sum_{i}^{} I_{\text{i}} \text{v}_{\text{i}}/\sum_{i}^{} I_{\text{i}}$, where $i$ is the channel number) for each protocluster with a red dashed line. We note that for calculating $V_{\text{\text{sys}}}$, we consider values that are greater than 3 times the RMS noise ($\sigma_{\text{rms}}$).  The null-to-null velocity width for most of the spectra ranges from $\sim$ 15 to $\sim$ 30 km s$^{-1}$. For W43-MM1, W43-MM2 and W43-MM3, we observed cloud components at $\sim$ 80 km s$^{-1}$, $\sim$ 115 km s$^{-1}$, which were reported as diffuse extended clouds of the W43 complex \citep{2011A&A...529A..41N,2014A&A...571A..32M}. For this reason, we ignore these clouds while calculating the systemic velocity ($V_{\text{sys}}$) of the protoclusters. In the W51-IRS2 region, we observe two clouds below and above $\sim$ 55.0 km s$^{-1}$, which we have marked with a black dashed line in Fig. \ref{fig:fig1}. Due to the presence of two prominent clouds in the average spectra, the $V_{\text{sys}}$ value is not at the peak of the spectra; rather it is slightly shifted from the peak. Likewise, in the W51-E region, we notice two broad clouds below and above 65.0 km s$^{-1}$, which we have also marked with a black dashed line in Fig. \ref{fig:fig1}. Earlier studies reported these as interacting clouds in W51 complex \citep{2010ApJS..190...58K,2017arXiv170206627G}. For this reason, we have taken these clouds into account. The systemic velocities in these 15 regions have already been reported by \cite{2022A&A...662A...8M} based on other surveys conducted in these regions. As we study the C$^{18}$O (2$-$1) line, we calculate the systemic velocities ($V_{\text{sys}}$) in these areas independently. We compare the values with the velocities reported by \cite{2022A&A...662A...8M} and notice that for most regions the difference is below $\sim$ 2 km s$^{-1}$. However, \cite{2022A&A...662A...8M} mentioned the same $V_{\text{sys}}$ values for the protoclusters W43-MM1, W43-MM2, and W43-MM3, which are at $+$ 97 km s$^{-1}$. Here we obtain +92.0 and +93.2 km s$^{-1}$ for the W43-MM2 and W43-MM3 regions. Likewise, for the W51-E and W51-IRS2 regions that reside in the W51 complex, we obtain two different velocities at +57.8 and +59.3 km s$^{-1}$, respectively. Earlier, \cite{2023A&A...678A.194C} examined the $V_{\text{sys}}$ in the 15 protoclusters from the mean line-of-sight values of the DCN (3$-$2) cores. They also observed similar differences in $V_{\text{sys}}$ between W43-MM1, W43-MM2, and W43-MM3 protoclusters residing in the W43 main complex, and W51-E and W51-IRS2 protoclusters residing in the W51 complex.c\\

\section{Moment maps of C$^{18}$O ($J$ = 2$-$1) line emission}\label{section_3}

We show the integrated intensity (moment 0) and integrated intensity-weighted velocity dispersion (moment 2) maps for 15 protoclusters in Figs. \ref{fig:fig2} and \ref{fig:fig3}.  We also show the integrated intensity-weighted peak velocity map (moment 1) for 15 protoclusters in the Appendix \ref{Appendix0}. For calculating the moment maps, we take all the values in each pixel that are above 4.5$\sigma_{\text{rms}}$ values. We adopt this cutoff to aviod artefact in the moment maps. In Appendix \ref{Appendix1}, we describe how we calculate the noise value for each pixel. In the W43-MM1 region, we observe a central ridge-shaped structure that may have been the result of the interaction of clouds \citep{2016A&A...595A.122L}. In the W51-E region, we notice a filament-like structure from south-west (SW) to north-east (NE) direction. In the G008.67 region, we notice two structures. One main structure is in the SW direction in the field of view, and another one is in the NE direction, which is relatively less prominent compared to the former one. Likewise, in the G353.41 region, we notice a filament-like structure from the SW to NE direction \citep{2024A&A...689A..74A}. In the G333.60 region, we observe two filament-like structures and maybe these structures are formed due to the feedback of the central H {\sc ii} region \citep{2024ApJS..274...15G}. In the G010.62 region, we notice spiral arm-like structures which were also observed in the moment 0 map of the DCN (3-2) line \citep{2023A&A...678A.194C}. This phenomenon is possibly due to the result of rotation \citep{2017A&A...597A..70L}. In the moment 2 maps in Fig. \ref{fig:fig3}, we also notice a complex structure in the W51-IRS2 region, and the velocity dispersion is relatively high compared to the other regions. For most of the regions, the velocity dispersion values for most pixels lie between 1 and 3 km s$^{-1}$, whereas in W51-IRS2, the values are between 8 and 10 km s$^{-1}$. This is possibly due to separate clouds in the same spatial position, which artificially increases the velocity dispersion in the pixels. This is supported by the average spectrum of the C$^{18}$O line in the W51-IRS2 region (see Section \ref{section_2}).  This effect may also be caused in other regions such as G338.93, the central position of G327.29, and the central area of the G010.62 region. For this reason, we calculate the velocity dispersion from the model spectra (obtained after decomposing the spectra into multi-Gaussian components) for analyzing the spatial distribution of $M_{\text{s}}$ in these regions (see Section \ref{section_4}).

\section{Estimation of turbulence}\label{section_4}

We study the turbulence in the protoclusters using two different methods. The first is to use sonic Mach number ($M_{\text{s}}$) analysis and second one is using size-linewidth relation. To calculate the spatial distribution of $M_{\text{s}}$ in the protoclusters, we require the spatial distribution of the kinetic temperature ($T_{\text{k}}$) of the gas. As we do not have direct information of $T_{\text{k}}$, which is obtained from the analysis of the ammonia spectral line \citep{2006A&A...447..929P,2023A&A...674A..46W}, we assume that the system is in local thermodynamic equilibrium (LTE) and the gas temperature ($T_{\text{k}}$) and the dust temperature ($T_{\text{d}}$) are almost the same. This assumption is considerable within the protoclusters and has been considered in previous studies \citep{2010ApJ...715...18S,2019MNRAS.490.4489S,2022ApJ...936...80S}.  We therefore first smooth the C$^{18}$O (2$-$1) image cubes into the same $T_{\text{d}}$ images, which are at 2.5$''$ resolution, and then regrid to the same $T_{\text{d}}$ cubes, so that both C$^{18}$O (2$-$1) and $T_{\text{d}}$ image cubes become identical. After that, we decompose the pixel-wise C$^{18}$O (2$-$1) spectra using the \texttt{Gausspy+} module. To obtain identical image cubes, we use the Python module \texttt{radio\_beam} and CASA task \texttt{imregrid}.  We note that the spectral resolution in the W43-MM1 region is 0.17 km s$^{-1}$, whereas in the rest of the regions it is 0.33 km s$^{-1}$. Therefore, we first smoothed the spectral axis of the W43-MM1 region, so the spectral resolution is 0.34 km s$^{-1}$, which is similar to the other cubes. We then fit the pixel-wise spectra into multi-Gaussian components. For extraction of the structure from the position-position-velocity (PPV) cube, we also follow the same procedure in the W43-MM1 region. In the following section, we discuss the fitting procedure using \texttt{Gausspy+} module. \\

\subsection{Decomposition of spectral profile}\label{sub_section5.1}

\texttt{Gausspy+} module is based on the automated Gaussian decomposition method (AGD) based on machine learning \citep{2019A&A...628A..78R}.  This module is a superior version of the original \texttt{Gausspy} module \citep{Lindner2015L}.  Using this technique, we can set the maximum jump in the number of components between two neighboring pixels, so that certain jumps in the number of components do not occur. This parameter is called \texttt{max\_jump\_comps} and in our case we set it to 2. Two smoothing parameters are there called \texttt{decompose.alpha1} and \texttt{decompose.alpha2} and these parameters are calculated automatically using some observed spectra. We can set how many spectra it takes to find out these parameters, which is called \texttt{training.n\_spectra}. These two smoothing parameters determine the smoothness of the spectrum during the decomposition process. {\color{black}{It is important to note that smoothing is spectral smoothing instead of spatial smoothing. It is necessary to make spectral smoothing and derivatives (up to four orders) of the spectra for finding the peak positions of the decomposed components accurately. The original unsmooth spectra are then decomposed based on the guessing of the peak positions from the smoothed spectra. }} For different types of spectra, e.g., weak and narrow components, it is better to use these two parameters. Otherwise, one smoothing parameter is sufficient to find the components.\\

{\color{black} The F$_{1}$ score parameter in the fitting procedure measures the accuracy of the decomposition in the training set. We first mask the pixels whose signal-to-noise ratio (SNR) is below 10 and check the accuracy factor F$_{1}$ from those spectra. Next, we examine spectra with SNRs between 10 and 20 and between 20 and 30. We notice that for spectra whose SNR $<$ 10, the accuracy factor drastically drops below 65\%. Thus, we first mask pixels whose SNR is less than 10. We have performed this procedure rather than checking the F$_{1}$ score without any condition because in this way the fit accuracy for lower SNR data is obtained more accurately. Apart from that in some regions, we also notice that below SNR = 10, the spectra are very noisy and the \texttt{Gausspy+} module fits the spectra with one component. When we plot the spatial distribution of the velocity dispersion, we notice that the outskirts of the region have a drastically increase in velocity dispersion. We then realize that this is due to the decomposition error. Using this cutoff, there is only a very small fraction of total pixels masked in each 2.5$''$ image cube.}\\

Consequently, after masking those pixels, we fit the pixel-wise spectra into multi-Gaussian components. We finally obtain the peak intensity, center velocity, and full-width-half maximum (FWHM) for each component. We note that the center velocity and the FWHM are in channel number units. Therefore, it is necessary to convert these into velocity units. We can restrict the maximum number of components during the fitting using the parameter \texttt{max\_ncomps}. In the work of \cite{2019A&A...628A..78R}, it was recommended to set it to $'$\texttt{none}$'$ or to add any value with caution. However, if the line has a major optical depth problem ($\tau_{\nu} >>$ 1), after adjusting it to $'$\texttt{none}$'$, it will unnecessarily include multiple components for a single spectrum.  {\color{black}Therefore, it is important to examine whether the C$^{18}$O (2$-$1) line has a major optical depth ($\tau_{\nu}$) effect. Optical depth of C$^{18}$O (2$-$1) line is obtained empirically as a function of the density of the molecular hydrogen column density [$N$(H$_{2}$)] from the work of \cite{2022ApJ...936...80S}, which is $\text{log}_{10}(\tau_{\text{C}^{18}\text{O}})$ = 0.6 ($\text{log}$$_{10}$ [$N$(H$_{2}$)] -24.0). This relationship was obtained from the study of the infrared dark cloud (IRDC) G014.492-00.13. After assuming the same excitation temperature ($T_{\text{ex}}$) and the filling factor ($f$) of both  C$^{17}$O (2$-$1) and C$^{18}$O (2$-$1) lines and taking the relationship of the optical depth between C$^{17}$O (2$-$1) and C$^{18}$O (2$-$1) lines, they fitted pixel-wise spectra and compared the brightness temperature ($T_{\text{B}}$) of the species. From that, they obtained the optical depth map of C$^{18}$O (2$-1$) line. This empirical law was then derived based on the column density map of the hydrogen molecule [$N$(H$_{2}$)] and the optical depth ($\tau_{\nu}$) of the C$^{18}$O (2$-$1) line. From this law, it is found that for $N$(H$_{2}$) = 10$^{23.5}$ cm$^{-2}$, $\tau_{\nu}$ = 0.50 and for $N$(H$_{2}$) = 10$^{24}$ cm$^{-2}$, $\tau_{\nu}$ = 1.0. Consequently, from the column density maps of hydrogen molecules [$N$(H$_{2}$)] of these 15 protoclusters, we also examine the number of pixels with $N$(H$_{2}$) $>$ 10$^{23.5}$ cm$^{-2}$ or $\tau_{\nu}$ $>$ 0.5 and $N$(H$_{2}$) $>$ 10$^{24}$ cm$^{-2}$ or $\tau_{\nu}$ $>$ 1.0. For the case of $\tau_{\nu}$ $>$ 0.5, we notice that for most protoclusters, it is below 6\%, however, in two cases G327.29 and G012.80, these are 12\% and 17\%, respectively. In the case of $\tau_{\nu}$ $>$ 1.0, we notice that for most protoclusters it is below 2\%, however, in the cases of G012.80 and W51-E, these are 7.8\% and 3.0\% respectively. Thus, we assume that the optical depth ($\tau_{\nu}$) does not have a significant impact on the multi-Gaussian fitting. This module contains several other parameters, e.g., \texttt{significance}, \texttt{snr\_noise\_spike}, \texttt{refit\_rchi2}. We have varied these parameters and checked whether these changes affect the final results. However, we did not find any changes in the decomposed spectra. {\color{black}In our analysis, we set the minimum FWHM equal to 2 (in channel number units), which is the minimum criteria for proper sampling of a Gaussian profile.}  We also check whether noise has an effect on the number of decomposed spectra in these regions, which we discuss in the Appendix \ref{Appendix2}. In addition, we also show some model spectra in Appendix \ref{Appendix3} in the G333.60 region which are obtained from the \texttt{Gausspy+} module.}\\

\begin{figure*}
    \centering
    \includegraphics[width=0.4\linewidth]{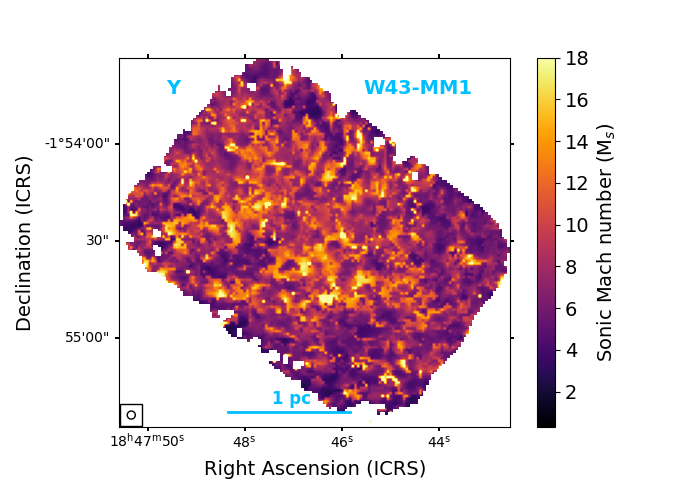}
    \includegraphics[width=0.4\linewidth]{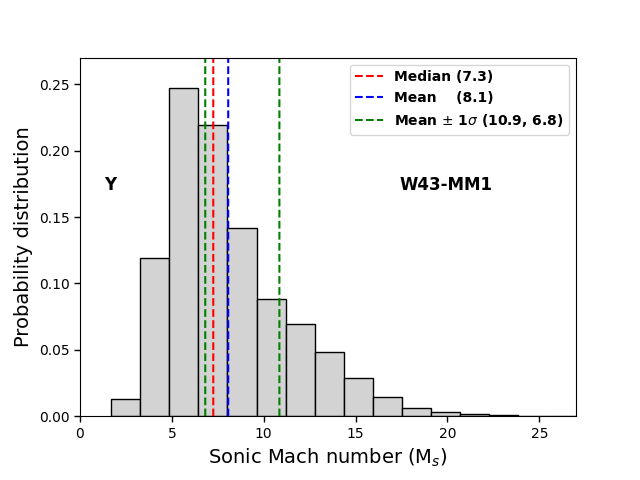}
    \caption{Left: Sonic Mach number map in the W43-MM1 protocluster. Right: Histogram distribution of the sonic Mach number shown on the left panel. Symbol Y in the figures indicate the young protocluster (see Section \ref{section_0}). We display the remainder of the 15 protoclusters in Appendix~\ref{A:sonic}.}
    \label{fig:fig4}
\end{figure*}

\begin{figure}
    \centering
    \includegraphics[width=1.0\linewidth]{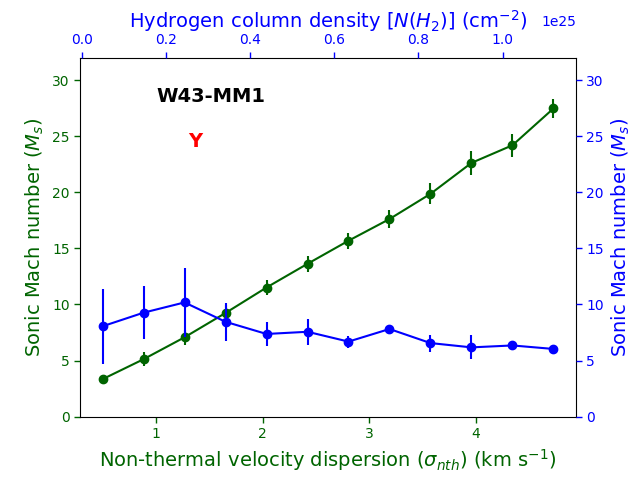}
    \caption{Correlation between (a) sonic Mach number ($M_{\text{s}}$) and non-thermal velocity dispersion ($\sigma_{\text{nth}}$) (green color) and (b) sonic Mach number ($M_{\text{s}}$) and hydrogen column density [$N({\text{H}_{2}})$] (blue color) for W43-MM1. Symbol Y in the figure indicates the young protocluster (see Section \ref{section_0}). We present the reminder of the 15 protoclusters in Appendix~\ref{A:correlation_column_density}.}
    \label{fig:fig5}
\end{figure}

\begin{figure*}
    \centering
     
    \includegraphics[width=2.35in,height=1.85in,angle=0]{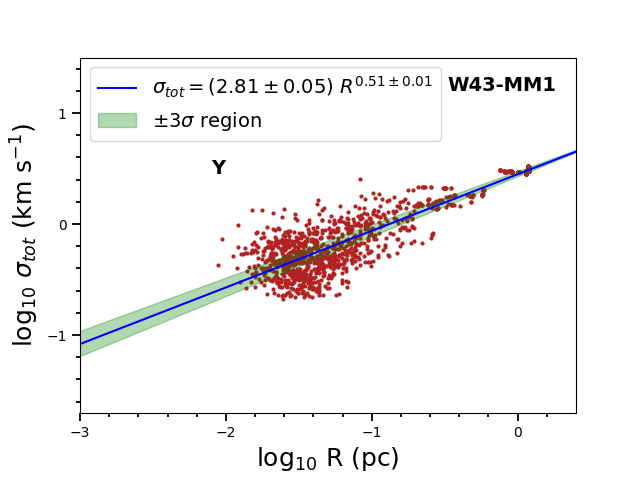}\includegraphics[width=2.35in,height=1.85in,angle=0]{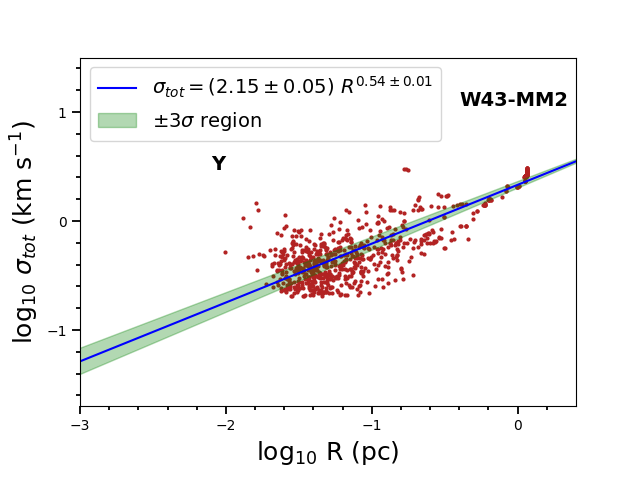} \includegraphics[width=2.35in,height=1.85in,angle=0]{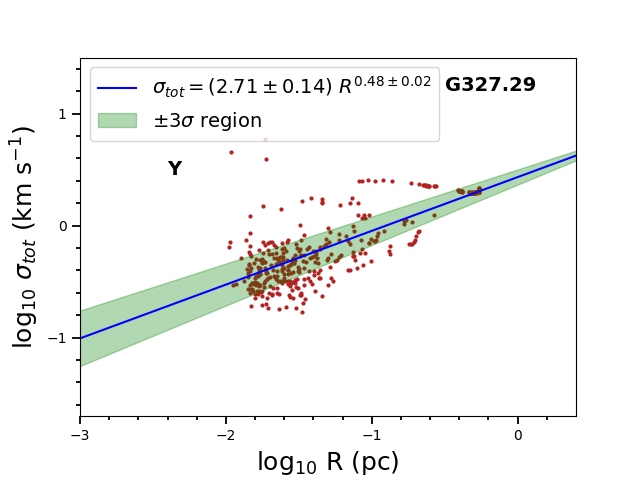}\\
	\includegraphics[width=2.35in,height=1.85in,angle=0]{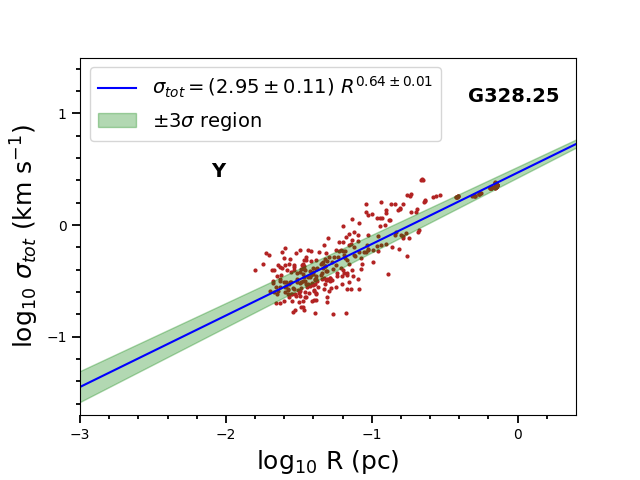}
    \includegraphics[width=2.35in,height=1.85in,angle=0]{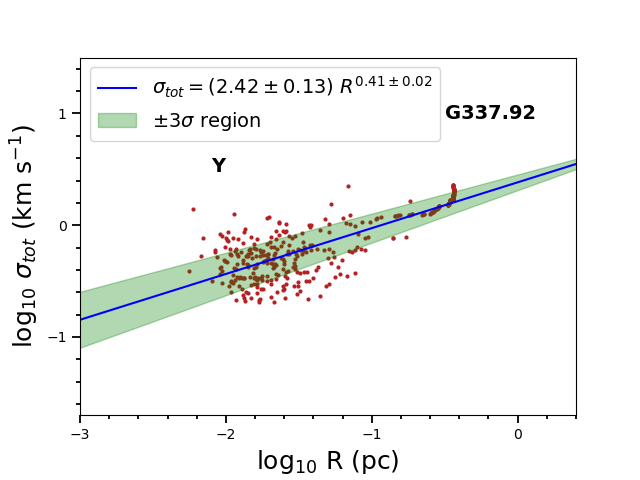}
    \includegraphics[width=2.35in,height=1.85in,angle=0]{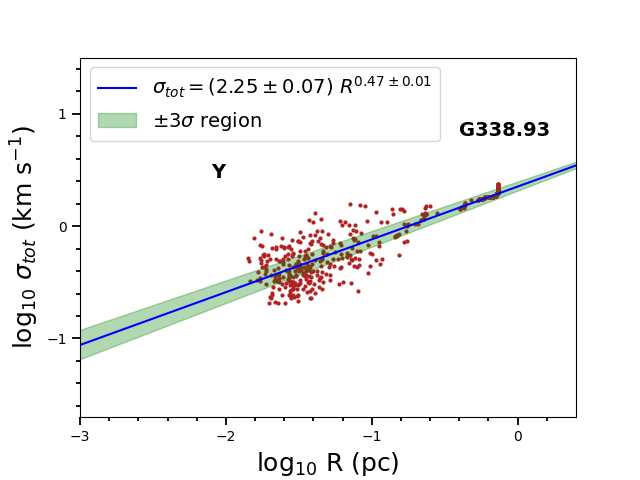}\\
    \includegraphics[width=2.35in,height=1.85in,angle=0]{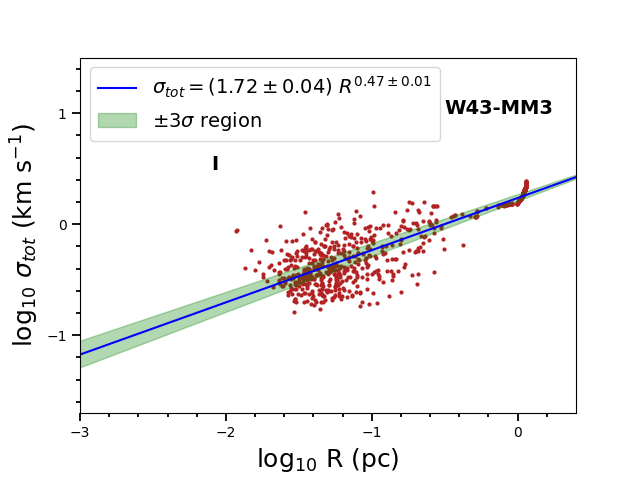}
    \includegraphics[width=2.35in,height=1.85in,angle=0]{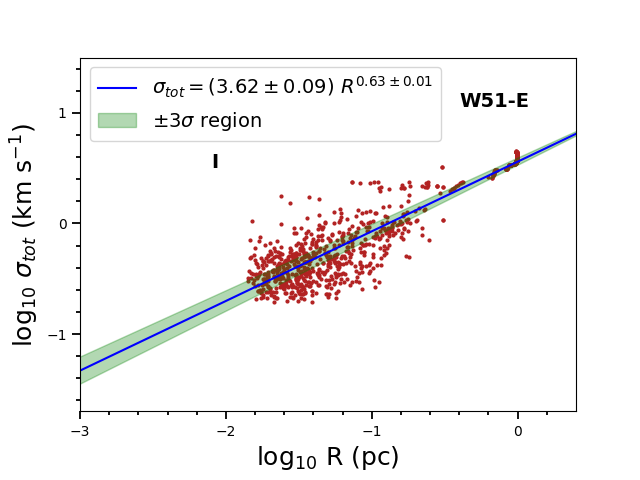}
    \includegraphics[width=2.35in,height=1.85in,angle=0]{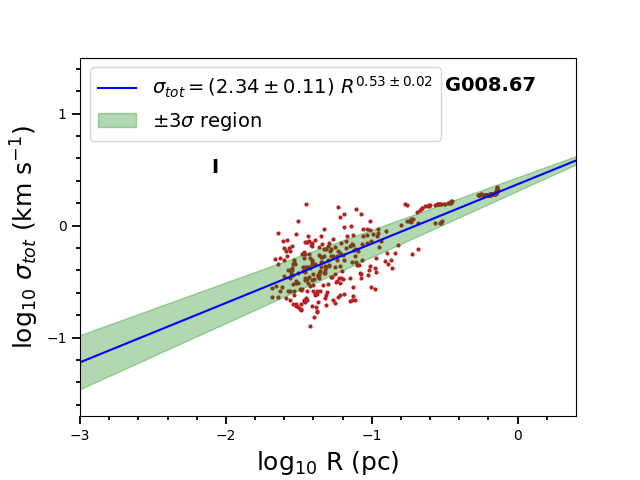}\\
    \includegraphics[width=2.35in,height=1.85in,angle=0]{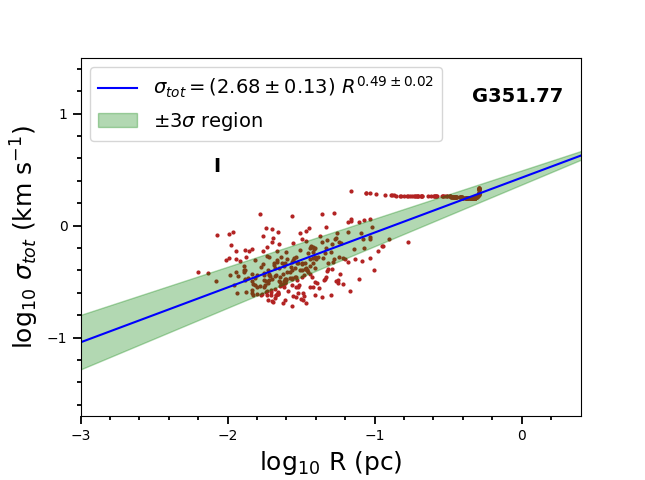}
    \includegraphics[width=2.35in,height=1.85in,angle=0]{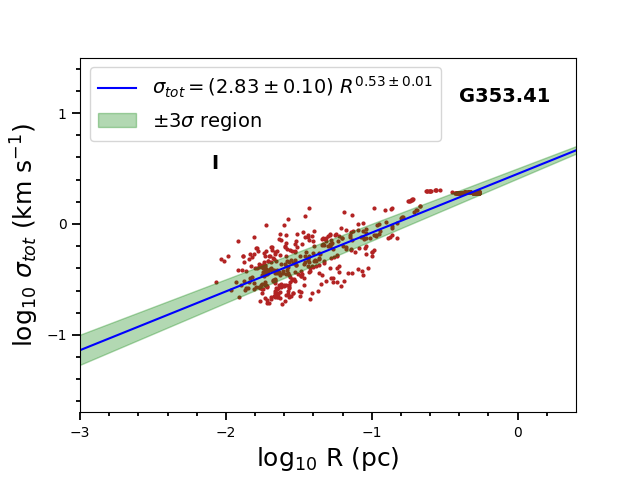}
    \includegraphics[width=2.35in,height=1.85in,angle=0]{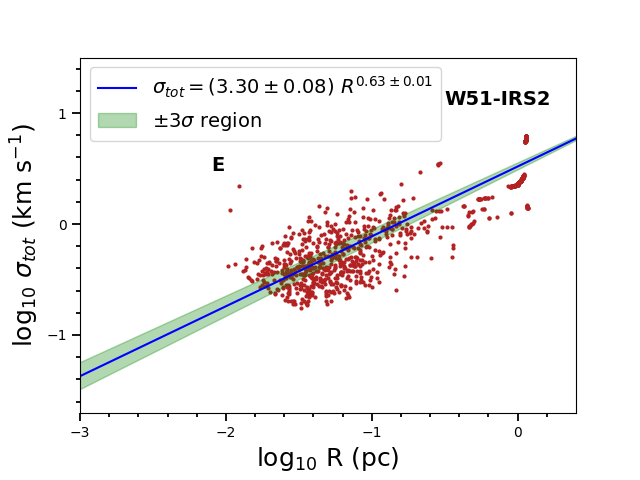}\\
    \includegraphics[width=2.35in,height=1.85in,angle=0]{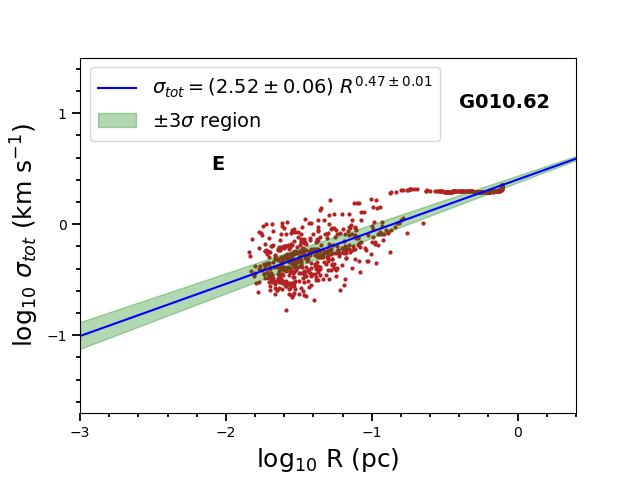}
    \includegraphics[width=2.35in,height=1.85in,angle=0]{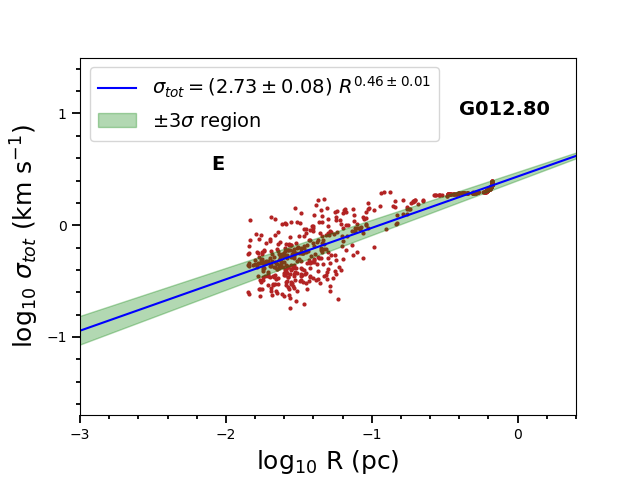}
    \includegraphics[width=2.35in,height=1.85in,angle=0]{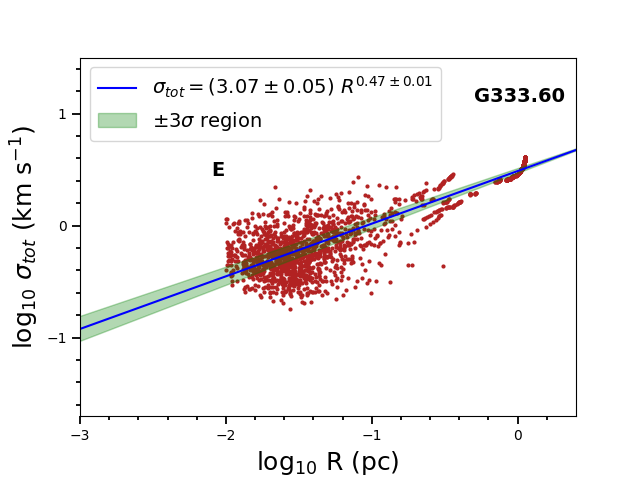}
   \caption{Correlation between velocity dispersion ($\sigma_{\text{tot}}$) and the plane-of-sky projected radius ($R$) for 15 protoclusters. The red dots represent structures derived using the \texttt{astrodendro} module. The solid blue lines represent the fitted lines for the correlation and the green shaded areas represent the $\pm$3$\sigma$ regions around the mean fitted values.  Symbols Y, I and E in the figures indicate young, intermediate and evolved protoclusters respectively (see Section \ref{section_0}).}
   \label{fig:fig6}
\end{figure*}


\subsection{Sonic Mach number ($M_{s}$) distribution}\label{sonic_mach_number}

After obtaining the pixel-wise decomposed components, we calculate the effective velocity dispersion ($\sigma_{\text{eff}}$) for each pixel using the formula mentioned in Appendix \ref{Appendix5}. We would like to mention that presence of two isolated components in a pixel artificially increases the velocity dispersion in that pixel, and so is the values for $M_{\text{s}}$. Thus, we use a different formula for calculating the velocity dispersion in each pixel, which is described in the Appendix \ref{Appendix5}. This formula is based on the weight of each component according to their integrated intensity. If multiple Gaussian components are present in a pixel after decomposition of the spectrum, we calculate the effective velocity dispersion ($\sigma_{\text{eff}}$) for that pixel based on the integrated intensity of the components.  After calculating the $\sigma_{\text{eff}}$ values in each pixel and taking the temperature information from the dust temperature ($T_{\text{d}}$) image, we measure the 1-D non-thermal velocity dispersion ($\sigma_{\text{nth}}$) using the following formula:

\begin{equation}
  \hspace{26mm}\sigma_{\text{nth}} =\sqrt{\sigma_{\text{eff}}^{2}- \sigma_{\text{th}}^2,}  
\end{equation}

where $\sigma_{\text{th}}$ is the thermal velocity dispersion equal to $\sqrt{k_{\text{B}}T_{\text{k}}/m_{c^{18}\text{O}}}$}; $k_{\text{B}}$ is the Boltzmann constant = 1.38 $\times$ 10$^{-16}$ erg K$^{-1}$; $T_{\text{k}}$ is the kinetic temperature; ${m_{c^{18}\text{O}}}$ is the mass of the C$^{18}$O molecule = 30$m_{\text{H}}$, where $m_{\text{H}}$ = 1.67 $\times$ 10$^{-24}$ g. In our analysis, we assume $T_{\text{d}}$ = $T_{\text{k}}$ (see Section \ref{section_4}). {\color{black}We note that the values of $T_{\text{d}}$ that we use in our analysis is the molecular hydrogen column density weighted temperature for a single line-of-sight. Likewise, $\sigma_{\text{eff}}$ that we calculate for each pixel is also the column density weighted velocity dispersion for a single line-of-sight. Thus, we finally obtain the gas mass-weighted Mach number ($M_{\text{s}}$) by the formula}:

\begin{equation}\label{eqn: eqn 1}
  \hspace{26mm} M_{\text{s}} = \frac{\sqrt{3}~\sigma_{\text{nth}}}{c_{\text{s}}},    
\end{equation}

where $c_{\text{s}}$ is the isothermal sound speed equal to {$\sqrt{k_{\text{B}} T_{\text{d}}/\mu m_{\text{H}}}$}; $\mu$ is the mean molecular weight of the gas = 2.35$m_{\text{H}}$ \citep{2013A&A...550A.135A,2020A&A...642A..68S,2020ApJ...896..110L}. We would like to mention that, when the beam size of the telescope is relatively large, apart from turbulence, other non-thermal effects like rotation and infall can also broaden the width of the spectral line. However, in our case, the angular resolution is $\sim$ 0.05 pc (at 2.5$''$ angular resolution), which is down to the core scale. Consequently, velocity gradients caused by infall or rotation toward the cores are well resolved \citep{2024A&A...689A..74A,2024arXiv241009843S}. On the other hand, C$^{18}$O line traces a relatively low density regime compared to other high density tracers such as DCN or N$_{2}$H$^{+}$. Thus, infall toward the cores will not significantly affect the linewidth of the C$^{18}$O line. However, the large-scale velocity gradient may have a significant effect on the linewidth. For that we check the large-scale velocity gradient in the protoclusters on the position-velocity (PV) diagram. We notice that only in the two protoclusters G008.67 and W43-MM1, large velocity gradient is observed in the C$^{18}$O (2$-$1) line. We calculate the velocity gradient for these two protoclusters and obtain 1.65 $\pm$ 0.01 km s$^{-1}$ pc$^{-1}$ for the G008.67 protocluster and 2.99 $\pm$ 0.01 km s$^{-1}$ pc$^{-1}$ for the W43-MM1 protocluster. The detailed calculation for the velocity gradient is discussed in Appendix \ref{Appendix6}. The distance of the G008.67 protocluster is $\sim$ 3.4 kpc and the distance of the W43-MM1 protocluster is $\sim$ 5.5 kpc. Consequently, 2.5$''$ corresponds to 0.04 pc and 0.06 pc in the G008.67 and W43-MM1 protoclusters, respectively. Now for the $\sim$ 1.5 km s$^{-1}$ velocity dispersion of the C$^{18}$O (2$-$1) line, it only affects 4 to 11\% of the line width. As a result, we assume that the nonthermal component is caused primarily by turbulence rather than large-scale rotation or infall. In addition, we also check whether beam smoothing to 2.5$''$ resolution has any significant effect on the Mach number analysis. We notice that there is no noticeable effect of beam smoothing. This is further supported by the negligible impact of the large-scale velocity gradient on line width. We discuss this in detail in Appendix \ref{Appendix7}.\\

In Fig. \ref{fig:fig4}, we show the spatial distribution and histogram plots of $M_{\text{s}}$ for 15 protoclusters. {\color{black} We note that the values of $M_{\text{s}}$ for all these protoclusters range between $\sim$ 1 and $\sim$ 25 with median values ranging from $\sim$ 5 to $\sim$ 8 and mean values between $\sim$ 5 and $\sim$ 9. The pattern of the histogram plot of $M_{\text{s}}$ is quite similar in all regions. It shows a peak within $\sim$ 4 to $\sim$ 7 and extends up to $\sim$ 25. Although the pattern is quite similar, we, however, did not find any significant increase or decrease in the mean and median values of $M_{\text{s}}$ according to the evolutionary stage of the protoclusters.
We also examined whether the distribution of Mach numbers changes after excluding the  H {\sc ii} emission and SiO (5$-$4) outflow regions. However, we did not find any noticeable changes in Mach numbers. As these regions are quite small compared to the field of view, it does not affect the distribution of the Mach number. Moreover, it supports our findings that Mach numbers do not vary according to the evolutionary stage of the protocluster. In two protoclusters, G337.92 and G351.77, there appears to be a slight concentration of $M_{\text{s}}$ in the central region. However, in other cases, we did not find any such effect. The fluctuation of the Mach number is there in the entire field of view.

\begin{figure*}
    \centering
    
    \includegraphics[width=1.0\linewidth]{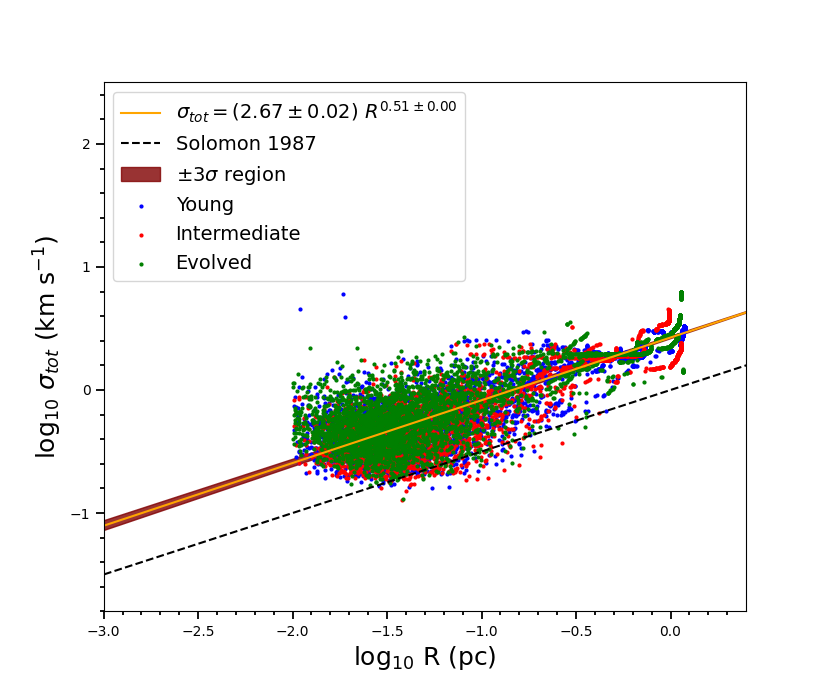}
    \caption{Correlation between velocity dispersion ($\sigma_{\text{tot}}$) and the plane-of-sky projected radius ($R$) for all 15 protoclusters. Blue, red and green dots represent the structures for young, intermediate and evolved protoclusters derived using the \texttt{astrodendro} module. The solid orange line represents the fitted line for the correlation and the maroon shaded areas represent the $\pm$3$\sigma$ regions around the mean fitted value. Black dashed line obtained from the earlier work of \cite{1987ApJ...319..730S}.}
    \label{fig:fig7}
\end{figure*}

{\color{black}

In Fig. \ref{fig:fig5}, we show the correlation between $M_{s}$ and $\sigma_{\text{nth}}$, and the correlation between $M_{\text{s}}$ and $N(\text{H}_2)$ for W43-MM1 protocluster. The rest are shown in Appendix~\ref{A:correlation_column_density}. For studying the correlation, we first grouped the data in 12 equal intervals according to the values of $\sigma_{\text{nth}}$ and $N(\text{H}_2)$. We note that the number of pixels gradually decreases with increasing $N(\text{H}_2)$. Thus, the mean and $\sigma$ values are calculated from fewer data points for higher $N(\text{H}_2)$. For example, only about 6\%  pixels contain $N(\text{H}_2)$ $>$ 10$^{23.5}$ cm$^{-2}$ in almost all regions, while it decreases to below 1.8\%  when $N(\text{H}_2)$ $>$ 10$^{24}$ cm$^{-2}$.  From the figure we see that there is no significant correlation between $M_{\text{s}}$ and $N(\text{H}_2)$ except for the G351.77 region, where a slight correlation is tentatively observed.\\

Here also, we compare the non-thermal line width and Mach number between C$^{18}$O line and DCN cores to check whether the turbulence inside the protoclusters varies towards the dense cores. For that, we obtain the velocity dispersion and temperature values of the DCN cores from \cite{2023A&A...678A.194C}. For our analysis, we only consider the cores fitted with a single Gaussian as mentioned in \cite{2023A&A...678A.194C}. The number of DCN cores for each protocluster is listed in Table \ref{tab:table3}. We calculate the mean (and 1 $\sigma$ dispersion) values of the non-thermal velocity dispersion ($\sigma_{\text{nth, DCN}}$) and the Mach number ($M_{\text{s, DCN}}$) of DCN cores for each protocluster. We list the values of the non-thermal velocity dispersion ($\sigma_{\text{nth, C$^{18}$O}}$) and the Mach number ($M_{\text{s, C$^{18}$O}}$) calculated from C$^{18}$O line  in Table \ref{tab:table3}. For calculating $\sigma_{\text{nth, DCN}}$, we take the value of the mass of the DCN molecule as 30 m$_{\text{H}}$, same as for C$^{18}$O molecule. We then calculate $M_{\text{s, DCN}}$ from Eqn. \ref{eqn: eqn 1}, the same as was calculated for the C$^{18}$O line (see Table \ref{tab:table3}). We would like to mention that spectral line observations are broadened by the spectrometer response function \citep{Koch_2018}. Earlier studies corrected the channel resolution effect using the formula: $\Delta_{\text{ch}}$/2$\sqrt{\text{2 ln 2}}$, where $\Delta_{\text{ch}}$ is the channel resolution of the observation \citep{2020ApJ...896..110L,2023A&A...675A..53B}. This effect is negligible for the C$^{18}$O line where the line width is relatively large compared to the DCN line. However, when we check it for the DCN line, we also observe that the average non-thermal velocity dispersion decreases only by 0.02 km s$^{-1}$ and the average Mach number decreases only by 0.10 from their uncorrected values. Therefore, this correction does not have a significant impact on the final results. From the values of $\sigma_{\text{nth, DCN}}$, we notice that the mean values range between 0.42 and 0.96 and the mean values of $M_{\text{s, DCN}}$ vary between 2.2 and 5.5. We did not find any significant difference according to the evolutionary stage of the protoclusters. Likewise, the mean $\sigma_{\text{nth, C$^{18}$O}}$ ranges from 0.91 to 1.58 and the mean $M_{\text{s,C$^{18}$O}}$ ranges from 5.3 to 8.7. On average, the mean value of the $\sigma_{\text{nth, DCN}}$ and $M_{\text{s, DCN}}$ is almost half that of the C$^{18}$O values. It indicates that, although turbulence is high in the low density regimes inside the protoclusters, it decreases towards the high density dense cores.\\

\begin{table*}
	\caption{Comparison between the turbulence in the protoclusters obtained from C$^{18}$O (2$-$1) lines and the turbulence of the DCN (3$-$2) cores inside the protoclusters.}
	\begin{tabular}{ c c c c c c c }
        \hline
		Source &  \hspace{5mm} Evolutionary &  \hspace{5mm} No. of DCN cores &  \hspace{5mm} $\sigma_{\text{nth, DCN}}$    &\hspace{5mm}   $\sigma_{\text{nth, C$^{18}$O}}$   &   \hspace{5mm}  $M_{\text{s, DCN}}$  & \hspace{5mm} $M_{\text{s, C$^{18}$O}}$\\ [0.5 ex]
		name &    \hspace{5mm} Stage        &  \hspace{5mm} (Single spectra) &  \hspace{5mm} (km s$^{-1}$)                 & \hspace{5mm} (km s$^{-1}$)                       &  \hspace{5mm}                         &   \\ [0.5 ex]
        \hline
		W43-MM1   & \hspace{5mm}  Y & \hspace{5mm} 15 & \hspace{5mm}  0.42 (0.23) & \hspace{5mm} 1.43$^{0.50}_{0.22}$    & \hspace{5mm}  2.2 (1.1)      & \hspace{5mm}  8.1$^{2.8}_{1.3}$   \\ [0.5 ex]    
        W43-MM2   & \hspace{5mm}  Y & \hspace{5mm} 18 & \hspace{5mm}  0.57 (0.22) & \hspace{5mm} 1.45$^{0.42}_{0.22}$    & \hspace{5mm}  3.6 (1.5)      & \hspace{5mm}  8.5$^{2.4}_{1.3}$   \\ [0.5 ex]   
        G327.29   & \hspace{5mm}  Y & \hspace{5mm} 13 & \hspace{5mm}  0.54 (0.15) & \hspace{5mm} 1.11$^{0.42}_{0.17}$    & \hspace{5mm}  3.1 (0.9)      & \hspace{5mm}  6.5$^{2.4}_{1.0}$   \\ [0.5 ex]
        G328.25   & \hspace{5mm}  Y & \hspace{5mm} 3  & \hspace{5mm}  0.83 (0.52) & \hspace{5mm} 1.04$^{0.43}_{0.15}$    & \hspace{5mm}  4.0 (2.5)      & \hspace{5mm}  6.0$^{2.4}_{0.9}$   \\ [0.5 ex]   
        G337.92   & \hspace{5mm}  Y & \hspace{5mm} 8  & \hspace{5mm}  0.96 (0.54) & \hspace{5mm} 1.11$^{0.65}_{0.19}$    & \hspace{5mm}  5.5 (2.7)      & \hspace{5mm}  6.6$^{3.5}_{1.1}$   \\ [0.5 ex]   
        G338.93   & \hspace{5mm}  Y & \hspace{5mm} 18 & \hspace{5mm}  0.67 (0.29) & \hspace{5mm} 1.48$^{0.41}_{0.20}$    & \hspace{5mm}  3.6 (1.4)      & \hspace{5mm}  8.6$^{2.1}_{1.2}$   \\ [0.5 ex]   
        W43-MM3   & \hspace{5mm}  I & \hspace{5mm} 9  & \hspace{5mm}  0.58 (0.30) & \hspace{5mm} 1.35$^{0.49}_{0.20}$    & \hspace{5mm}  3.1 (1.8)      & \hspace{5mm}  7.9$^{2.9}_{1.2}$   \\ [0.5 ex]   
        W51-E     & \hspace{5mm}  I & \hspace{5mm}  7 & \hspace{5mm}  0.75 (0.31) & \hspace{5mm} 1.08$^{0.42}_{0.14}$    & \hspace{5mm}  4.2 (1.7)      & \hspace{5mm}  6.5$^{2.7}_{0.8}$   \\ [0.5 ex]   
        G008.67   & \hspace{5mm}  I & \hspace{5mm} 9  & \hspace{5mm}  0.61 (0.23) & \hspace{5mm} 1.19$^{0.42}_{0.18}$    & \hspace{5mm}  3.3 (1.3)      & \hspace{5mm}  7.5$^{2.5}_{1.1}$   \\ [0.5 ex]   
        G351.77   & \hspace{5mm}  I & \hspace{5mm} 4  & \hspace{5mm}  0.57 (0.16) & \hspace{5mm} 1.24$^{0.66}_{0.21}$    & \hspace{5mm}  2.8 (0.8)      & \hspace{5mm}  7.0$^{3.7}_{1.1}$   \\ [0.5 ex]   
        G353.41   & \hspace{5mm}  I & \hspace{5mm} 14 & \hspace{5mm}  0.57 (0.29) & \hspace{5mm} 0.91$^{0.33}_{0.12}$    & \hspace{5mm}  3.0 (1.3)      & \hspace{5mm}  5.7$^{2.1}_{0.7}$   \\ [0.5 ex]   
        W51-IRS2  & \hspace{5mm}  E & \hspace{5mm} 55 & \hspace{5mm}  0.47 (0.23) & \hspace{5mm} 1.05$^{0.37}_{0.14}$    & \hspace{5mm}  2.4 (0.8)      & \hspace{5mm}  5.6$^{1.9}_{0.7}$   \\ [0.5 ex]   
        G010.62   & \hspace{5mm}  E & \hspace{5mm} 28 & \hspace{5mm}  0.43 (0.16) & \hspace{5mm} 1.43$^{0.45}_{0.22}$    & \hspace{5mm}  2.3 (0.8)      & \hspace{5mm}  8.4$^{2.7}_{1.2}$   \\ [0.5 ex]   
        G012.80   & \hspace{5mm}  E & \hspace{5mm} 37 & \hspace{5mm}  0.52 (0.29) & \hspace{5mm} 0.99$^{0.39}_{0.12}$    & \hspace{5mm}  2.7 (1.1)      & \hspace{5mm}  5.3$^{2.1}_{0.7}$   \\ [0.5 ex]   
        G333.60   & \hspace{5mm}  E & \hspace{5mm} 28 & \hspace{5mm}  0.51 (0.17) & \hspace{5mm} 1.58$^{0.72}_{0.26}$    & \hspace{5mm}  2.6 (0.7)      & \hspace{5mm}  8.7$^{4.3}_{1.3}$   \\ [0.5 ex]   
		\hline	
			
	\end{tabular}

	\vspace{2mm}
	\label {tab:table3}

\textbf{Notes.} Col. 1: Protocluster names. Col. 2: Evolutionary stages of the protoclusters. Symbols Y, I and E denote the young, intermediate, and evolved protoclusters (see Section \ref{section_0}). Col. 3: Number of DCN (3$-$2) cores which show single type Gaussian spectra \citep{2023A&A...678A.194C}. Col. 4: Mean (1$\sigma$) non-thermal velocity dispersion of DCN cores ($\sigma_{\text{nth, DCN}}$). Col. 5: Mean (1$\sigma$) non-thermal velocity dispersion of C$^{18}$O (2$-$1) line ($\sigma_{\text{nth, C$^{18}$O}}$). Col. 6:  Mean (1$\sigma$) sonic Mach number ($M_{\text{s, DCN}}$) for DCN cores.  Col 6: Mean (1$\sigma$) sonic Mach number ($M_{\text{s, C$^{18}$O}}$) for C$^{18}$O (2$-$1) line. 

\end{table*}

\begin{table*}
	\caption{Properties of the structures obtained from $\texttt{Astrodendro}$ module.}
	\begin{tabular}{  c c c c c }
        \hline
		Protocluster & \hspace{18.39mm} Evolutionary stage & \hspace{18.39mm} Number of structures & \hspace{18.39mm} $\sigma_{\text{tot}}$ = $A$ (km s$^{-1}$) ($\frac{R}{\text{pc}})^{p}$\\ [0.5 ex]
		
		name & \hspace{18.39mm} & \hspace{18.39mm} (${N}$) & \hspace{18.39mm}  {\color{black}Value of $A$ (km s$^{-1}$) and $p$} \\ [0.5 ex]      
		\hline
		
		W43-MM1  & \hspace{18mm} Y & \hspace{18mm}       {\color{black}{1330}}  & \hspace{18mm} {\color{black}2.81 $\pm$ 0.05, 0.51 $\pm$ 0.01} \\ [0.5 ex]     
		W43-MM2  & \hspace{18mm} Y & \hspace{18mm}       {\color{black}899}  & \hspace{18mm} {\color{black}2.15 $\pm$ 0.05, 0.54 $\pm$ 0.01} \\ [0.5 ex]
		G327.29  & \hspace{18mm} Y & \hspace{18mm}  \,\,\,{\color{black}429}  & \hspace{18mm} {\color{black}2.71 $\pm$ 0.14, 0.48 $\pm$ 0.02} \\ [0.5 ex]
		G328.25  & \hspace{18mm} Y & \hspace{18mm}  \,\,\,{\color{black}369}  & \hspace{18mm} {\color{black}2.95 $\pm$ 0.11, 0.64 $\pm$ 0.01} \\ [0.5 ex]
		G337.92  & \hspace{18mm} Y & \hspace{18mm}  \,\,\,{\color{black}359}  & \hspace{18mm} {\color{black}2.42 $\pm$ 0.13, 0.41 $\pm$ 0.02} \\ [0.5 ex]
		G338.93  & \hspace{18mm} Y & \hspace{18mm}       {\color{black}472}  & \hspace{18mm} {\color{black}2.25 $\pm$ 0.07, 0.47 $\pm$ 0.01} \\ [0.5 ex]    
		W43-MM3  & \hspace{18mm} I & \hspace{18mm}       {\color{black}723}  & \hspace{18mm} {\color{black}1.72 $\pm$ 0.04, 0.57 $\pm$ 0.01} \\ [0.5 ex]
		W51-E    & \hspace{18mm} I & \hspace{18mm}       {\color{black}887}  & \hspace{18mm} {\color{black}3.62 $\pm$ 0.09, 0.63 $\pm$ 0.01} \\ [0.5 ex]
		G008.67  & \hspace{18mm} I & \hspace{18mm}  \,\,\,{\color{black}379}  & \hspace{18mm} {\color{black}2.34 $\pm$ 0.11, 0.53 $\pm$ 0.02} \\ [0.5 ex]
		G351.77  & \hspace{18mm} I & \hspace{18mm}  \,\,\,{\color{black}402}  & \hspace{18mm} {\color{black}2.68 $\pm$ 0.13, 0.49 $\pm$ 0.02} \\ [0.5 ex]
		G353.41  & \hspace{18mm} I & \hspace{18mm}  \,\,\,{\color{black}490}  & \hspace{18mm} {\color{black}2.83 $\pm$ 0.10, 0.53 $\pm$ 0.01} \\ [0.5 ex]
		W51-IRS2 & \hspace{18mm} E & \hspace{18mm}       {\color{black}962}  & \hspace{18mm} {\color{black}3.30 $\pm$ 0.08, 0.63 $\pm$ 0.01} \\ [0.5 ex]
		G010.62  & \hspace{18mm} E & \hspace{18mm}       {\color{black}721}  & \hspace{18mm} {\color{black}2.52 $\pm$ 0.06, 0.47 $\pm$ 0.01} \\ [0.5 ex]
		G012.80  & \hspace{18mm} E & \hspace{18mm}       {\color{black}593}  & \hspace{18mm} {\color{black}2.73 $\pm$ 0.08, 0.46 $\pm$ 0.01} \\ [0.5 ex]
		G333.60  & \hspace{18mm} E & \hspace{18mm}       {\color{black}1945}  & \hspace{18mm} {\color{black}3.07 $\pm$ 0.05, 0.47 $\pm$ 0.01} \\ [0.5 ex]
		\hline
	\end{tabular}
	
	\vspace{2mm}
	\label {tab:table4}

\textbf{Notes.} Col. 1: Protocluster names.  Col. 2: Evolutionary stages of the protoclusters. Symbols Y, I and E in the denote young, intermediate, and evolved regions (see Section \ref{section_0}). Col. 3: Number of independent structures. Col. 4: Values of $A$ and $p$ obtained from correlation between velocity dispersion ($\sigma_{\text{tot}}$) and size ($R$). 

\end{table*}

\subsection{Structure decomposition}\label{s:structure-decompositio}
An alternative method to determine the nature of turbulence in molecular clouds is to study the relationship between the non-thermal line width with the size of the structures in the protoclusters \citep{2022MNRAS.511.4480L,2022MNRAS.516.1983S}. It is generally believed that interstellar molecular clouds have hierarchical complex structures that range from relatively diffuse structures to dense clumpy structures. These various structures are thought to have been formed by the cascade of eddies generated by interstellar turbulence. Hence, it is essential to extract these structures in order to understand their nature. To accomplish this, we use the Python package \texttt{astrodendro} \citep{2008ApJ...679.1338R,2012MNRAS.425..720S}, which properly decomposes the structures and provides us with various information such as velocity-dispersion ($\sigma_{\text{tot}}$), area ($S$), size ($R_{\text{obs}}$), position angle (PA), major ($R_{\text{maj}}$) and minor ($R_{\text{min}}$) sizes of the structures in the plane of the sky.  In the following, we discuss the structure extraction in detail using the \texttt{astrodendro} module.\\


Using the \texttt{astrodendro} module, the structure is extracted from the position-position-velocity (PPV) cube. After extracting these structures, it computes the intensity-weighted velocity dispersion ($\sigma_{\text{tot}}$) for each structure using the formula:\\

\begin{equation}
\hspace{26mm}\sigma_{\text{tot}} = \left[ \frac{\sum_{}I_{\text{v}}(\text{v}-\bar{\text{v}})^{2}}{\sum_{}I_{\text{v}}}  \right]^{1/2},
\end{equation}

where $\bar{\text{v}}$ is the intensity-weighted mean velocity, which is equal to {$\sum_{}I_{\text{v}} \text{v}/\sum_{}I_{\text{v}}$}. Similarly, it also estimates the major ($R_{\text{maj}}$) and minor ($R_{\text{min}}$) sizes of the structures (which are ellipses) in the plane of the sky. As input, we have to insert the PPV cube, from which it extracts the structures. 
We first calculate the noise map using the line free channels and then convert each velocity channel for each pixel into zero whose amplitude is {\color{black}$<$ 2.5$\sigma_{\text{rms}}$}. Now, the modified cube represents a cube only with significant fluxes. There are three main parameters that control the extraction of the structure. These are: \texttt{min\_value}, \texttt{min\_delta}, and  \texttt{min\_npix}. As we use the modified cube, we set the \texttt{min\_value} as close to zero. The second parameter is \texttt{min\_delta}. It determines the threshold beyond which it is considered as a structure. We take this value at {\color{black}2.5$\sigma_{\text{rms}}$}. The third parameter is \texttt{min\_npix}. This parameter denotes the minimum pixels that are required in the position-position-velocity (x,y,v) space so that it is considered as a single component. We consider minimum {\color{black}6} channels in the velocity space (v) and for the spatial dimension (x,y), we set the value  2$a$ which is equal to {$[2 \pi/4ln2].{[FWHM_{\text{x,beam}}.FWHM_{\text{y,beam}}/(pixel\_size)^{2}}$]}. Here, $FWHM_{\text{x,beam}}$ and $FWHM_{\text{y,beam}}$ are the full width at half maximum of the major and minor axes of the beam. After obtaining the $R_{\text{maj}}$ and $R_{\text{min}}$ values, we calculate the spherical radius, $R_{\text{obs}}$ which is equal to 1.91 times the effective RMS size, $R_{\text{rms}}~(\sqrt{R_{\text{maj}} R_{\text{min}} }$)  (\cite{2023ApJ...949...63O}, and references therein).  We correct the convolved beam from $R_{\text{obs}}$ to obtain the deconvolved size $R$, which is equal to $\sqrt{R_{\text{obs}}^{2} - \theta_{\text{beam}}^{2}}$.  Finally, we convert this $R$ into the spatial scale (in pc) after converting the plane-of-sky projected angle into the spatial scale using the source's distance.\\

We show the correlation between $\sigma_{\text{tot}}$ and $R$ for the 15 protoclusters in Fig. \ref{fig:fig6}. The details of the decomposed structures are mentioned in Table \ref{tab:table4}.  $\sigma_{\text{tot}}$ vs $R$ is shown in Fig. \ref{fig:fig6} on a logarithmic scale for better visual, as the upper end is crowded with a large number of data points. We fit the size-linewidth relation using the Python module \texttt{scipy.optimize.curvefit}. For all cases, the error in the power-law index is {\color{black}either 0.01 or 0.02} and the main error is in the intensity value. Errors in the intensity and power-law index values are rounded up to two decimal places. After obtaining the similar value in the power-law index error, we recheck errors especially in the power-law index using the \texttt{bootstrapping} method. For that, we run 10$^{4}$ samples in each of these regions and randomly select both the velocity dispersion and the length scale and we recover the same results as reported above.\\


{\color{black}We note that we analyzed the size-linewidth relation at the original angular resolution of the observation.} Consequently, in all of these regions, the length scales of the structures vary from $\sim$ 0.01 pc to $\sim$ 1.0 pc, with most of them lying below 0.1 pc. This indicates the existence of a relatively large number of small-scale structures inside the protoclusters. The lower and upper length scale limits of the structures are determined by the resolution of the ALMA telescope and the large scale obtained from the spatial extension of this observation, respectively. The number of decomposed structures within these protoclusters varies between  {\color{black} 359 and 1945}. In the young protoclusters W43-MM1 and  W43-MM2, the number of structures is greater than {\color{black}850}. Similarly, for protoclusters G327.29, G328.25, G337.92 {\color{black}and G338.93} the number of structures is greater than  {\color{black}350 but less than 500}. Likewise, in the intermediate protoclusters, W43-MM3 and W51-E, the number of structures is above {\color{black}700}, whereas for the other three protoclusters, G008.67, G351.77 and G353.41, the number of structures is below {\color{black}700}. However, in the evolved protoclusters W51-IRS2, G010.62, G012.80, and G333.60, the number of structures {\color{black}varies between 593 and 1945}. {\color{black}In  G333.60 region, number is structure is 1945 which is relatively large and more than twice the number of structures found in most of the other regions}. The possible reason for the relatively large number of structures in G333.60 region could be due to the feedback effect of the H {\sc ii} region, the surrounding environment inside the protoclusters is very complex and small, clumpy structures are produced as a result of this.\\

In addition, we did not observe any noticeable differences in intensity ($A$) or power-law index ($p$) according to the evolutionary stages. Values of $A$ and $p$ of the protoclusters vary between {\color{black}1.72 and 3.62} and between {\color{black}0.41 and 0.64}, respectively. We note that \texttt{Astrodendro} or any other structure extraction module exhibits a bias with respect to contaminated structures. When extracting small-scale structures (leaves), there is a contamination of large-scale structures (branch and trunk) and vice versa. As a result, large velocity dispersion values are added to the small-scale structures, and vice versa. Therefore, $A$ is higher and $p$ is lower than the original value. Thus, we can say that the fitted $A$ value represents the upper limit and the fitted $p$ value represents the lower limit of the actual value. {\color{black}Apart from the contamination issue, velocity dispersion of the large wing-like structures may reduce significantly due to the value of \texttt{min\_value} = 2.5$\sigma_{\text{rms}}$. This is also considered as a caveat in the \texttt{Astrodendro} module. We have checked this effect in Appendix \ref{A:structure_intensity_cut} for four protoclusters.} We also point out that velocity dispersion is caused by thermal and non-thermal motions. However, from the analysis of the sonic Mach number ($M_{\text{s}}$), it is evident that the contribution of the thermal velocity dispersion ($\sigma_{\text{th}}$) is very small. Thus, the thermal velocity dispersion ($\sigma_{\text{th}}$) does not have any significant impact on the intensity ($A$) and the power-law index ($p$) of the size-linewidth relation.\\

We also examined the size-linewidth relation after excluding the H {\sc ii} and SiO (5$-$4) outflow emission regions. We have taken the polygon regions of H {\sc ii} and SiO (5$-$4) from previous ALMA-IMF studies \citep{2024ApJ...960...48T,2024ApJS..274...15G}. We check this because in the H {\sc ii} region due to strong stellar radiation, the velocity dispersion is high, which may not be associated with turbulence. In addition, the expansion of the velocity in the outflow region is mainly caused by strong velocity gradients, which may not be the result of turbulence. We check this for the size-linewidth relationship as the structures are extracted from the iso-intensity contours, and thus one isolated intensity structure may cover a significant portion of the outflow or H{\sc ii} region. However, we notice that after excluding these regions, the intensity and the power-law index are similar to the full region analysis and within the 3$\sigma$ errors of the original. One primary reason is that C$^{18}$O (2$-$1) line emission spreads throughout the field of view of the protoclusters. Consequently, only a small fraction of the area is excluded and does not significantly alter the value of the intensity ($A$) and the power-law index ($p$).} \\

{\color{black}Fig. \ref{fig:fig7} shows the correlation between the size and linewidth of these 15 protoclusters. The blue dots represent structures for young protoclusters, whereas the red and green dots represent structures for intermediate and evolved protoclusters. All three types of protoclusters have the same spatial extent and velocity dispersion values. The total number of structures for young, intermediate, and evolved protoclusters is 3858, 2881, and 4219, respectively. We obtain a fit value of 2.67 $\pm$ 0.02 km s$^{-1}$ and a p value of 0.51 $\pm$ 0.00. We also plot the result obtained from the work of \cite{1987ApJ...319..730S}. They studied the PCA analysis of the CO molecule in molecular clouds and obtained the value of A = 1.0 km s$^{-1}$ and the value of p = 0.50. Although the value of $p$ is similar to our value, the value of $A$ is significantly different from our results. We assume that the discrepancy is mainly caused by the percentage of overlapping structures in the field of view.}


\section{Discussion}
\label{s:discussion}

\subsection{Sonic Mach number}

In Section~\ref{sonic_mach_number}, we studied the sonic Mach number ($M_{\text{s}}$) in the 15 massive protoclusters. The $M_{\text{s}}$ behaves similarly in all protoclusters, extending up to $\sim$25, with mean and median values ranging from $\sim$ 5 to $\sim$ 9 and $\sim$ 5 to $\sim$ 8, respectively. Such high levels of turbulence must affect the stellar-formation processes.  Moreover, as shown by \cite{2022A&A...662A...8M}, these 15 protoclusters span different evolutionary stages among the star-forming process. It seems that turbulence is maintained over time. This could be the effect of feedback processes, such as outflows, supernova explosions, or large-scale galactic interactions \citep{2006ApJ...653.1266J,2009MNRAS.392..294A,2023A&A...674A..75N}.\\

It is interesting to note that similar values of $M_{\text{s}}$ were also reported in younger regions, namely in infrared dark clouds (\citealt{2023A&A...674A..46W}, \citealt{2011A&A...533A..85L},\citealt{2015AJ....150..159D}) and in more evolved regions such as Orion B, a nearby high-mass molecular cloud \citep{2017A&A...599A..99O}. \cite{2023A&A...674A..46W} studied the infrared dark cloud G35.20-0.74 N with the Jansky Very Large Array (JVLA) using emission lines from NH$_{3}$ (1,1) to (7,7) (similar angular resolution to our $M_{\text{s}}$ analysis). If we convert their sonic Mach number for the three-dimensional system, that is, multiplying their estimates by $\sqrt{3}$, we find the mean, median, and maximum $M_{\text{s}}$ of 6.4, 4.8 and 21.0, respectively. These values are similar to those we obtain in our 15 massive protoclusters. Similarly, \cite{2015AJ....150..159D} studied infrared dark clouds (IRDC) using NH$_{3}$ (1,1) and (2,2) lines after combining JVLA and the Green Bank Telescope (GBT) (similar angular resolution to our $M_{\text{s}}$ analysis). They noticed that the turbulence inside the clumps is highly supersonic (with $M_{\text{s}}$ $\approx$ 5 $-$ 9), which may support the clumps against gravity. In addition to these studies, \cite{2017A&A...599A..99O} examined the turbulence in the nearby high-mass molecular cloud Orion B using various CO isotopes with the IRAM 30 meter telescope. The spatial resolution in this study was 0.05 pc, which is similar to our $M_{\text{s}}$ analysis. They also obtained the histogram plot of $M_{\text{s}}$ similar to our results. The mean value of $M_{\text{s}}$ is $\sim$ 6.5 and extends to $\sim$ 20. In addition, they also showed that the distribution of turbulence energy into solenoidal and compressive modes may affect star formation efficiency (SFE). The results of all of these studies, including ours, indicate that turbulence may play a significant role in the dynamics of molecular clouds. Overall, the dynamical effect of turbulence in these 15 protoclusters is beyond the scope of this work. Following the calculation of the magnetic field in these regions, by dust polarization or Zeeman measurements, we will investigate in detail the effect of these opposing forces (magnetic field and turbulence) on gravity inside the protoclusters.\\

In addition to the estimation of the Mach number in Section~\ref{sonic_mach_number}, we also study the correlation between the Mach number and the hydrogen column density [$N(\text{H}_2)$] for 15 massive protoclusters. We notice that there is no significant correlation exists between them. {\color{black}  The reason is C$^{18}$O line emission traces dense as well as relative low density regimes. The maximum amount of gas, however, appears in the low density regime after including the total power data of ALMA telescope. Thus, in the relatively low density regime probed by C$^{18}$O line does not have any significant variation in turbulence.} However, comparing the non-thermal velocity dispersion values between C$^{18}$O line and DCN cores, we note that the DCN cores have almost half the non-thermal line width and sonic Mach number compared to the C$^{18}$O line. It indicates that turbulence decreases towards dense cores. This result is in line with previous studies \citep{2020ApJ...896..110L,2022ApJ...925..144S,2023A&A...674A..46W,2023A&A...675A..53B}. For example, \cite{2022ApJ...925..144S} examined the IRDC G14.492-00.139 with relatively low density tracer C$^{18}$O (2$-$1) and several high density tracers like N$_{2}$D$^{+}$ (3$-$2), DCO$^{+}$ (3$-$2), DCN (3$-$2) towards the dense cores and noticed that low density tracer C$^{18}$O  has almost three times larger line width compared to the high density tracers mentioned above, whose kinematics are associated with dense cores. From their analysis they concluded that a low-density turbulent envelope surrounds less turbulent dense cores. Likewise, in a study by \citep{2023A&A...674A..46W} showed that most of the cores in the evolved infrared dark cloud G35.20-0.74 N located towards the local minima of the sonic Mach number. In addition, towards the high-mass molecular clouds I18308 and I19220, prestellar cores are formed where the turbulence is mostly transonic, which is low compared to the surrounding molecular clouds where the turbulence is highly supersonic \citep{2018ApJ...855....9L}. Furthermore, in massive infrared dark cloud NGC 6334S, \cite{2020ApJ...896..110L} showed that the turbulence in the dense cores is subsonic to transonic. In addition to that, in IRDC G028.37+00.07, \cite{2023A&A...675A..53B} showed that in dense cores turbulence is transonic in nature. As a consequence, turbulence may not be sufficient to support the cores against gravity unless strong magnetic fields play an important role. In nearby low-mass star-forming molecular clouds such as Taurus, where cores are well resolved, previous studies showed that turbulence decreases towards the periphery of the cores and within the deep interior of the cores, turbulence is subsonic in nature \citep{1998ApJ...504..223G,2007ApJ...671.1839S,2021A&A...648A.114C,2021ApJ...912....7P, 2022MNRAS.516..185K,2023PASA...40...53K}. {\color{black}There are two possible reasons for this. First one is the scale-dependent velocity dispersion of the turbulent eddies. The typical size of the DCN core is $\sim$ 0.015 pc \citep{2023A&A...678A.194C}, whereas the typical size of the  C$^{18}$O (2$-$1) structures is $\sim$ 0.05 pc. {\color{black}C$^{18}$O line traces dense as well as relatively low density regimes  which contain larger structures. Therefore, the median values of the size is higher in C$^{18}$O line compared to DCN line and in turn the larger velocity dispersion.}} The second one is at the periphery of the prestellar cores, when the effective magnetic Reynolds number ($R_{\text{m}}) \sim$ 1, neutral particles drift through the magnetic field, and due to friction between the neutral particles and the ions, a significant amount of turbulence dissipates \citep{2008ApJ...677.1151L,2004fost.book.....S,2019FrASS...6....5H}. The question of whether the turbulence dissipates in ions or neutral is a matter of debate. Few studies show that turbulence only dissipates in ions at the periphery of the cores. The dissipation of turbulence in neutral occurs at the Kolmogorov scale which is deep inside the core \citep{2008ApJ...677.1151L,2010ApJ...720..603H}. As an alternative, few studies have shown that dissipation of turbulence occurs in neutrals rather than ions at the periphery of the core \citep{2021ApJ...912....7P,2024A&A...690L...5P}. This is mostly valid in the prestellar cores. The situation for protostellar cores may be complex because of feedback effects of the young protostars. However, previous studies have shown that there is very little difference in the Mach number between starless quiescent cores and protostellar cores \citep{2013MNRAS.432.3288S,2021MNRAS.503.4601B}.\\

\subsection{Size-linewidth correlation}

In addition to the $M_{s}$ analysis in the protoclusters, we studied the size-linewidth correlation for 15 massive protoclusters individually, which we fit with power-laws (see Section~\ref{s:structure-decompositio}). We obtained power-law index ($p$) values ranging {\color{black}from 0.41 to 0.64}. There was no significant difference between the evolutionary stages of the protocluster. {\color{black}As we did find any prominent trend according to the evolutionary stage of the protoclusters, we also plotted the size-linewidth relation after adding all the structures for 15 protoclusters. We obtain that the value of $p$ is 0.51 $\pm$0.00.} Previous studies have observed similar values of the power law index in molecular clouds where turbulence is supersonic.  For example, \cite{1987ApJ...319..730S} investigated the size-linewidth relation for 273 molecular clouds using the CO molecular line observed by the Five College Radio Astronomy Observatory (FCRAO). They found that velocity dispersion ($\sigma_{\text{tot}}$) varies with length scale ($R$) with a value of $p$ = 0.50, which is steeper than 0.38 obtained from the work of \cite{1981MNRAS.194..809L}. In a similar study using $^{12}$CO ($J$=1$-$0) line, \cite{2004ApJ...615L..45H} reported a size-linewidth relationship for 27 giant molecular clouds and obtained $p$ equal to 0.56. \cite{2012MNRAS.425..720S} studied size-linewidth relations in the dense interstellar medium of the Central Molecular Zone using the tracers like N$_{2}$H$^{+}$, HCN, H$^{13}$CN, HCO$^{+}$ and found that the power-law index varies between 0.62 and 0.79. {\color{black}Likewise, \cite{2022MNRAS.511.4480L} studied the size-linewidth relation in the IRDC G034.43$+$00.24 using H$^{13}$CO$^{+}$ (1$-$0) line and observed that $p$ = 0.5. Similarly, \cite{2016ApJ...822...52R} studied the size-linewidth relationship of molecular clouds using CO line emissions in the inner and outer Galaxy and found power law index values of 0.52 and 0.49, respectively.} In addition, \cite{2023MNRAS.525..962P} studied the Galactic center clouds using HNCO (4$_{04}$-3$_{03}$) line and found that the power-law index is 0.68 $\pm$ 0.04, which is identical to their simulated results, which is 0.69 $\pm$ 0.03. Furthermore, several other studies have revealed that the power-law index of molecular clouds is steeper than the Kolmogorov law of turbulence ($p$ = 0.33) and varies between 0.50 and 0.70, similar to our study \citep{2017A&A...603A..89K,2021arXiv210713323Y,2022MNRAS.513..638Z,2023ApJ...949...63O}. This value of the power-law index is also obtained from theoretical studies. According to theoretical studies, incompressible hydrodynamic turbulence follows the Kolmogorov scaling relation based on which the non-thermal velocity dispersion ($\sigma_{\text{nth}}$) varies with the length scale with a power-law index ($p$) of 0.33 \citep{1941DoSSR..30..301K,1995tlan.book.....F}. However, for compressible media, the velocity power law is steeper than the Kolmogorov law of turbulence  ($p$ = 0.33), closer to the Burger law of turbulence, in which the power-law index is around 0.50 \citep{BURGERS1948171,1983ApJ...272L..45F, 2007ApJ...666L..69K,2023PASA...40...46K}. \\

}

\section{Conclusions}\label{s:concl}
{\color{black}
In this work, we studied the turbulence in the fifteen massive galactic protoclusters (with masses $\sim$ 1.0 $-$ 25.0 $\times$ 10$^{3}~$\(\textup{M}_\odot\) within $\sim$ 2.5 $\times$ 2.5 pc$^{2}$) observed by the ALMA-IMF large program, using the C$^{18}$O (2$-$1) emission line. The main findings of this study are as follows.\\

(1) The probability distribution function (PDF) of the sonic Mach number ($M_{\text{s}}$) in all these regions exhibits a peak between $\sim$ 4 and $\sim$ 7 and extends to $\sim$ 25. This kind of pattern was also observed in previous studies in high-mass star-forming regions. The values of $M_{\text{s}}$ demonstrate that the turbulence in the density regimes traced by C$^{18}$O (2$-$1) line emission is supersonic in nature. The fluctuations in Mach number are homogeneously distributed within the protoclusters, and we observed no clear differences between the young and the more evolved protoclusters. \\

(2) We did not find a correlation between $M_{\text{s}}$ and hydrogen column density [$N(\text{H}_{2})$]. {\color{black}The reason is that C$^{18}$O line emission traces dense as well as relative low density environments. However, the maximum amount of gas is spread in the low density regime, which we obtain after including the total power data from the ALMA telescope. Thus, in the relatively low density regime probed by C$^{18}$O line does not have any significant variation in turbulence.}\\


(3) We compared the non-thermal line widths of the C$^{18}$O (2$-1$) line and the DCN (3$-$2) cores. They exhibit mean non-thermal line-widths of $\sim$ 1.2 and $\sim$0.6\,km s$^{-1}$, respectively and mean Mach numbers of $\sim$ 7.1 and $\sim$ 3.3 respectively. These observations show a strong decrease in turbulence towards the dense cores.\\ 

(4) We extracted the structures (leaves, branches, and truncks) from the position-position-velocity (PPV) cubes of the C$^{18}$O (2$-$1) lines using \texttt{Astrodendro} module. The size-linewidth relations of the structures are well suited by a power law in the form $A (\text{km s}^{-1}$) $(\frac{R}{\text{pc}})^p$, where $A$ is the intensity and $R$ is the radius of the structure. Among the 15 protoclusters, $A$ and $p$ vary between {\color{black}1.72 and 3.62 and between 0.41 and 0.64} respectively. We did not find any noticeable increase or decrease in $A$ and $p$ according to the evolutionary stage of the protoclusters. The values of $p$ obtained from this study are steeper than the standard Kolmogorov law of turbulence, as expected for compressible media.\\

Overall, this work examines the turbulence of 15 massive protoclusters using C$^{18}$O (2$-$1) and DCN (3$-$2) lines. We notice that turbulence in the relatively low density regime traced by C$^{18}$O line  is relatively higher compared to the high density DCN cores. In the future, this study will help us with a better understanding of the role of turbulence in molecular clouds.\\

 \begin{acknowledgements} {\color{black}We greatly acknowledge to the referee for helping us to improve the manuscript significantly.} This paper makes use of the following ALMA data: ADS/JAO.ALMA 2017.1.01355.L. ALMA is a partnership of ESO (representing its member states), NSF (USA) and NINS (Japan), together with NRC (Canada), MOST and ASIAA (Taiwan), and KASI (Republic of Korea), in co-operation with the Republic of Chile. The Joint ALMA Observatory is operated by ESO, AUI/NRAO and NAOJ. The project leading to this publication has received support from ORP, which is funded by the European Union$'$s Horizon 2020 research and innovation program under grant agreement No. 101004719 [ORP]. This project has received funding from the
European Research Council (ERC) via the ERC Synergy Grant \textit{ECOGAL} (grant 855130) and from the French Agence Nationale de la Recherche (ANR) through
the project \textit{COSMHIC} (ANR-20-CE31-0009).  A.K. would like to thank Guido Garay for useful discussions during the 2nd Centre of Excellence in Astrophysics and Related Technologies (CATA) annual meeting held on Santiago, Chile in 2023. A.K. and L.B. gratefully acknowledge support from ANID BASAL project FB210003.  {\color{black}A.K. would also like to acknowledge Fondecyt postdoctoral fellowship (project id: 3250070, 2025).} A.S.  gratefully acknowledges support by the Fondecyt Regular (project code 1220610), and ANID BASAL project FB210003. R.A. gratefully acknowledges support from ANID Beca Doctorado Nacional 21200897. P.S. was partially supported by a Grant-in-Aid for Scientific Research (KAKENHI Number JP22H01271 and JP23H01221) of JSPS.  R.G.M. acknowledges support from UNAM-PAPIIT project IN108822 and from CONAHCyT Ciencia de Frontera project ID 86372. Part of this work was performed at the high-performance computers at IRyA-UNAM. T.C. and M.B. have received financial support from the French State
in the framework of the IdEx Université de Bordeaux Investments for the future Program. A.G. acknowledges support from the NSF under grants AST 2008101 and CAREER 2142300. M.B. is a postdoctoral fellow in the University of Virginia’s VICO collaboration and is funded by grants from the NASA Astrophysics Theory Program (grant num- ber 80NSSC18K0558) and the NSF Astronomy \& Astrophysics program (grant number 2206516). G.B. acknowledges support from the PID2023-146675NB-100 (MCI-AEI-FEDER,UE) program. H.-L. Liu is supported by Yunnan Fundamental Research Project (grant No. 202301AT070118, 202401AS070121). This work is supported by the China-Chile Joint Research Fund (CCJRF No. 2312). CCJRF is provided by Chinese Academy of Sciences South America Center for Astronomy (CASSACA) and established by National Astronomical Observatories, Chinese Academy of Sciences (NAOC) and Chilean Astronomy Society (SOCHIAS) to support China-Chile collabora-
tions in astronomy. P. García is sponsored by the Chinese Academy of Sciences CAS), through a grant to the CAS South America Center for Astronomy (CASSACA).
\end{acknowledgements}

\subsection*{\textbf{Telescope}} The Atacama Large Millimeter/submillimeter Array (ALMA)

\subsection*{\textbf{Software}} Astropy \citep{2013A&A...558A..33A}, CASA (Common Astronomy Software Applications package- National Radio Astronomical Observatory) (version 6.5.1.23) \citep{2007ASPC..376..127M}, Matplotlib \citep{4160265}, Gausspy+ \citep{2019A&A...628A..78R}, Astrodendro \citep{2008ApJ...679.1338R}.

\bibliographystyle{aa} 
\bibliography{a.bib} 

\begin{thebibliography}{115}
\expandafter\ifx\csname natexlab\endcsname\relax\def\natexlab#1{#1}\fi

\bibitem[{{Agertz} {et~al.}(2009){Agertz}, {Lake}, {Teyssier}, {Moore}, {Mayer}, \& {Romeo}}]{2009MNRAS.392..294A}
{Agertz}, O., {Lake}, G., {Teyssier}, R., {et~al.} 2009, \mnras, 392, 294

\bibitem[{{{\'A}lvarez-Guti{\'e}rrez} {et~al.}(2021){{\'A}lvarez-Guti{\'e}rrez}, {Stutz}, {Law}, {Reissl}, {Klessen}, {Leigh}, {Liu}, \& {Reeves}}]{2021ApJ...908...86A}
{{\'A}lvarez-Guti{\'e}rrez}, R.~H., {Stutz}, A.~M., {Law}, C.~Y., {et~al.} 2021, \apj, 908, 86

\bibitem[{{{\'A}lvarez-Guti{\'e}rrez} {et~al.}(2024){{\'A}lvarez-Guti{\'e}rrez}, {Stutz}, {Sandoval-Garrido}, {Louvet}, {Motte}, {Galv{\'a}n-Madrid}, {Cunningham}, {Sanhueza}, {Bonfand}, {Bontemps}, {Gusdorf}, {Ginsburg}, {Csengeri}, {Reyes}, {Salinas}, {Baug}, {Bronfman}, {Busquet}, {D{\'\i}az-Gonz{\'a}lez}, {Fernandez-Lopez}, {Guzm{\'a}n}, {Koley}, {Liu}, {Olguin}, {Valeille-Manet}, \& {Wyrowski}}]{2024A&A...689A..74A}
{{\'A}lvarez-Guti{\'e}rrez}, R.~H., {Stutz}, A.~M., {Sandoval-Garrido}, N., {et~al.} 2024, \aap, 689, A74

\bibitem[{{Ao} {et~al.}(2013){Ao}, {Henkel}, {Menten}, {Requena-Torres}, {Stanke}, {Mauersberger}, {Aalto}, {M{\"u}hle}, \& {Mangum}}]{2013A&A...550A.135A}
{Ao}, Y., {Henkel}, C., {Menten}, K.~M., {et~al.} 2013, \aap, 550, A135

\bibitem[{{Armante} {et~al.}(2024{\natexlab{a}}){Armante}, {Gusdorf}, {Louvet}, {Motte}, {Pouteau}, {Lesaffre}, {Galv{\'a}n-Madrid}, {Dell'Ova}, {Bonfand}, {Nony}, {Brouillet}, {Cunningham}, {Ginsburg}, {Men'shchikov}, {Bontemps}, {D{\'\i}az Gonz{\'a}lez}, {Csengeri}, {Fern{\'a}ndez-L{\'o}pez}, {Gonz{\'a}lez}, {Herpin}, {Liu}, {Sanhueza}, {Stutz}, \& {Valeille-Manet}}]{2024arXiv240109203A}
{Armante}, M., {Gusdorf}, A., {Louvet}, F., {et~al.} 2024{\natexlab{a}}, arXiv e-prints, arXiv:2401.09203

\bibitem[{{Armante} {et~al.}(2024{\natexlab{b}}){Armante}, {Gusdorf}, {Louvet}, {Motte}, {Pouteau}, {Lesaffre}, {Galv{\'a}n-Madrid}, {Dell'Ova}, {Bonfand}, {Nony}, {Brouillet}, {Cunningham}, {Ginsburg}, {Men'shchikov}, {Bontemps}, {D{\'\i}az-Gonz{\'a}lez}, {Csengeri}, {Fern{\'a}ndez-L{\'o}pez}, {Gonz{\'a}lez}, {Herpin}, {Liu}, {Sanhueza}, {Stutz}, \& {Valeille-Manet}}]{2024A&A...686A.122A}
{Armante}, M., {Gusdorf}, A., {Louvet}, F., {et~al.} 2024{\natexlab{b}}, \aap, 686, A122

\bibitem[{{Astropy Collaboration} {et~al.}(2013){Astropy Collaboration}, {Robitaille}, {Tollerud}, {Greenfield}, {Droettboom}, {Bray}, {Aldcroft}, {Davis}, {Ginsburg}, {Price-Whelan}, {Kerzendorf}, {Conley}, {Crighton}, {Barbary}, {Muna}, {Ferguson}, {Grollier}, {Parikh}, {Nair}, {Unther}, {Deil}, {Woillez}, {Conseil}, {Kramer}, {Turner}, {Singer}, {Fox}, {Weaver}, {Zabalza}, {Edwards}, {Azalee Bostroem}, {Burke}, {Casey}, {Crawford}, {Dencheva}, {Ely}, {Jenness}, {Labrie}, {Lim}, {Pierfederici}, {Pontzen}, {Ptak}, {Refsdal}, {Servillat}, \& {Streicher}}]{2013A&A...558A..33A}
{Astropy Collaboration}, {Robitaille}, T.~P., {Tollerud}, E.~J., {et~al.} 2013, \aap, 558, A33

\bibitem[{{Barnes} {et~al.}(2021){Barnes}, {Henshaw}, {Fontani}, {Pineda}, {Cosentino}, {Tan}, {Caselli}, {Jim{\'e}nez-Serra}, {Law}, {Avison}, {Bigiel}, {Feng}, {Kong}, {Longmore}, {Moser}, {Parker}, {S{\'a}nchez-Monge}, \& {Wang}}]{2021MNRAS.503.4601B}
{Barnes}, A.~T., {Henshaw}, J.~D., {Fontani}, F., {et~al.} 2021, \mnras, 503, 4601

\bibitem[{{Barnes} {et~al.}(2023){Barnes}, {Liu}, {Zhang}, {Tan}, {Bigiel}, {Caselli}, {Cosentino}, {Fontani}, {Henshaw}, {Jim{\'e}nez-Serra}, {Kalb}, {Law}, {Longmore}, {Parker}, {Pineda}, {S{\'a}nchez-Monge}, {Lim}, \& {Wang}}]{2023A&A...675A..53B}
{Barnes}, A.~T., {Liu}, J., {Zhang}, Q., {et~al.} 2023, \aap, 675, A53

\bibitem[{{Bonfand} {et~al.}(2024){Bonfand}, {Csengeri}, {Bontemps}, {Brouillet}, {Motte}, {Louvet}, {Ginsburg}, {Cunningham}, {Galv{\'a}n-Madrid}, {Herpin}, {Wyrowski}, {Valeille-Manet}, {Stutz}, {Di Francesco}, {Gusdorf}, {Fern{\'a}ndez-L{\'o}pez}, {Lefloch}, {Liu}, {Sanhueza}, {{\'A}lvarez-Guti{\'e}rrez}, {Olguin}, {Nony}, {Lopez-Sepulcre}, {Dell'Ova}, {Pouteau}, {Jeff}, {Chen}, {Armante}, {Towner}, {Bronfman}, \& {Kessler}}]{2024arXiv240215023B}
{Bonfand}, M., {Csengeri}, T., {Bontemps}, S., {et~al.} 2024, arXiv e-prints, arXiv:2402.15023

\bibitem[{{Brouillet} {et~al.}(2022){Brouillet}, {Despois}, {Molet}, {Nony}, {Motte}, {Gusdorf}, {Louvet}, {Bontemps}, {Herpin}, {Bonfand}, {Csengeri}, {Ginsburg}, {Cunningham}, {Galv{\'a}n-Madrid}, {Maud}, {Busquet}, {Bronfman}, {Fern{\'a}ndez-L{\'o}pez}, {Jeff}, {Lefloch}, {Pouteau}, {Sanhueza}, {Stutz}, \& {Valeille-Manet}}]{2022A&A...665A.140B}
{Brouillet}, N., {Despois}, D., {Molet}, J., {et~al.} 2022, \aap, 665, A140

\bibitem[{Burgers(1948)}]{BURGERS1948171}
Burgers, J. 1948, in Advances in Applied Mechanics, Vol.~1, A Mathematical Model Illustrating the Theory of Turbulence, ed. R.~{Von Mises} \& T.~{Von Kármán} (Elsevier), 171--199

\bibitem[{{Busquet} {et~al.}(2016){Busquet}, {Estalella}, {Palau}, {Liu}, {Zhang}, {Girart}, {de Gregorio-Monsalvo}, {Pillai}, {Anglada}, \& {Ho}}]{2016ApJ...819..139B}
{Busquet}, G., {Estalella}, R., {Palau}, A., {et~al.} 2016, \apj, 819, 139

\bibitem[{{Choudhury} {et~al.}(2021){Choudhury}, {Pineda}, {Caselli}, {Offner}, {Rosolowsky}, {Friesen}, {Redaelli}, {Chac{\'o}n-Tanarro}, {Shirley}, {Punanova}, \& {Kirk}}]{2021A&A...648A.114C}
{Choudhury}, S., {Pineda}, J.~E., {Caselli}, P., {et~al.} 2021, \aap, 648, A114

\bibitem[{{Csengeri} {et~al.}(2017){Csengeri}, {Bontemps}, {Wyrowski}, {Megeath}, {Motte}, {Sanna}, {Wienen}, \& {Menten}}]{2017A&A...601A..60C}
{Csengeri}, T., {Bontemps}, S., {Wyrowski}, F., {et~al.} 2017, \aap, 601, A60

\bibitem[{{Cunningham} {et~al.}(2023){Cunningham}, {Ginsburg}, {Galv{\'a}n-Madrid}, {Motte}, {Csengeri}, {Stutz}, {Fern{\'a}ndez-L{\'o}pez}, {{\'A}lvarez-Guti{\'e}rrez}, {Armante}, {Baug}, {Bonfand}, {Bontemps}, {Braine}, {Brouillet}, {Busquet}, {D{\'\i}az-Gonz{\'a}lez}, {Di Francesco}, {Gusdorf}, {Herpin}, {Liu}, {L{\'o}pez-Sepulcre}, {Louvet}, {Lu}, {Maud}, {Nony}, {Olguin}, {Pouteau}, {Rivera-Soto}, {Sandoval-Garrido}, {Sanhueza}, {Tatematsu}, {Towner}, \& {Valeille-Manet}}]{2023A&A...678A.194C}
{Cunningham}, N., {Ginsburg}, A., {Galv{\'a}n-Madrid}, R., {et~al.} 2023, \aap, 678, A194

\bibitem[{{Dell'Ova} {et~al.}(2024){Dell'Ova}, {Motte}, {Gusdorf}, {Pouteau}, {Men'shchikov}, {D{\'\i}az-Gonz{\'a}lez}, {Galv{\'a}n-Madrid}, {Lesaffre}, {Didelon}, {Stutz}, {Towner}, {Marsh}, {Whitworth}, {Armante}, {Bonfand}, {Nony}, {Valeille-Manet}, {Bontemps}, {Csengeri}, {Cunningham}, {Ginsburg}, {Louvet}, {{\'A}lvarez-Guti{\'e}rrez}, {Brouillet}, {Salinas}, {Sanhueza}, {Nakamura}, {Nguyen Luong}, {Baug}, {Fern{\'a}ndez-L{\'o}pez}, {Liu}, \& {Olguin}}]{2024A&A...687A.217D}
{Dell'Ova}, P., {Motte}, F., {Gusdorf}, A., {et~al.} 2024, \aap, 687, A217

\bibitem[{{Dewangan}(2022)}]{2022MNRAS.513.2942D}
{Dewangan}, L.~K. 2022, \mnras, 513, 2942

\bibitem[{{Dirienzo} {et~al.}(2015){Dirienzo}, {Brogan}, {Indebetouw}, {Chandler}, {Friesen}, \& {Devine}}]{2015AJ....150..159D}
{Dirienzo}, W.~J., {Brogan}, C., {Indebetouw}, R., {et~al.} 2015, \aj, 150, 159

\bibitem[{{Federrath}(2015)}]{2015MNRAS.450.4035F}
{Federrath}, C. 2015, \mnras, 450, 4035

\bibitem[{{Federrath}(2018)}]{2018PhT....71f..38F}
{Federrath}, C. 2018, Physics Today, 71, 38

\bibitem[{{Federrath} \& {Klessen}(2012)}]{2012ApJ...761..156F}
{Federrath}, C. \& {Klessen}, R.~S. 2012, \apj, 761, 156

\bibitem[{{Fleck}(1983)}]{1983ApJ...272L..45F}
{Fleck}, R.~C., J. 1983, \apjl, 272, L45

\bibitem[{{Frisch}(1995)}]{1995tlan.book.....F}
{Frisch}, U. 1995, {Turbulence. The legacy of A.N. Kolmogorov}

\bibitem[{{Galv{\'a}n-Madrid} {et~al.}(2024){Galv{\'a}n-Madrid}, {D{\'\i}az-Gonz{\'a}lez}, {Motte}, {Ginsburg}, {Cunningham}, {Menten}, {Armante}, {Bonfand}, {Braine}, {Csengeri}, {Dell'Ova}, {Louvet}, {Nony}, {Rivera-Soto}, {Sanhueza}, {Stutz}, {Wyrowski}, {{\'A}lvarez-Guti{\'e}rrez}, {Baug}, {Bontemps}, {Bronfman}, {Fern{\'a}ndez-L{\'o}pez}, {Gusdorf}, {Koley}, {Liu}, {Salinas}, {Towner}, \& {Whitworth}}]{2024ApJS..274...15G}
{Galv{\'a}n-Madrid}, R., {D{\'\i}az-Gonz{\'a}lez}, D.~J., {Motte}, F., {et~al.} 2024, \apjs, 274, 15

\bibitem[{{Ginsburg}(2017)}]{2017arXiv170206627G}
{Ginsburg}, A. 2017, arXiv e-prints, arXiv:1702.06627

\bibitem[{{Ginsburg} {et~al.}(2022){Ginsburg}, {Csengeri}, {Galv{\'a}n-Madrid}, {Cunningham}, {{\'A}lvarez-Guti{\'e}rrez}, {Baug}, {Bonfand}, {Bontemps}, {Busquet}, {D{\'\i}az-Gonz{\'a}lez}, {Fern{\'a}ndez-L{\'o}pez}, {Guzm{\'a}n}, {Herpin}, {Liu}, {L{\'o}pez-Sepulcre}, {Louvet}, {Maud}, {Motte}, {Nakamura}, {Nony}, {Olguin}, {Pouteau}, {Sanhueza}, {Stutz}, {Towner}, {ALMA-IMF Consortium}, {Armante}, {Battersby}, {Bronfman}, {Braine}, {Brouillet}, {Chapillon}, {Di Francesco}, {Gusdorf}, {Izumi}, {Joncour}, {Walker Lu}, {Men'shchikov}, {Menten}, {Moraux}, {Molet}, {Mundy}, {Nguyen Luong}, {Reyes-Reyes}, {Robitaille}, {Rosolowsky}, {Sandoval-Garrido}, {Svoboda}, {Tatematsu}, {Walker}, {Whitworth}, {Wu}, \& {Wyrowski}}]{2022A&A...662A...9G}
{Ginsburg}, A., {Csengeri}, T., {Galv{\'a}n-Madrid}, R., {et~al.} 2022, \aap, 662, A9

\bibitem[{{Goodman} {et~al.}(1998){Goodman}, {Barranco}, {Wilner}, \& {Heyer}}]{1998ApJ...504..223G}
{Goodman}, A.~A., {Barranco}, J.~A., {Wilner}, D.~J., \& {Heyer}, M.~H. 1998, \apj, 504, 223

\bibitem[{{Green} {et~al.}(2024){Green}, {Wong}, {Indebetouw}, {Nayak}, {Bolatto}, {Tarantino}, {Rubio}, {Madden}, \& {Hirschauer}}]{2024ApJ...966...51G}
{Green}, A., {Wong}, T., {Indebetouw}, R., {et~al.} 2024, \apj, 966, 51

\bibitem[{{Hacar} {et~al.}(2016){Hacar}, {Alves}, {Burkert}, \& {Goldsmith}}]{2016A&A...591A.104H}
{Hacar}, A., {Alves}, J., {Burkert}, A., \& {Goldsmith}, P. 2016, \aap, 591, A104

\bibitem[{{He} {et~al.}(2023){He}, {Liu}, {Tang}, {Qin}, {Zhou}, {Esimbek}, {Pan}, {Li}, {Zhao}, {Ji}, \& {Komesh}}]{2023ApJ...957...61H}
{He}, Y.-X., {Liu}, H.-L., {Tang}, X.-D., {et~al.} 2023, \apj, 957, 61

\bibitem[{{Hennebelle} \& {Inutsuka}(2019)}]{2019FrASS...6....5H}
{Hennebelle}, P. \& {Inutsuka}, S.-i. 2019, Frontiers in Astronomy and Space Sciences, 6, 5

\bibitem[{{Heyer} \& {Brunt}(2004)}]{2004ApJ...615L..45H}
{Heyer}, M.~H. \& {Brunt}, C.~M. 2004, \apjl, 615, L45

\bibitem[{{Hezareh} {et~al.}(2010){Hezareh}, {Houde}, {McCoey}, \& {Li}}]{2010ApJ...720..603H}
{Hezareh}, T., {Houde}, M., {McCoey}, C., \& {Li}, H.-b. 2010, \apj, 720, 603

\bibitem[{{Hofner} {et~al.}(2000){Hofner}, {Wyrowski}, {Walmsley}, \& {Churchwell}}]{2000ApJ...536..393H}
{Hofner}, P., {Wyrowski}, F., {Walmsley}, C.~M., \& {Churchwell}, E. 2000, \apj, 536, 393

\bibitem[{Hunter(2007)}]{4160265}
Hunter, J.~D. 2007, Computing in Science \& Engineering, 9, 90

\bibitem[{{Joung} \& {Mac Low}(2006)}]{2006ApJ...653.1266J}
{Joung}, M.~K.~R. \& {Mac Low}, M.-M. 2006, \apj, 653, 1266

\bibitem[{{Kang} {et~al.}(2010){Kang}, {Bieging}, {Kulesa}, {Lee}, {Choi}, \& {Peters}}]{2010ApJS..190...58K}
{Kang}, M., {Bieging}, J.~H., {Kulesa}, C.~A., {et~al.} 2010, \apjs, 190, 58

\bibitem[{{Kauffmann} {et~al.}(2017){Kauffmann}, {Pillai}, {Zhang}, {Menten}, {Goldsmith}, {Lu}, \& {Guzm{\'a}n}}]{2017A&A...603A..89K}
{Kauffmann}, J., {Pillai}, T., {Zhang}, Q., {et~al.} 2017, \aap, 603, A89

\bibitem[{Koch {et~al.}(2018)Koch, Rosolowsky, \& Leroy}]{Koch_2018}
Koch, E., Rosolowsky, E., \& Leroy, A.~K. 2018, Research Notes of the AAS, 2, 220

\bibitem[{{Koley}(2019)}]{2019MNRAS.483..593K}
{Koley}, A. 2019, \mnras, 483, 593

\bibitem[{{Koley}(2022)}]{2022MNRAS.516..185K}
{Koley}, A. 2022, \mnras, 516, 185

\bibitem[{{Koley}(2023{\natexlab{a}})}]{2023PASA...40...53K}
{Koley}, A. 2023{\natexlab{a}}, \pasa, 40, e053

\bibitem[{{Koley}(2023{\natexlab{b}})}]{2023PASA...40...46K}
{Koley}, A. 2023{\natexlab{b}}, \pasa, 40, e046

\bibitem[{{Koley} {et~al.}(2021){Koley}, {Roy}, {Menten}, {Jacob}, {Pillai}, \& {Rugel}}]{2021MNRAS.501.4825K}
{Koley}, A., {Roy}, N., {Menten}, K.~M., {et~al.} 2021, \mnras, 501, 4825

\bibitem[{{Koley} {et~al.}(2022){Koley}, {Roy}, {Momjian}, {Sarma}, \& {Datta}}]{2022MNRAS.516L..48K}
{Koley}, A., {Roy}, N., {Momjian}, E., {Sarma}, A.~P., \& {Datta}, A. 2022, \mnras, 516, L48

\bibitem[{{Kolmogorov}(1941)}]{1941DoSSR..30..301K}
{Kolmogorov}, A. 1941, Akademiia Nauk SSSR Doklady, 30, 301

\bibitem[{{Kong} {et~al.}(2021){Kong}, {Arce}, {Carpenter}, {Bally}, {Ossenkopf-Okada}, {S{\'a}nchez-Monge}, {Sargent}, {Suri}, {McGehee}, {Lis}, {Klessen}, {Mairs}, {Zucker}, {Smith}, {Nakamura}, {Pillai}, {Kauffmann}, \& {Zhang}}]{2021AJ....161..229K}
{Kong}, S., {Arce}, H.~G., {Carpenter}, J.~M., {et~al.} 2021, \aj, 161, 229

\bibitem[{{Kowal} \& {Lazarian}(2007)}]{2007ApJ...666L..69K}
{Kowal}, G. \& {Lazarian}, A. 2007, \apjl, 666, L69

\bibitem[{{Larson}(1981)}]{1981MNRAS.194..809L}
{Larson}, R.~B. 1981, \mnras, 194, 809

\bibitem[{{Leurini} {et~al.}(2011){Leurini}, {Pillai}, {Stanke}, {Wyrowski}, {Testi}, {Schuller}, {Menten}, \& {Thorwirth}}]{2011A&A...533A..85L}
{Leurini}, S., {Pillai}, T., {Stanke}, T., {et~al.} 2011, \aap, 533, A85

\bibitem[{{Li} \& {Houde}(2008)}]{2008ApJ...677.1151L}
{Li}, H.-b. \& {Houde}, M. 2008, \apj, 677, 1151

\bibitem[{{Li} {et~al.}(2020){Li}, {Zhang}, {Liu}, {Beuther}, {Palau}, {Girart}, {Smith}, {Hora}, {Lin}, {Qiu}, {Strom}, {Wang}, {Li}, \& {Yue}}]{2020ApJ...896..110L}
{Li}, S., {Zhang}, Q., {Liu}, H.~B., {et~al.} 2020, \apj, 896, 110

\bibitem[{{Lindner} {et~al.}(2015){Lindner}, {Vera-Ciro}, {Murray}, {Stanimirovi{\'c}}, {Babler}, {Heiles}, {Hennebelle}, {Goss}, \& {Dickey}}]{Lindner2015L}
{Lindner}, R.~R., {Vera-Ciro}, C., {Murray}, C.~E., {et~al.} 2015, \aj, 149, 138

\bibitem[{{Liu}(2017)}]{2017A&A...597A..70L}
{Liu}, H.~B. 2017, \aap, 597, A70

\bibitem[{{Liu} {et~al.}(2019){Liu}, {Stutz}, \& {Yuan}}]{2019MNRAS.487.1259L}
{Liu}, H.-L., {Stutz}, A., \& {Yuan}, J.-H. 2019, \mnras, 487, 1259

\bibitem[{{Liu} {et~al.}(2022){Liu}, {Tej}, {Liu}, {Goldsmith}, {Stutz}, {Juvela}, {Qin}, {Xu}, {Bronfman}, {Evans}, {Saha}, {Issac}, {Tatematsu}, {Wang}, {Li}, {Zhang}, {Baug}, {Dewangan}, {Wu}, {Zhang}, {Lee}, {Liu}, {Zhou}, \& {Soam}}]{2022MNRAS.511.4480L}
{Liu}, H.-L., {Tej}, A., {Liu}, T., {et~al.} 2022, \mnras, 511, 4480

\bibitem[{{Louvet} {et~al.}(2016){Louvet}, {Motte}, {Gusdorf}, {Nguy{\^e}n Luong}, {Lesaffre}, {Duarte-Cabral}, {Maury}, {Schneider}, {Hill}, {Schilke}, \& {Gueth}}]{2016A&A...595A.122L}
{Louvet}, F., {Motte}, F., {Gusdorf}, A., {et~al.} 2016, \aap, 595, A122

\bibitem[{{Louvet} {et~al.}(2024){Louvet}, {Sanhueza}, {Stutz}, {Men'shchikov}, {Motte}, {Galv{\'a}n-Madrid}, {Bontemps}, {Pouteau}, {Ginsburg}, {Csengeri}, {Di Francesco}, {Dell'Ova}, {Gonz{\'a}lez}, {Didelon}, {Braine}, {Cunningham}, {Thomasson}, {Lesaffre}, {Hennebelle}, {Bonfand}, {Gusdorf}, {{\'A}lverez-Guti{\'e}rrez}, {Nony}, {Busquet}, {Olguin}, {Bronfman}, {Salinas}, {Fernandez-Lopez}, {Moraux}, {Liu}, {Lu}, {Huei-Ru}, {Towner}, {Valeille-Manet}, {Brouillet}, {Herpin}, {Lefloch}, {Baug}, {Maud}, {L{\'o}pez-Sepulcre}, \& {Svoboda}}]{2024A&A...690A..33L}
{Louvet}, F., {Sanhueza}, P., {Stutz}, A., {et~al.} 2024, \aap, 690, A33

\bibitem[{{Lu} {et~al.}(2018){Lu}, {Zhang}, {Liu}, {Sanhueza}, {Tatematsu}, {Feng}, {Smith}, {Myers}, {Sridharan}, \& {Gu}}]{2018ApJ...855....9L}
{Lu}, X., {Zhang}, Q., {Liu}, H.~B., {et~al.} 2018, \apj, 855, 9

\bibitem[{{Mac Low} \& {Klessen}(2004)}]{2004RvMP...76..125M}
{Mac Low}, M.-M. \& {Klessen}, R.~S. 2004, Reviews of Modern Physics, 76, 125

\bibitem[{{Mallick} {et~al.}(2023){Mallick}, {Sharma}, {Dewangan}, {Ojha}, {Panwar}, \& {Baug}}]{2023JApA...44...34M}
{Mallick}, K.~K., {Sharma}, S., {Dewangan}, L.~K., {et~al.} 2023, Journal of Astrophysics and Astronomy, 44, 34

\bibitem[{{Mangum} \& {Shirley}(2015)}]{2015PASP..127..266M}
{Mangum}, J.~G. \& {Shirley}, Y.~L. 2015, \pasp, 127, 266

\bibitem[{{Marsh} {et~al.}(2006){Marsh}, {Velusamy}, \& {Ware}}]{2006AJ....132.1789M}
{Marsh}, K.~A., {Velusamy}, T., \& {Ware}, B. 2006, \aj, 132, 1789

\bibitem[{{Marsh} {et~al.}(2015){Marsh}, {Whitworth}, \& {Lomax}}]{2015MNRAS.454.4282M}
{Marsh}, K.~A., {Whitworth}, A.~P., \& {Lomax}, O. 2015, \mnras, 454, 4282

\bibitem[{{Marsh} {et~al.}(2017){Marsh}, {Whitworth}, {Lomax}, {Ragan}, {Becciani}, {Cambr{\'e}sy}, {Di Giorgio}, {Eden}, {Elia}, {Kacsuk}, {Molinari}, {Palmeirim}, {Pezzuto}, {Schneider}, {Sciacca}, \& {Vitello}}]{2017MNRAS.471.2730M}
{Marsh}, K.~A., {Whitworth}, A.~P., {Lomax}, O., {et~al.} 2017, \mnras, 471, 2730

\bibitem[{{McMullin} {et~al.}(2007){McMullin}, {Waters}, {Schiebel}, {Young}, \& {Golap}}]{2007ASPC..376..127M}
{McMullin}, J.~P., {Waters}, B., {Schiebel}, D., {Young}, W., \& {Golap}, K. 2007, in Astronomical Society of the Pacific Conference Series, Vol. 376, Astronomical Data Analysis Software and Systems XVI, ed. R.~A. {Shaw}, F.~{Hill}, \& D.~J. {Bell}, 127

\bibitem[{{Melchior} \& {Combes}(2016)}]{2016A&A...585A..44M}
{Melchior}, A.-L. \& {Combes}, F. 2016, \aap, 585, A44

\bibitem[{{Menon} {et~al.}(2021){Menon}, {Federrath}, {Klaassen}, {Kuiper}, \& {Reiter}}]{2021MNRAS.500.1721M}
{Menon}, S.~H., {Federrath}, C., {Klaassen}, P., {Kuiper}, R., \& {Reiter}, M. 2021, \mnras, 500, 1721

\bibitem[{{Motte} {et~al.}(2022){Motte}, {Bontemps}, {Csengeri}, {Pouteau}, {Louvet}, {Stutz}, {Cunningham}, {L{\'o}pez-Sepulcre}, {Brouillet}, {Galv{\'a}n-Madrid}, {Ginsburg}, {Maud}, {Men'shchikov}, {Nakamura}, {Nony}, {Sanhueza}, {{\'A}lvarez-Guti{\'e}rrez}, {Armante}, {Baug}, {Bonfand}, {Busquet}, {Chapillon}, {D{\'\i}az-Gonz{\'a}lez}, {Fern{\'a}ndez-L{\'o}pez}, {Guzm{\'a}n}, {Herpin}, {Liu}, {Olguin}, {Towner}, {Bally}, {Battersby}, {Braine}, {Bronfman}, {Chen}, {Dell'Ova}, {Di Francesco}, {Gonz{\'a}lez}, {Gusdorf}, {Hennebelle}, {Izumi}, {Joncour}, {Lee}, {Lefloch}, {Lesaffre}, {Lu}, {Menten}, {Mignon-Risse}, {Molet}, {Moraux}, {Mundy}, {Nguyen Luong}, {Reyes}, {Reyes Reyes}, {Robitaille}, {Rosolowsky}, {Sandoval-Garrido}, {Schuller}, {Svoboda}, {Tatematsu}, {Thomasson}, {Walker}, {Wu}, {Whitworth}, \& {Wyrowski}}]{2022A&A...662A...8M}
{Motte}, F., {Bontemps}, S., {Csengeri}, T., {et~al.} 2022, \aap, 662, A8

\bibitem[{{Motte} {et~al.}(2014){Motte}, {Nguy{\^e}n Luong}, {Schneider}, {Heitsch}, {Glover}, {Carlhoff}, {Hill}, {Bontemps}, {Schilke}, {Louvet}, {Hennemann}, {Didelon}, \& {Beuther}}]{2014A&A...571A..32M}
{Motte}, F., {Nguy{\^e}n Luong}, Q., {Schneider}, N., {et~al.} 2014, \aap, 571, A32

\bibitem[{{Motte} {et~al.}(2018){Motte}, {Nony}, {Louvet}, {Marsh}, {Bontemps}, {Whitworth}, {Men'shchikov}, {Nguyen Luong}, {Csengeri}, {Maury}, {Gusdorf}, {Chapillon}, {K{\"o}nyves}, {Schilke}, {Duarte-Cabral}, {Didelon}, \& {Gaudel}}]{2018NatAs...2..478M}
{Motte}, F., {Nony}, T., {Louvet}, F., {et~al.} 2018, Nature Astronomy, 2, 478

\bibitem[{{Motte} {et~al.}(2024){Motte}, {Pouteau}, {Nony}, {Dell'Ova}, {Gusdorf}, {Brouillet}, {Stutz}, {Bontemps}, {Ginsburg}, {Csengeri}, {Men'shchikov}, {Valeille-Manet}, {Louvet}, {Bonfand}, {Galv{\'a}n-Madrid}, {{\'A}lvarez-Guti{\'e}rrez}, {Armante}, {Bronfman}, {Chen}, {Cunningham}, {D{\'\i}az-Gonz{\'a}lez}, {Didelon}, {Fern{\'a}ndez-L{\'o}pez}, {Herpin}, {Kessler}, {Koley}, {Lefloch}, {Le Nestour}, {Liu}, {Moraux}, {Nguyen Luong}, {Olguin}, {Salinas}, {Sandoval-Garrido}, {Sanhueza}, {Veyry}, \& {Yoo}}]{2024arXiv241202011M}
{Motte}, F., {Pouteau}, Y., {Nony}, T., {et~al.} 2024, arXiv e-prints, arXiv:2412.02011

\bibitem[{{Myers}(1983)}]{1983ApJ...270..105M}
{Myers}, P.~C. 1983, \apj, 270, 105

\bibitem[{{Nakano}(1998)}]{1998ApJ...494..587N}
{Nakano}, T. 1998, \apj, 494, 587

\bibitem[{{Nguyen Luong} {et~al.}(2011){Nguyen Luong}, {Motte}, {Schuller}, {Schneider}, {Bontemps}, {Schilke}, {Menten}, {Heitsch}, {Wyrowski}, {Carlhoff}, {Bronfman}, \& {Henning}}]{2011A&A...529A..41N}
{Nguyen Luong}, Q., {Motte}, F., {Schuller}, F., {et~al.} 2011, \aap, 529, A41

\bibitem[{{Nishimura} {et~al.}(2015){Nishimura}, {Tokuda}, {Kimura}, {Muraoka}, {Maezawa}, {Ogawa}, {Dobashi}, {Shimoikura}, {Mizuno}, {Fukui}, \& {Onishi}}]{2015ApJS..216...18N}
{Nishimura}, A., {Tokuda}, K., {Kimura}, K., {et~al.} 2015, \apjs, 216, 18

\bibitem[{{Nony} {et~al.}(2023){Nony}, {Galv{\'a}n-Madrid}, {Motte}, {Pouteau}, {Cunningham}, {Louvet}, {Stutz}, {Lefloch}, {Bontemps}, {Brouillet}, {Ginsburg}, {Joncour}, {Herpin}, {Sanhueza}, {Csengeri}, {Towner}, {Bonfand}, {Fern{\'a}ndez-L{\'o}pez}, {Baug}, {Bronfman}, {Busquet}, {Di Francesco}, {Gusdorf}, {Lu}, {Olguin}, {Valeille-Manet}, \& {Whitworth}}]{2023A&A...674A..75N}
{Nony}, T., {Galv{\'a}n-Madrid}, R., {Motte}, F., {et~al.} 2023, \aap, 674, A75

\bibitem[{{Nony} {et~al.}(2020){Nony}, {Motte}, {Louvet}, {Plunkett}, {Gusdorf}, {Fechtenbaum}, {Pouteau}, {Lefloch}, {Bontemps}, {Molet}, \& {Robitaille}}]{2020A&A...636A..38N}
{Nony}, T., {Motte}, F., {Louvet}, F., {et~al.} 2020, \aap, 636, A38

\bibitem[{{Ohno} {et~al.}(2023){Ohno}, {Tokuda}, {Konishi}, {Matsumoto}, {Sewi{\l}o}, {Kondo}, {Sano}, {Tsuge}, {Zahorecz}, {Goto}, {Neelamkodan}, {Wong}, {Fukushima}, {Takekoshi}, {Muraoka}, {Kawamura}, {Tachihara}, {Fukui}, \& {Onishi}}]{2023ApJ...949...63O}
{Ohno}, T., {Tokuda}, K., {Konishi}, A., {et~al.} 2023, \apj, 949, 63

\bibitem[{{Orkisz} {et~al.}(2017){Orkisz}, {Pety}, {Gerin}, {Bron}, {Guzm{\'a}n}, {Bardeau}, {Goicoechea}, {Gratier}, {Le Petit}, {Levrier}, {Liszt}, {{\"O}berg}, {Peretto}, {Roueff}, {Sievers}, \& {Tremblin}}]{2017A&A...599A..99O}
{Orkisz}, J.~H., {Pety}, J., {Gerin}, M., {et~al.} 2017, \aap, 599, A99

\bibitem[{{Palau} {et~al.}(2021){Palau}, {Zhang}, {Girart}, {Liu}, {Rao}, {Koch}, {Estalella}, {Chen}, {Liu}, {Qiu}, {Li}, {Zapata}, {Bontemps}, {Ho}, {Beuther}, {Ching}, {Shinnaga}, \& {Ahmadi}}]{2021ApJ...912..159P}
{Palau}, A., {Zhang}, Q., {Girart}, J.~M., {et~al.} 2021, \apj, 912, 159

\bibitem[{{Pan} {et~al.}(2023){Pan}, {Liu}, \& {Qin}}]{2023MNRAS.519.3851P}
{Pan}, S., {Liu}, H.-L., \& {Qin}, S.-L. 2023, \mnras, 519, 3851

\bibitem[{{Paron} {et~al.}(2018){Paron}, {Areal}, \& {Ortega}}]{2018A&A...617A..14P}
{Paron}, S., {Areal}, M.~B., \& {Ortega}, M.~E. 2018, \aap, 617, A14

\bibitem[{{Petkova} {et~al.}(2023){Petkova}, {Kruijssen}, {Henshaw}, {Longmore}, {Glover}, {Sormani}, {Armillotta}, {Barnes}, {Klessen}, {Nogueras-Lara}, {Tress}, {Armijos-Abenda{\~n}o}, {Colzi}, {Federrath}, {Garc{\'\i}a}, {Ginsburg}, {Henkel}, {Mart{\'\i}n}, {Riquelme}, \& {Rivilla}}]{2023MNRAS.525..962P}
{Petkova}, M.~A., {Kruijssen}, J.~M.~D., {Henshaw}, J.~D., {et~al.} 2023, \mnras, 525, 962

\bibitem[{{Pillai} {et~al.}(2006){Pillai}, {Wyrowski}, {Menten}, \& {Kr{\"u}gel}}]{2006A&A...447..929P}
{Pillai}, T., {Wyrowski}, F., {Menten}, K.~M., \& {Kr{\"u}gel}, E. 2006, \aap, 447, 929

\bibitem[{{Pineda} {et~al.}(2021){Pineda}, {Schmiedeke}, {Caselli}, {Stahler}, {Frayer}, {Church}, \& {Harris}}]{2021ApJ...912....7P}
{Pineda}, J.~E., {Schmiedeke}, A., {Caselli}, P., {et~al.} 2021, \apj, 912, 7

\bibitem[{{Pineda} {et~al.}(2024){Pineda}, {Soler}, {Offner}, {Koch}, {Segura-Cox}, {Neri}, {Kuffmeier}, {Ivlev}, {Teresa Valdivia-Mena}, {Sipil{\"a}}, {Maureira}, {Caselli}, {Cunningham}, {Schmiedeke}, {Gieser}, {Chen}, \& {Spezzano}}]{2024A&A...690L...5P}
{Pineda}, J.~E., {Soler}, J.~D., {Offner}, S., {et~al.} 2024, \aap, 690, L5

\bibitem[{{Pouteau} {et~al.}(2022){Pouteau}, {Motte}, {Nony}, {Galv{\'a}n-Madrid}, {Men'shchikov}, {Bontemps}, {Robitaille}, {Louvet}, {Ginsburg}, {Herpin}, {L{\'o}pez-Sepulcre}, {Dell'Ova}, {Gusdorf}, {Sanhueza}, {Stutz}, {Brouillet}, {Thomasson}, {Armante}, {Baug}, {Bonfand}, {Busquet}, {Csengeri}, {Cunningham}, {Fern{\'a}ndez-L{\'o}pez}, {Liu}, {Olguin}, {Towner}, {Bally}, {Braine}, {Bronfman}, {Joncour}, {Gonz{\'a}lez}, {Hennebelle}, {Lu}, {Menten}, {Moraux}, {Tatematsu}, {Walker}, \& {Whitworth}}]{2022A&A...664A..26P}
{Pouteau}, Y., {Motte}, F., {Nony}, T., {et~al.} 2022, \aap, 664, A26

\bibitem[{{Rawat} {et~al.}(2024){Rawat}, {Samal}, {Eswaraiah}, {Wang}, {Elia}, {Panigrahy}, {Zavagno}, {Yadav}, {Walker}, {Jose}, {Ojha}, {Zhang}, \& {Dutta}}]{2024MNRAS.528.1460R}
{Rawat}, V., {Samal}, M.~R., {Eswaraiah}, C., {et~al.} 2024, \mnras, 528, 1460

\bibitem[{{Reyes-Reyes} {et~al.}(2024){Reyes-Reyes}, {Stutz}, {Megeath}, {Xu}, {{\'A}lvarez-Guti{\'e}rrez}, {Sandoval-Garrido}, \& {Liu}}]{2024MNRAS.529.2220R}
{Reyes-Reyes}, S.~D., {Stutz}, A.~M., {Megeath}, S.~T., {et~al.} 2024, \mnras, 529, 2220

\bibitem[{{Rice} {et~al.}(2016){Rice}, {Goodman}, {Bergin}, {Beaumont}, \& {Dame}}]{2016ApJ...822...52R}
{Rice}, T.~S., {Goodman}, A.~A., {Bergin}, E.~A., {Beaumont}, C., \& {Dame}, T.~M. 2016, \apj, 822, 52

\bibitem[{{Riener} {et~al.}(2019){Riener}, {Kainulainen}, {Henshaw}, {Orkisz}, {Murray}, \& {Beuther}}]{2019A&A...628A..78R}
{Riener}, M., {Kainulainen}, J., {Henshaw}, J.~D., {et~al.} 2019, \aap, 628, A78

\bibitem[{{Rosolowsky} {et~al.}(2008){Rosolowsky}, {Pineda}, {Kauffmann}, \& {Goodman}}]{2008ApJ...679.1338R}
{Rosolowsky}, E.~W., {Pineda}, J.~E., {Kauffmann}, J., \& {Goodman}, A.~A. 2008, \apj, 679, 1338

\bibitem[{{Sabatini} {et~al.}(2022){Sabatini}, {Bovino}, {Sanhueza}, {Morii}, {Li}, {Redaelli}, {Zhang}, {Lu}, {Feng}, {Tafoya}, {Izumi}, {Sakai}, {Tatematsu}, \& {Allingham}}]{2022ApJ...936...80S}
{Sabatini}, G., {Bovino}, S., {Sanhueza}, P., {et~al.} 2022, \apj, 936, 80

\bibitem[{{Sabatini} {et~al.}(2019){Sabatini}, {Giannetti}, {Bovino}, {Brand}, {Leurini}, {Schisano}, {Pillai}, \& {Menten}}]{2019MNRAS.490.4489S}
{Sabatini}, G., {Giannetti}, A., {Bovino}, S., {et~al.} 2019, \mnras, 490, 4489

\bibitem[{{Saha} {et~al.}(2022){Saha}, {Tej}, {Liu}, {Liu}, {Issac}, {Lee}, {Garay}, {Goldsmith}, {Juvela}, {Qin}, {Stutz}, {Li}, {Wang}, {Baug}, {Bronfman}, {Xu}, {Zhang}, \& {Eswaraiah}}]{2022MNRAS.516.1983S}
{Saha}, A., {Tej}, A., {Liu}, H.-L., {et~al.} 2022, \mnras, 516, 1983

\bibitem[{{Sakai} {et~al.}(2022){Sakai}, {Sanhueza}, {Furuya}, {Tatematsu}, {Li}, {Aikawa}, {Lu}, {Zhang}, {Morii}, {Nakamura}, {Takemura}, {Izumi}, {Hirota}, {Silva}, {Guzman}, {Sakai}, \& {Yamamoto}}]{2022ApJ...925..144S}
{Sakai}, T., {Sanhueza}, P., {Furuya}, K., {et~al.} 2022, \apj, 925, 144

\bibitem[{{S{\'a}nchez-Monge} {et~al.}(2013){S{\'a}nchez-Monge}, {Palau}, {Fontani}, {Busquet}, {Ju{\'a}rez}, {Estalella}, {Tan}, {Sep{\'u}lveda}, {Ho}, {Zhang}, \& {Kurtz}}]{2013MNRAS.432.3288S}
{S{\'a}nchez-Monge}, {\'A}., {Palau}, A., {Fontani}, F., {et~al.} 2013, \mnras, 432, 3288

\bibitem[{{Sandoval-Garrido} {et~al.}(2024){Sandoval-Garrido}, {Stutz}, {{\'A}lvarez-Guti{\'e}rrez}, {Galv{\'a}n-Madrid}, {Motte}, {Ginsburg}, {Cunningham}, {Reyes-Reyes}, {Redaelli}, {Bonfand}, {Salinas}, {Koley}, {Braine}, {Bronfman}, {Busquet}, {Csengeri}, {Di Francesco}, {Fern{\'a}ndez-L{\'o}pez}, {Garcia}, {Gusdorf}, {Liu}, \& {Sanhueza}}]{2024arXiv241009843S}
{Sandoval-Garrido}, N.~A., {Stutz}, A.~M., {{\'A}lvarez-Guti{\'e}rrez}, R.~H., {et~al.} 2024, arXiv e-prints, arXiv:2410.09843

\bibitem[{{Sanhueza} {et~al.}(2010){Sanhueza}, {Garay}, {Bronfman}, {Mardones}, {May}, \& {Saito}}]{2010ApJ...715...18S}
{Sanhueza}, P., {Garay}, G., {Bronfman}, L., {et~al.} 2010, \apj, 715, 18

\bibitem[{{Sanhueza} {et~al.}(2024){Sanhueza}, {Liu}, {Morii}, {Girart}, {Zhang}, {Stephens}, {Jackson}, {Cortes}, {Koch}, {Cyganowski}, {Saha}, {Beuther}, {Zhang}, {Beltran}, {Cheng}, {Olguin}, {Lu}, {Choudhury}, {Pattle}, {andez-Lopez}, {Hwang}, {Kang}, {Karoly}, {Ginsburg}, {Lyo}, {Taniguchi}, {Jiao}, {Eswaraiah}, {Luo}, {Wang}, {Commercon}, {Li}, {Xu}, {Chen}, {Zapata}, {Chung}, {Nakamura}, {Panigrahy}, \& {Sakai}}]{2024arXiv241208790S}
{Sanhueza}, P., {Liu}, J., {Morii}, K., {et~al.} 2024, arXiv e-prints, arXiv:2412.08790

\bibitem[{{Schnee} {et~al.}(2007){Schnee}, {Caselli}, {Goodman}, {Arce}, {Ballesteros-Paredes}, \& {Kuchibhotla}}]{2007ApJ...671.1839S}
{Schnee}, S., {Caselli}, P., {Goodman}, A., {et~al.} 2007, \apj, 671, 1839

\bibitem[{{Shetty} {et~al.}(2012){Shetty}, {Beaumont}, {Burton}, {Kelly}, \& {Klessen}}]{2012MNRAS.425..720S}
{Shetty}, R., {Beaumont}, C.~N., {Burton}, M.~G., {Kelly}, B.~C., \& {Klessen}, R.~S. 2012, \mnras, 425, 720

\bibitem[{{Solomon} {et~al.}(1987){Solomon}, {Rivolo}, {Barrett}, \& {Yahil}}]{1987ApJ...319..730S}
{Solomon}, P.~M., {Rivolo}, A.~R., {Barrett}, J., \& {Yahil}, A. 1987, \apj, 319, 730

\bibitem[{{Stahler} \& {Palla}(2004)}]{2004fost.book.....S}
{Stahler}, S.~W. \& {Palla}, F. 2004, {The Formation of Stars}

\bibitem[{{Stutz}(2018)}]{2018MNRAS.473.4890S}
{Stutz}, A.~M. 2018, \mnras, 473, 4890

\bibitem[{{Stutz} \& {Gould}(2016)}]{2016A&A...590A...2S}
{Stutz}, A.~M. \& {Gould}, A. 2016, \aap, 590, A2

\bibitem[{{Syed} {et~al.}(2020){Syed}, {Wang}, {Beuther}, {Soler}, {Rugel}, {Ott}, {Brunthaler}, {Kerp}, {Heyer}, {Klessen}, {Henning}, {Glover}, {Goldsmith}, {Linz}, {Urquhart}, {Ragan}, {Johnston}, \& {Bigiel}}]{2020A&A...642A..68S}
{Syed}, J., {Wang}, Y., {Beuther}, H., {et~al.} 2020, \aap, 642, A68

\bibitem[{{Towner} {et~al.}(2024){Towner}, {Ginsburg}, {Dell'Ova}, {Gusdorf}, {Bontemps}, {Csengeri}, {Galv{\'a}n-Madrid}, {Louvet}, {Motte}, {Sanhueza}, {Stutz}, {Bally}, {Baug}, {Chen}, {Cunningham}, {Fern{\'a}ndez-L{\'o}pez}, {Liu}, {Lu}, {Nony}, {Valeille-Manet}, {Wu}, {{\'A}lvarez-Guti{\'e}rrez}, {Bonfand}, {Di Francesco}, {Nguyen-Luong}, {Olguin}, \& {Whitworth}}]{2024ApJ...960...48T}
{Towner}, A.~P.~M., {Ginsburg}, A., {Dell'Ova}, P., {et~al.} 2024, \apj, 960, 48

\bibitem[{{Ungerechts} {et~al.}(1997){Ungerechts}, {Bergin}, {Goldsmith}, {Irvine}, {Schloerb}, \& {Snell}}]{1997ApJ...482..245U}
{Ungerechts}, H., {Bergin}, E.~A., {Goldsmith}, P.~F., {et~al.} 1997, \apj, 482, 245

\bibitem[{{Wang} \& {Wang}(2023)}]{2023A&A...674A..46W}
{Wang}, C. \& {Wang}, K. 2023, \aap, 674, A46

\bibitem[{{Wang} {et~al.}(2020){Wang}, {Koch}, {Galv{\'a}n-Madrid}, {Lai}, {Liu}, {Lin}, \& {Pattle}}]{2020ApJ...905..158W}
{Wang}, J.-W., {Koch}, P.~M., {Galv{\'a}n-Madrid}, R., {et~al.} 2020, \apj, 905, 158

\bibitem[{{Yun} {et~al.}(2021){Yun}, {Lee}, {Evans}, {Offner}, {Heyer}, {Cho}, {Gaches}, {Yang}, {Chen}, {Choi}, {Lee}, {Baek}, {Choi}, {Kim}, {Kang}, {Lee}, \& {Tatematsu}}]{2021arXiv210713323Y}
{Yun}, H.-S., {Lee}, J.-E., {Evans}, Neal~J., I., {et~al.} 2021, arXiv e-prints, arXiv:2107.13323

\bibitem[{{Zhou} {et~al.}(2022){Zhou}, {Li}, \& {Chen}}]{2022MNRAS.513..638Z}
{Zhou}, J.-X., {Li}, G.-X., \& {Chen}, B.-Q. 2022, \mnras, 513, 638

\end{thebibliography}

\noindent\rule{\linewidth}{0.4pt}

\hspace{-5mm}$^{1}$Departamento de Astronom\'{i}a, Universidad de Concepci\'{o}n, Casilla 160-C, Concepci\'{o}n, Chile\\
              \vspace{0mm}
              \hspace{-1mm}$^{2}$Franco-Chilean Laboratory for Astronomy, IRL 3386, CNRS and Universidad de Chile, Santiago, Chile\\
              \hspace{-1mm}$^{3}$Univ. Grenoble Alpes, CNRS, IPAG, 38000 Grenoble, France\\
              \hspace{-1mm}$^{4}$Department of Astronomy, University of Florida, PO Box 112055, USA\\
               \hspace{-1mm}$^{5}$Instituto de Radioastronom$\acute{\text{i}}$a y Astrof$\acute{\text{i}}$sica, Universidad Nacional Aut$\acute{\text{o}}$noma de M$\acute{\text{e}}$xico, Morelia, Michoac$\acute{\text{a}}$n 58089, M$\acute{\text{e}}$xico\\
               $^{6}$National Astronomical Observatory of Japan, National Institutes of
              Natural Sciences, 2-21-1 Osawa, Mitaka, Tokyo 181-8588, Japan\\
              $^{7}$Department of Astronomical Science, SOKENDAI (The Graduate
              University for Advanced Studies), 2-21-1 Osawa, Mitaka, Tokyo
              181-8588, Japan\\
              $^{8}$S. N. Bose National Centre for Basic Sciences, Sector-III, Salt Lake, Kolkata 700106, India\\
              $^{9}$Departament de F\'isica Qu\`antica i Astrof\'isica (FQA), Universitat de Barcelona (UB), Mart\'i i Franqu\`es 1, 08028 Barcelona, Catalonia, Spain\\
              $^{10}$Institut de Ci\`encies del Cosmos (ICCUB), Universitat de Barcelona, Mart\'i i Franqu\`es 1, 08028 Barcelona, Catalonia, Spain\\
              $^{11}$Institut d'Estudis Espacials de Catalunya (IEEC), Esteve Terradas 1, Edifici RDIT, Ofic. 212 Parc Mediterrani de la Tecnologia (PMT) Campus del Baix LLobregat – UPC 08860 Castelldefels (Barcelona), Catalonia, Spain\\
              $^{12}$Laboratoire d’Astrophysique de Bordeaux, Univ. Bordeaux, CNRS, UMR 5804, F-33615 Pessac, France\\
              $^{13}$School of Physics and Astronomy, Yunnan University, Kunming, 650091, People’s Republic of China\\
              $^{14}$Laboratoire d’Astrophysique de Bordeaux, Univ. Bordeaux, CNRS,B18N, allée Geoffroy Saint-Hilaire, 33615 Pessac, France\\
              $^{15}$Laboratoire de Physique de l’$\acute{\text{E}}$cole Normale Sup$\acute{\text{e}}$rieure, ENS, Universit$\acute{\text{e}}$ PSL, CNRS, Sorbonne Université, Universit$\acute{\text{e}}$ Paris Cit$\acute{\text{e}}$, F-75005, Paris, France\\
              $^{16}$Observatoire de Paris, PSL University, Sorbonne Universit$\acute{\text{e}}$, LERMA, 75014, Paris, France\\
              $^{17}$Instituto Argentino de Radioastronomía (CCT-La Plata, CONICET;UNLP; CICPBA), C.C. No. 5, 1894, Villa Elisa, Buenos Aires, Argentina\\
              $^{18}$SKA Observatory, Jodrell Bank, Lower Withington, Macclesfield SK11 9FT, United Kingdom\\
              $^{19}$Astronomy Department, Universidad de Chile, Casilla 36-D, Santiago, Chile\\
              $^{20}$Departments of Astronomy and Chemistry, University of Virginia, Charlottesville, VA 22904, USA\\

              \hspace{-5mm}Corresponding author: A. Koley \\
              
              \hspace{-5mm}\email{atanuphysics15@gmail.com}\\
            
\noindent\rule{\linewidth}{0.4pt}

\begin{appendix}

\onecolumn

\appendix

\section{Integrated intensity weighted velocity (moment 1) maps of 15 massive protoclusters}\label{Appendix0}
 We show the integrated intensity weighted velocity (moment 1) maps for 15 massive protoclusters in Fig. \ref{fig:fig8}. We note that all the moment 1 maps have been constructed by subtracting the systemic velocity ($V$$_{\text{sys}}$), which result in a more prominent appearance of the clouds with negative and positive velocity (with respect to the $V$$_{\text{sys}}$). We will study these maps in the upcoming C$^{18}$O (2$-$1) kinematic paper (Koley et al. in prep).

\begin{figure*}[h!]
	\centering 
	\includegraphics[width=2.5in,height=1.8in,angle=0]{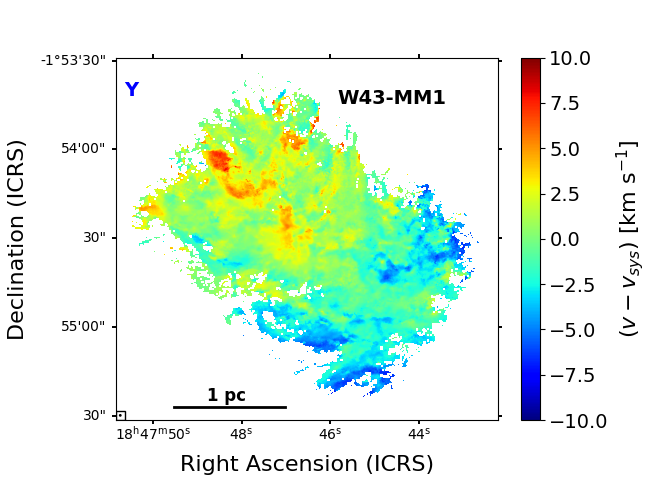}\includegraphics[width=2.5in,height=1.8in,angle=0]{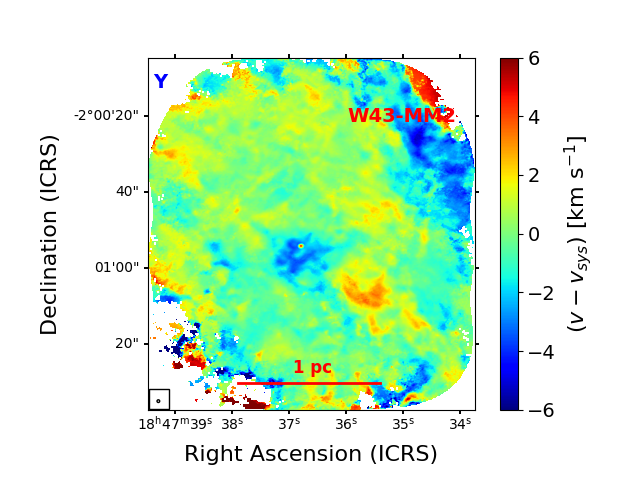}\includegraphics[width=2.5in,height=1.8in,angle=0]{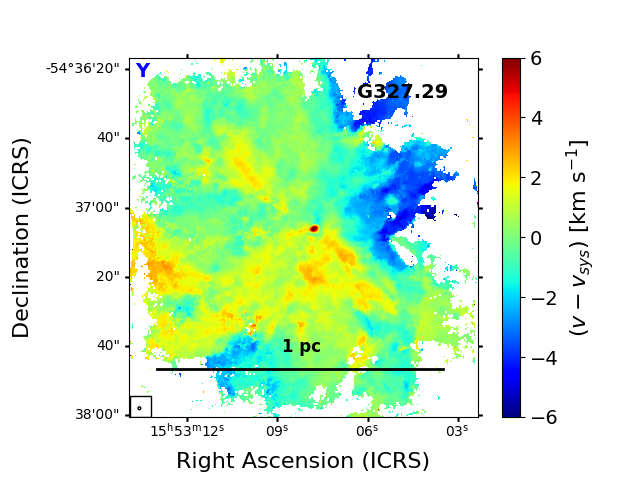}\\\includegraphics[width=2.5in,height=1.8in,angle=0]{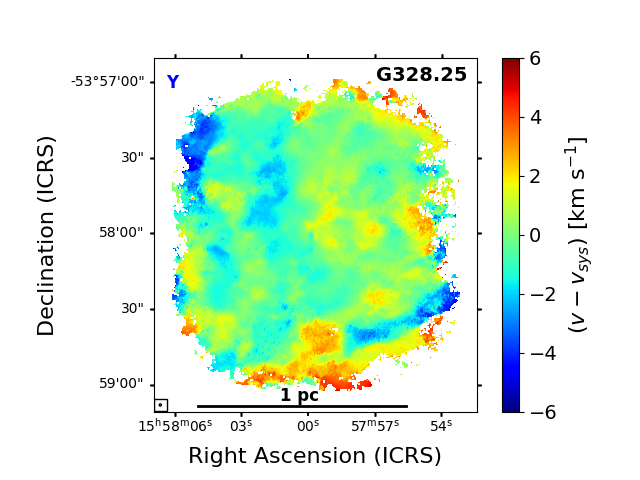}\includegraphics[width=2.5in,height=1.8in,angle=0]{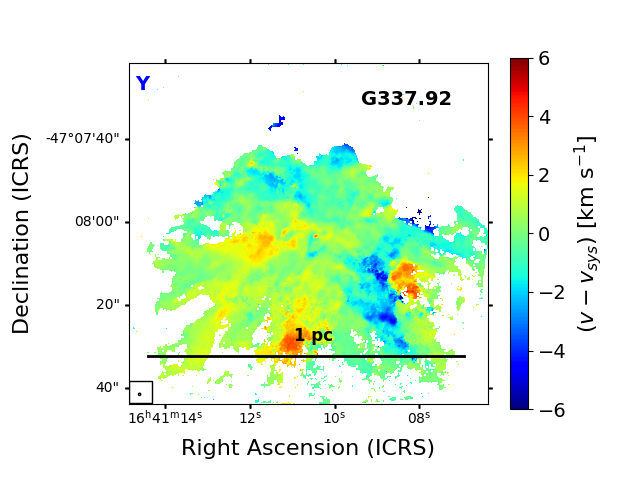}\includegraphics[width=2.5in,height=1.8in,angle=0]{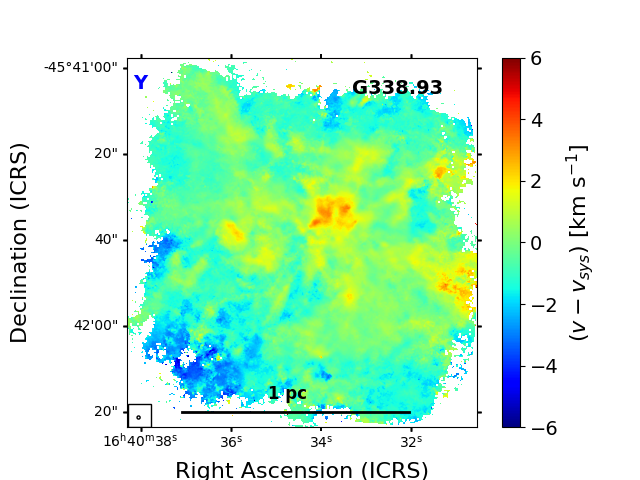}\\\includegraphics[width=2.5in,height=1.8in,angle=0]{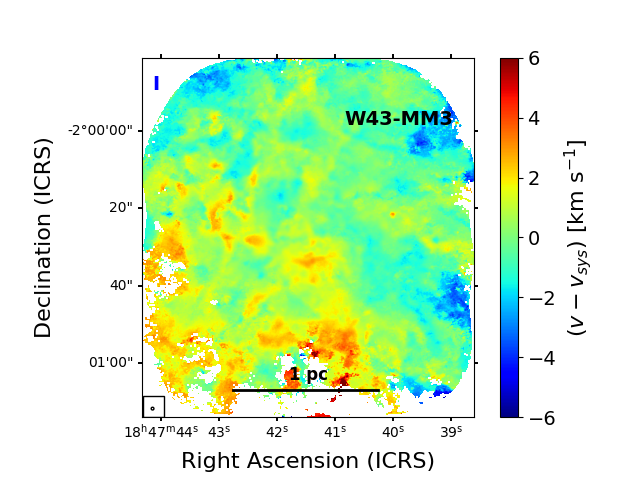}\includegraphics[width=2.5in,height=1.8in,angle=0]{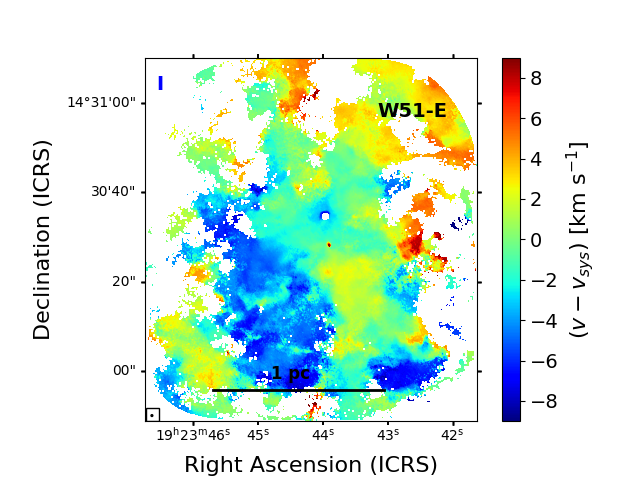}\includegraphics[width=2.5in,height=1.8in,angle=0]{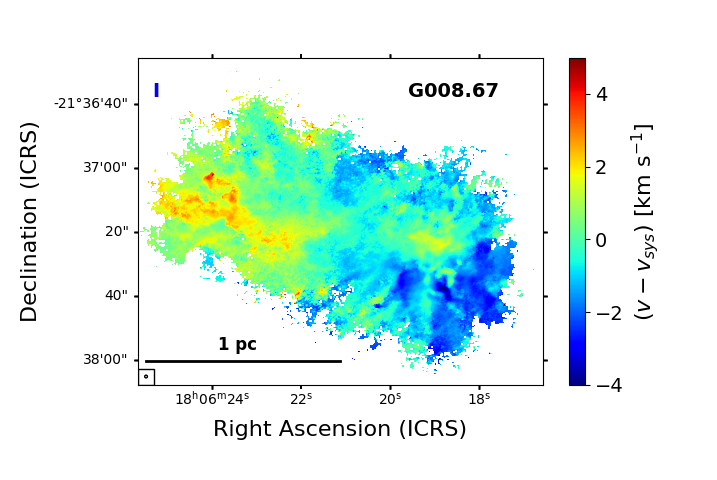}\\
	\includegraphics[width=2.5in,height=1.8in,angle=0]{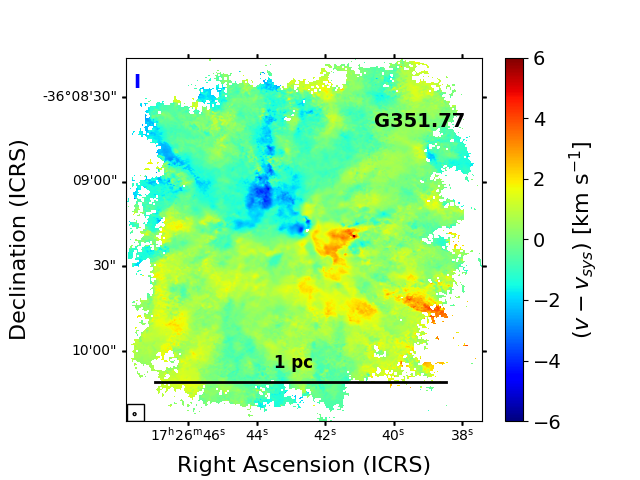}\includegraphics[width=2.5in,height=1.8in,angle=0]{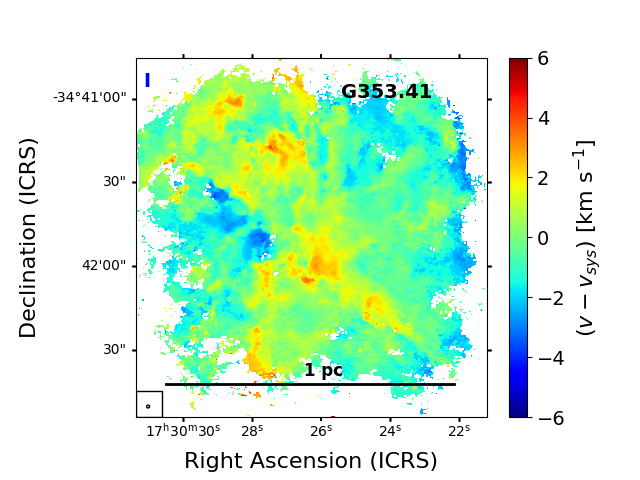}\includegraphics[width=2.5in,height=1.8in,angle=0]{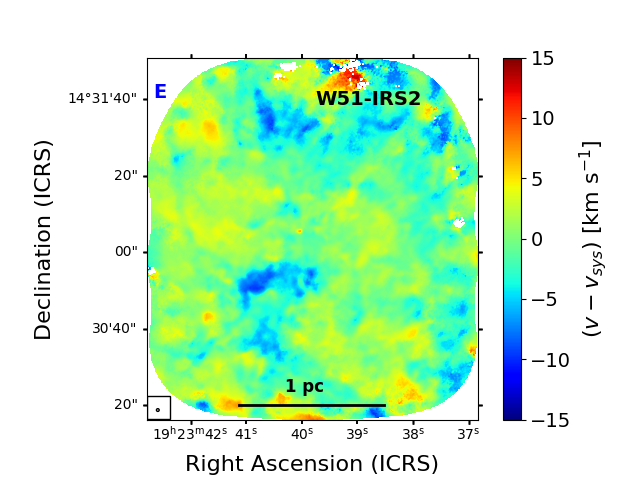}\\
	\includegraphics[width=2.5in,height=1.8in,angle=0]{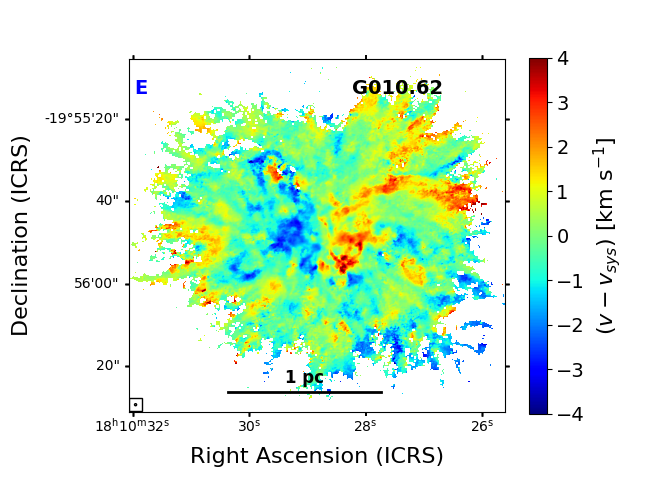}\includegraphics[width=2.5in,height=1.8in,angle=0]{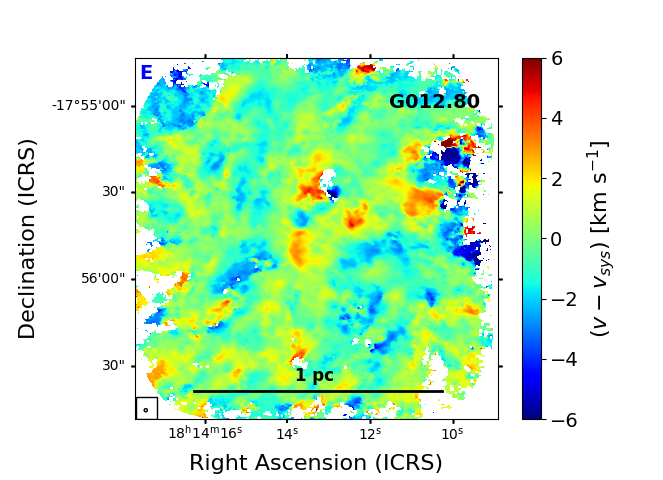}\includegraphics[width=2.5in,height=1.8in,angle=0]{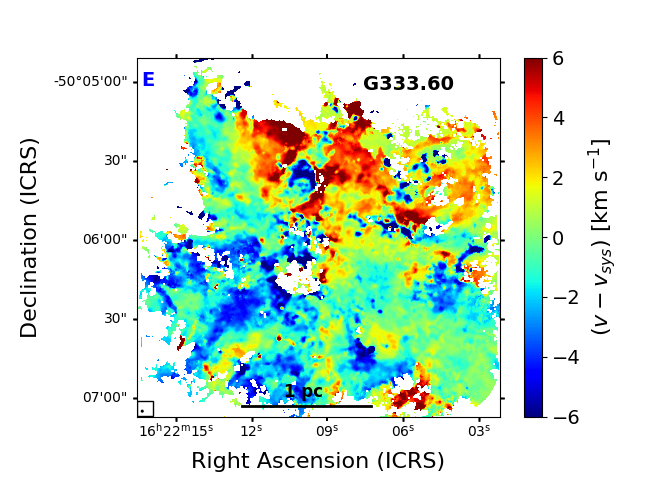}
	\caption{Integrated intensity weighted velocity (moment 1) maps of the C$^{18}$O ($J$=2$-$1) lines for 15 protoclusters. Symbols Y, I and E in the figures indicate young, intermediate and evolved protoclusters respectively (see Section \ref{section_0}).}
	\label{fig:fig8}
\end{figure*}

\section{Noise estimation }\label{Appendix1}
The noise distribution is calculated for 15 regions using the line-free channels. We take the maximum line-free channels and calculate the noise for each pixel in the region. In Fig.\ref{fig:fig9},  we show the noise distribution maps for G008.67 and G353.41 regions. The noise level is almost constant in the central areas and gradually increases as one moves towards the edge. This is caused by the spatial variation in the sensitivity of the primary beam. These maps are scaled-up versions of their corresponding primary beam response functions (\texttt{.pb} image cube produced during CASA task $\texttt{TCLEAN}$). In Table \ref{tab:table2}, we mention the noise values for each protocluster, which are based on the central portion of the noise maps, where the level of noise is almost constant.

\begin{figure*}
	\centering 
	\includegraphics[width=3.6in,height=2.6in,angle=0]{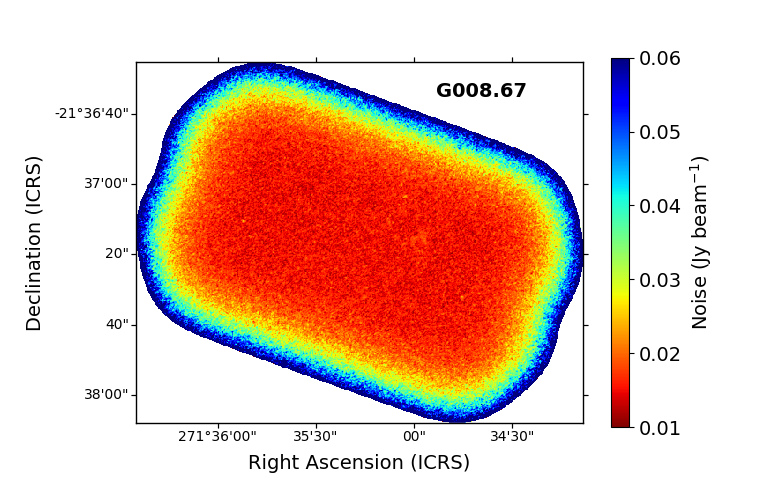}\includegraphics[width=3.6in,height=2.6in,angle=0]{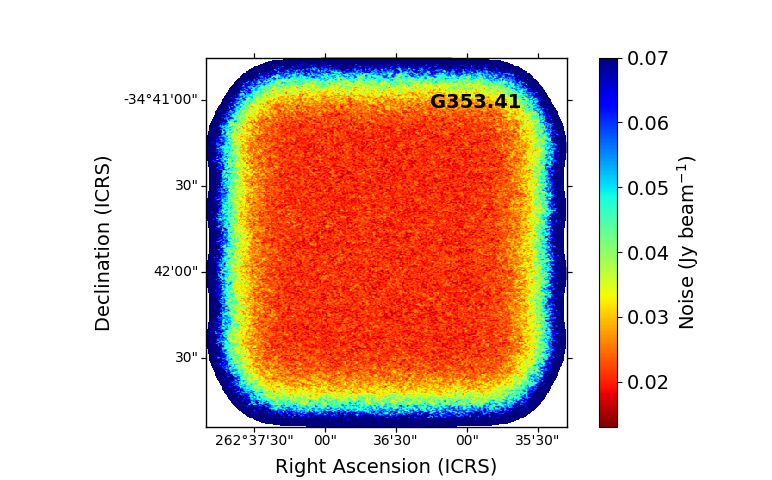}
	\caption{Left figure is the noise map (per channel) of the G008.67 region. Right figure is the noise map (per channel) of the G353.41 region.}
	\label{fig:fig9}
\end{figure*}

\section{Examining the effect of noise on the number of decomposed components}\label{Appendix2}
We examine whether the level of noise affects the number of decomposed components in these regions. For this purpose, we choose one protocluster G010.62.  In this region, we first decompose the pixel-wise spectra using the original noise map and obtain the distribution of the number of components (${n}$). We then carefully check the original noise map and include random noise in each pixel of the model cube or noise free data cube. The noise value that has been included in each pixel of the model cube is obtained from the edge side of the original noise map. As an example, in the upper panel of Fig. \ref{fig:fig10}, we illustrate the noise map and the number of decomposed components ($n$) of the G010.62 region. The number of maximum Gaussian components ($n$) in the central region of G010.62 is four and decreases towards the edge where the value of $n$ is one. We check the noise map and notice that the value of the noise towards the central region is $\sim$ 0.035 Jy beam$^{-1}$ and the gradually increases towards the the edge side, where it is $\sim$ 0.053 Jy beam$^{-1}$. We take this noise value from the edge side and include it in each pixel of the model cube and create a data cube. In this data cube, the noise in each pixel is 0.053 Jy beam$^{-1}$. Thereafter, we decompose that cube into multi-Gaussian components with the same signal-to-noise (SNR) cutoff as was done for the original one and obtain the distribution of the number of components ($n$) in this region. The constant noise map and the resulting distribution of the components are displayed on the left and right sides of the lower panel of Fig. \ref{fig:fig10}. We now compare the distribution of $n$ for both the cases and notice that due to the increase in noise, the number of components in few regions is less compared to the original one. Consequently, our conclusion is that the increase in the number of components towards the centre of the protocluster and the decrease towards the outer side may not always be physical, but may also be related to the level of noise.

\begin{figure*}
	\centering 
	\includegraphics[width=3.6in,height=2.6in,angle=0]{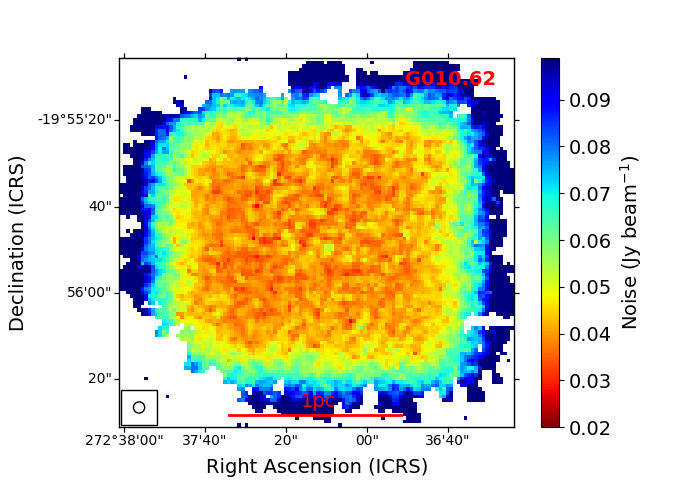}\includegraphics[width=3.6in,height=2.6in,angle=0]{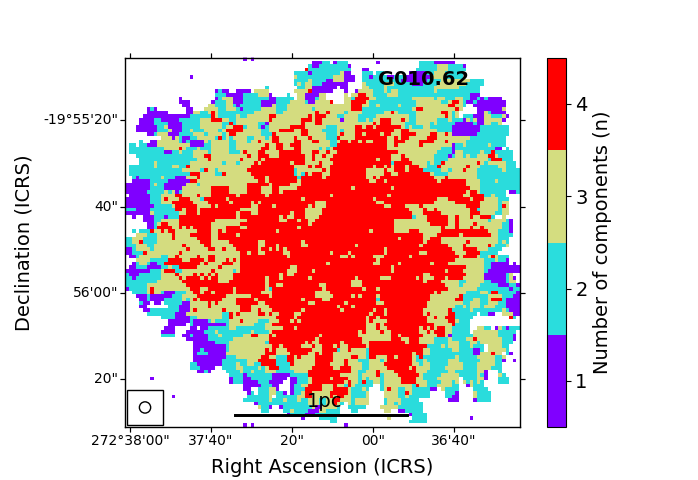}\\
    \includegraphics[width=3.6in,height=2.6in,angle=0]{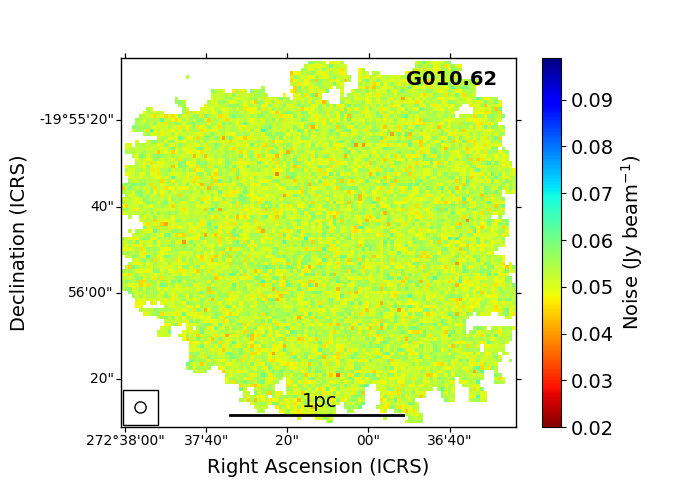}\includegraphics[width=3.6in,height=2.6in,angle=0]{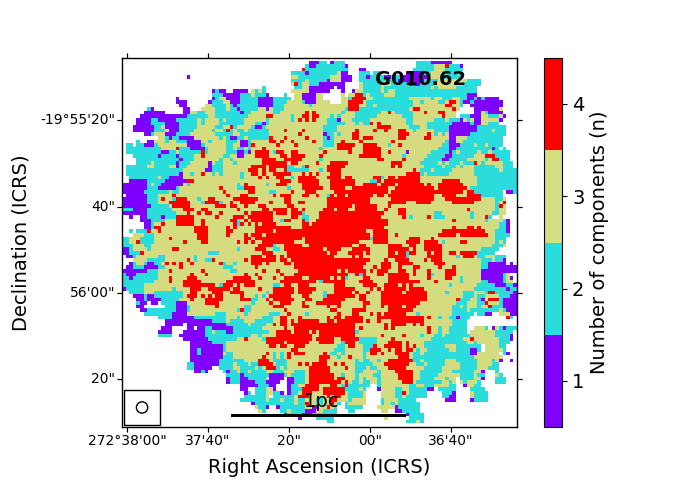}
	
	\caption{Upper left figure is the noise map of the G010.62 region (at 2.5$''$ resolution, same as the dust temperature cube). Upper right figure is the distribution of number of decomposed components ($n$) in the G010.62 region. Lower left figure is the constant noise map at $\sim$ 0.053 Jy beam$^{-1}$. Lower right figure is the distribution of number of decomposed components ($n$) when the noise level in each pixel is equal to $\sim$ 0.053 Jy beam$^{-1}$.}
	\label{fig:fig10}
\end{figure*}
\onecolumn


\section{Model spectra in G333.60 region}\label{Appendix3}

We have shown a few sample of model spectra in  Fig. \ref{fig:fig11} in the region of G333.60. Here, the observed C$^{18}$O ($J$ = 2$-$1) spectra are indicated by black solid lines, while decomposed spectra and the model spectra are indicated by red solid lines. We have also shown the residuals for each of the spectra.

\begin{figure*}[h!]
	\centering 
	\includegraphics[width=7.2in,height=7.8in,angle=0]{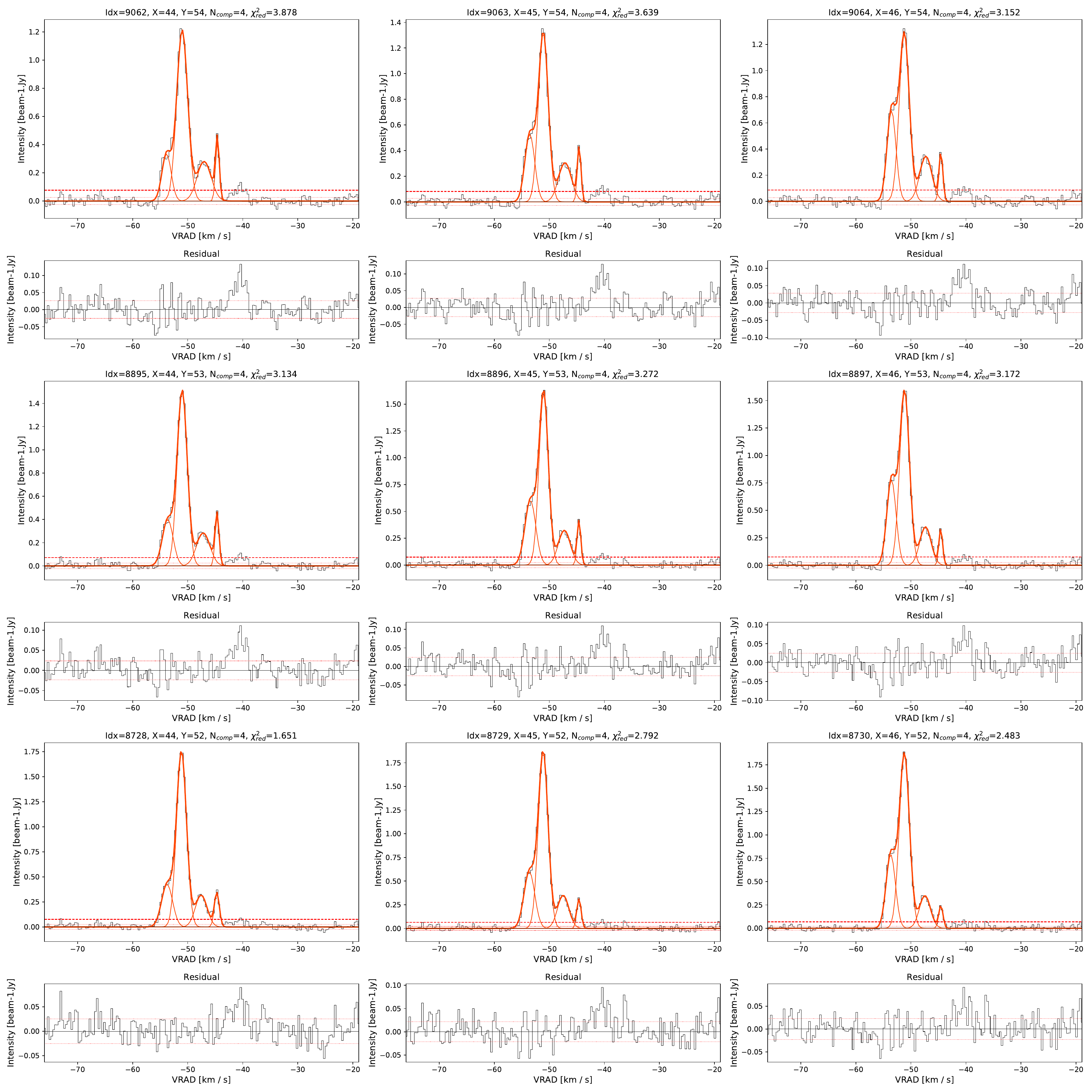}
	
	\caption{First, third, and fifth rows show the observed and model C$^{18}$O ($J$ = 2$-$1) spectra in G333.60 region. Black solid line denotes the observed spectra, whereas, red sold lines represent the decomposed and resultant model spectra in G333.60 region.  In each of the figures two red dotted horizontal lines denote the $\pm$1$\sigma$ value of the noise. In addition one solid horizontal red line represents the $+$3$\sigma$ noise level of the spectra. In the header of each of the figures idx represents the spectra number that is obtained from the \texttt{Gausspy+} module. X and Y denote the pixel number in the image cube, N$_{\text{comp}}$ is the number of decomposed components and $\chi_{\text{red}}^{2}$ is the reduced chi-squared value after fitting.  Second, fourth, and sixth rows represent the residuals of the corresponding spectra after being fitted with model components. Here also two red dotted horizontal lines in each figure denotes the $\pm$1$\sigma$ value of the noise.}
	\label{fig:fig11}
\end{figure*}


\onecolumn

	


\section{Integrated intensity weighted velocity dispersion}\label{Appendix5}

The traditional intensity-weighted velocity dispersion ($\sigma_{\text{v}}$) of a position-position-velocity (PPV) image cube is obtained by:

\begin{equation}\label{eqn_1}
	\displaystyle    \sigma_{\text{v}} = \sqrt{\frac{\sum_{i} ^{} I_{i}({v}_{i}-M_{1})^{2}~d{v}_{i}} {\sum_{i} ^{} I_{i}~d{v}_{i} }}
\end{equation}

Here, $I_{i}$ is the intensity of the $i^{th}$ channel, $M_{1}$ is the intensity-weighted centre velocity (moment 1) map of the spectra. $M_{1}$ is calculated by
the formula:

\begin{equation}
	\displaystyle    M_{1} = \frac{\sum_{i} ^{} I_{i} {v}_{i}~d{v}_{i}} {\sum_{i} ^{} I_{i}~d{v}_{i} }
\end{equation}

However, $\sigma_{\text{v}}$ does not represent the effective velocity dispersion for a spectra when the spectra contains multiple components and these components are separated from each other or more specifically they are disjoint components. In this case, it will always \textbf{overestimate} the  $\sigma_{\text{v}}$ from the original value. For this reason, after obtaining the decomposed model components using \texttt{Gausspy+} module, we calculate the effective velocity dispersion ($\sigma_{\text{eff}}$) by the formula:\\

\begin{equation}
	\displaystyle    \sigma_{\text{eff}} = \sqrt{\frac{\sum_{i} ^{} I_{i,1}(v_{i,1}-v_{c,1})^{2}~dv_{i,1} + \sum_{i} ^{} I_{i,2}(v_{i,2}-v_{c,2})^{2}~dv_{i,2} + .......} {\sum_{i} ^{} I_{i,1}~dv_{i,1} + \sum_{i} ^{} I_{i,2}~dv_{i,2} +....... }}
\end{equation}

\vspace{-4.3mm}

\begin{equation}
	\displaystyle     \sigma_{\text{eff}} = \sqrt{\frac{\sum_{i} ^{} I_{i,1}(v_{i,1}-v_{c,1})^{2}~dv_{i,1} } {\sum_{i} ^{} I_{i,1}~dv_{i,1} + \sum_{i} ^{} I_{i,2}~dv_{i,2} +....... } + \frac{\sum_{i} ^{} I_{i,2}(v_{i,2}-v_{c,2})^{2}~dv_{i,2} } {\sum_{i} ^{} I_{i,1}~dv_{i,1} + \sum_{i} ^{} I_{i,2}~dv_{i,2} +....... } + ......}
\end{equation}

\vspace{-4.3mm}



\begin{equation}
	\displaystyle  \sigma_{\text{eff}} = \sqrt{\Biggl(\frac{\sum_{i} ^{} I_{i,1}~dv_{i,1}}{\sum_{i} ^{} I_{i,1}~dv_{i,1} + \sum_{i} ^{} I_{i,2}~dv_{i,2} +..}\Biggl)\Biggl(\frac{\sum_{i} ^{} I_{i,1}(v_{i,1}-v_{c,1})^{2}~dv_{i,1} } {\sum_{i} ^{} I_{i,1}~dv_{i,1} }\Biggl) + \Biggl(\frac{\sum_{i} ^{} I_{i,2}~dv_{i,2}}{\sum_{i} ^{} I_{i,1}~dv_{i,1} + \sum_{i} ^{} I_{i,2}~dv_{i,2} +..}\Biggl)\Biggl(\frac{\sum_{i} ^{} I_{i,2}(v_{i,2}-v_{c,2})^{2}~dv_{i,2} } {\sum_{i} ^{} I_{i,2}~dv_{i,2} }\Biggl) + ..}
\end{equation}

\begin{equation}\label{eqn_2}
	\displaystyle  \sigma_{\text{eff}} = \sqrt{\Biggl(\frac{w_{1}}{w}~\sigma_{1}^{2} + \frac{w_{2}}{w}~\sigma_{2}^{2}  + ..\Biggl)}
\end{equation}

Where $I_{i,1}$, $I_{i,2}$ are the intensities of the first and second components for the $i^{th}$ channel; $v_{i,1}$, $v_{i,2}$ are the velocities of the first and second components for the $i^{th}$ channel; $v_{c,1}$, $v_{c,2}$ are the centre velocities of the first and second components;$w= \Biggl(\sum_{i,1} ^{} I_{i,1}~dv_{i,1} + \sum_{i,2} ^{} I_{i,2}~dv_{i,2} +.. \Biggl)$ = total integrated intensity; $w_{1}= \sum_{i,1} ^{} I_{i,1}~dv_{i,1}$ = 
integrated intensity of the first component; $w_{2}= \sum_{i,2} ^{} I_{i,2}~dv_{i,2}$ = integrated intensity of the second component; $\sigma_{1}$ and $\sigma_{2}$ are the intensity-weighted velocity dispersion for the first and second components. Previously, \cite{2023MNRAS.519.3851P} used a similar kind of formula for effective velocity dispersion to avoid overestimation of velocity dispersion. However, it was not exactly the same as our effective velocity dispersion ($\sigma_{\text{eff}}$). \\

In the average spectra of two regions, W51-E and W51-IRS2, we observed broad separate clouds (see Section \ref{section_2}). In these regions, we observe a prominent overestimation of velocity dispersion if we calculate it using Eqn. \ref{eqn_1} instead of Eqn. \ref{eqn_2}.
In Fig. \ref{fig:fig12}, we show the ratio of $\sigma_{\text{eff}}$ and $\sigma_{\text{v}}$ for these two regions and notice that for most pixels the ratio is below 0.50.\\

\onecolumn

\begin{figure*}
	\centering 
	\includegraphics[width=3.6in,height=2.6in,angle=0]{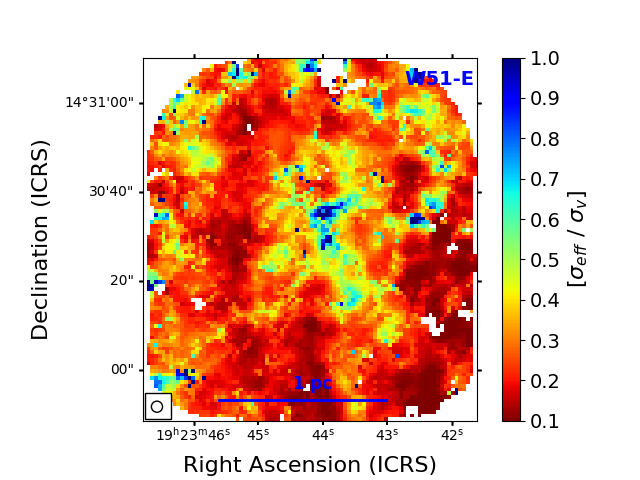}\includegraphics[width=3.6in,height=2.6in,angle=0]{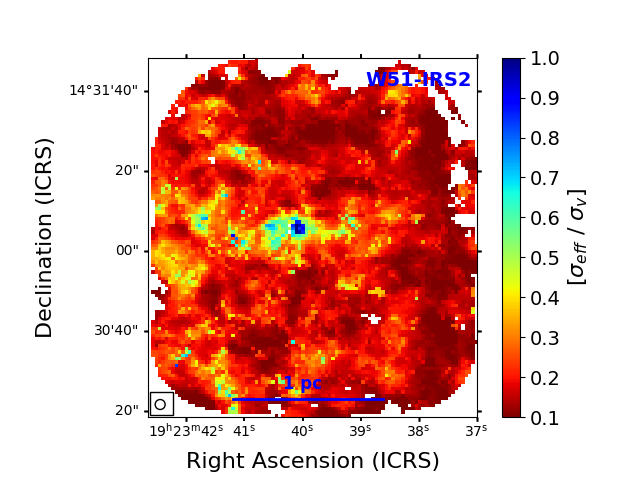}
	
	\caption{Left figure is the ratio of the effective velocity dispersion ($\sigma_{\text{eff}}$) and the traditional velocity dispersion ($\sigma_{\text{v}}$) for W51-E region. Right figure is the ratio of the effective velocity dispersion ($\sigma_{\text{eff}}$) and the traditional velocity dispersion ($\sigma_{\text{v}}$) for W51-IRS2 region.}
	\label{fig:fig12}
\end{figure*}

{\color{black}
\section{Calculation of position-velocity (PV) diagram towards an inclined angle }\label{Appendix6}
In a few cases, we notice that integrated intensity structures (moment 0 map) are inclined on the plane of the sky rather than towards the vertical or horizontal direction on the sky plane. Therefore, to study the position-velocity (PV) diagram properly, it was necessary to rotate the axes in parallel and perpendicular directions of the integrated intensity maps. In the following, we have outlined the necessary steps that were required to do this with an example shown in Figure \ref{fig:fig13}. Suppose that an integrated intensity structure (shown in orange color) is present on the sky plane with an inclination angle $\theta$ with respect to the horizontal axis (Right Ascension). If one rotates this structure with an angle $-\theta$, the structure looks like the violet dashed line, and the pixel positions of the lower left corner and the upper right corner are \texttt{x\_initial\_0, y\_initial\_0} and \texttt{x\_final, y\_final} respectively. The increment between two successive pixels along the horizontal and vertical directions of the image plane is $\Delta x$ and $\Delta y$. To obtain the information of the equidistant points within the inclined intensity structure (which are mentioned with red solid squares), we increase the pixel value successively from the lower left corner in two directions, which are parallel and perpendicular of the inclined structure. These directions are indicated by green and blue color arrows.\\  

We start from the lower left hand pixel \texttt{x\_initial\_0, y\_initial\_0} and count all the pixels round[\texttt{x\_initial\_0+$\Delta y$.cos(90+$\theta$)},\texttt{y\_initial\_0+$\Delta y$.sin(90+$\theta$)}] by increasing $\Delta y$ with one unit from 0 to (\texttt{y\_final - y\_initial\_0}). After that, we change the initial pixel value:  \texttt{x\_initial\_0} $\rightarrow$ \texttt{x\_initial\_0+$\Delta x$.cos($\theta$)}  and \texttt{y\_initial\_0} $\rightarrow$ \texttt{y\_initial\_0+$\Delta x$.sin($\theta$)} and count all the equidistant points as before. This same procedure goes on and each time the initial pixel value is modified by increasing the $\Delta x$ with unity and this step will go on up to (\texttt{x\_final - x\_initial\_0}). In this way all the equidistant points that are enclosed within the inclined intensity structure are counted.\\

We calculated the large-scale velocity gradient in 15 protoclusters using the above procedure. First, we decompose the pixel-wise spectra into multi-Gaussian components. After that we create image planes for center velocities and the integrated intensities for each of the components. Then we plot the center velocities of all of these components in two different directions. We give the color of each component according to its integrated intensity. We performed this analysis for all 15 protoclusters by varying the inclination angle from 0 to 360$^{\circ}$. However, only in two protoclusters G008.67 and W43-MM1, we obtain a large-scale velocity gradient. These are shown in Figures \ref{fig:fig14} and \ref{fig:fig15}. While fitting, we give more weight to the components with high integrated intensities. The velocity gradients that we obtain are 1.95 $\pm$ 0.01 and 2.99 $\pm$ 0.01 km s$^{-1}$ pc$^{-1}$ for these two protoclusters. We fitted velocity gradients up to half of the image plane in the W43-MM1 protocluster because after that point there is no prominent velocity gradient observed.\\

}

\begin{figure*}
	\centering 
	\includegraphics[width=3.0in,height=2.8in,angle=0]{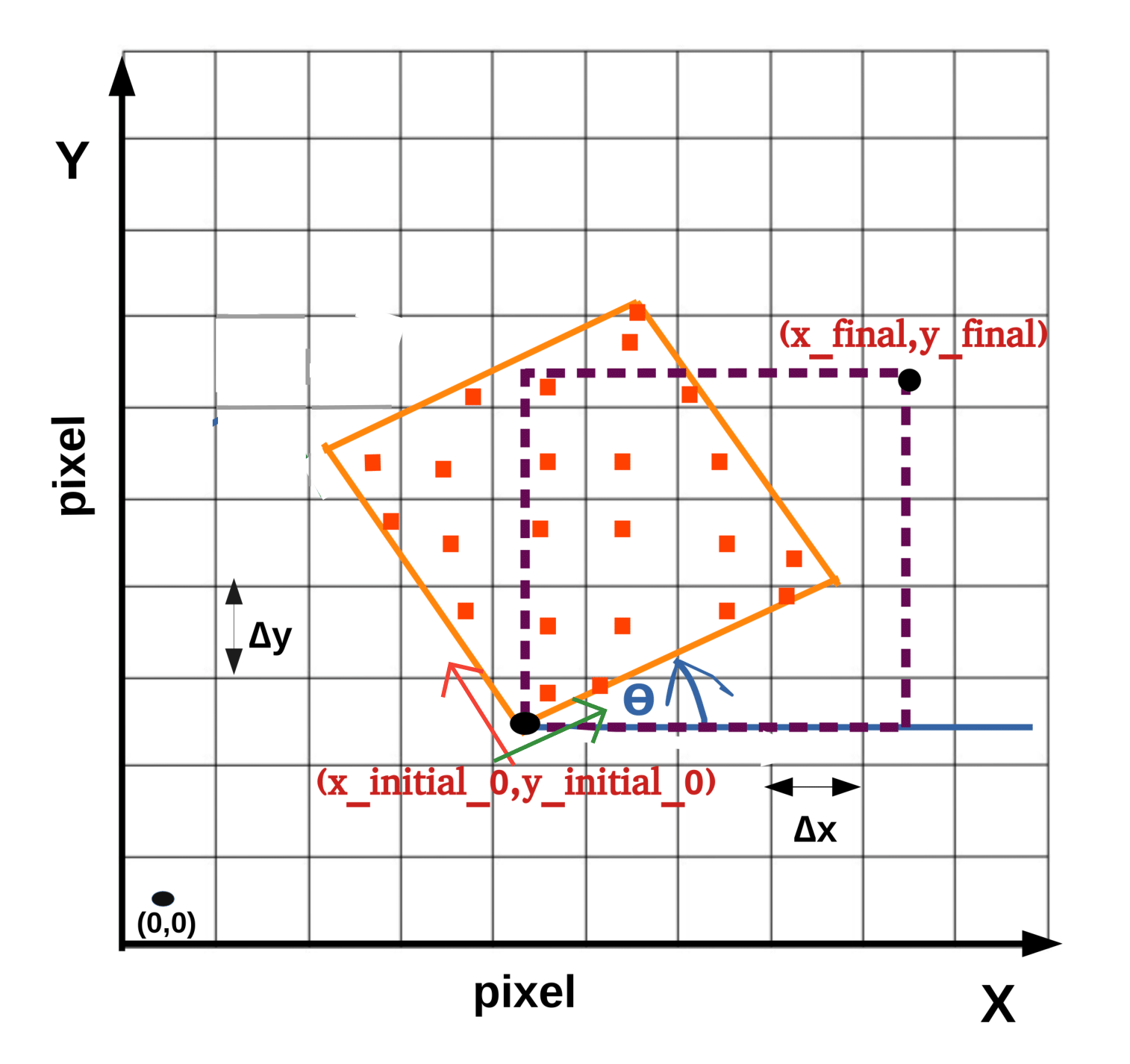}
	
	\caption{{\color{black}Inclined (with an angle $\theta$) integrated intensity structure (orange square) on the two dimensional image plane. The same integrated intensity structure shown in violet dashed line when rotate with an angle - $\theta$. The red solid squares are the pixels which are enclosed within the inclined integrated intensity. Green and red arrows denote the two directions parallel and perpendicular to the inclined structure.}}
	\label{fig:fig13}
\end{figure*}

\begin{figure*}
    \centering
    \includegraphics[width=2.3in, height=2.0in]{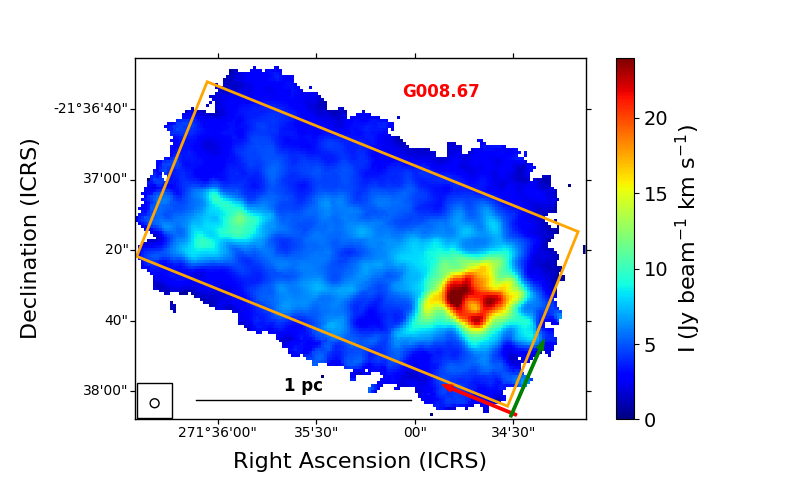}
    \includegraphics[width=2.3in, height=2.0in]{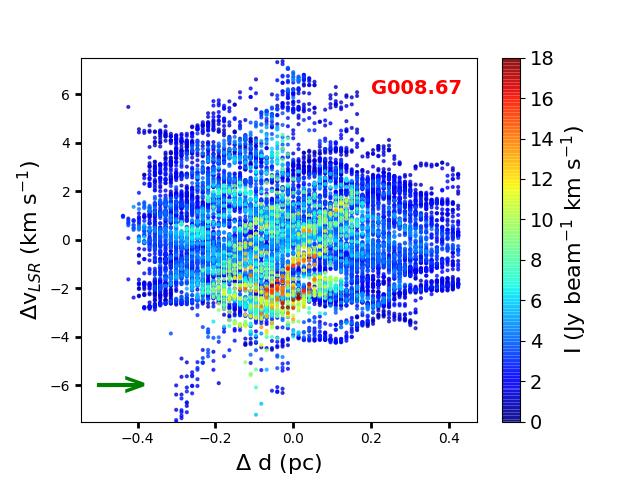} \includegraphics[width=2.3in, height=2.0in]{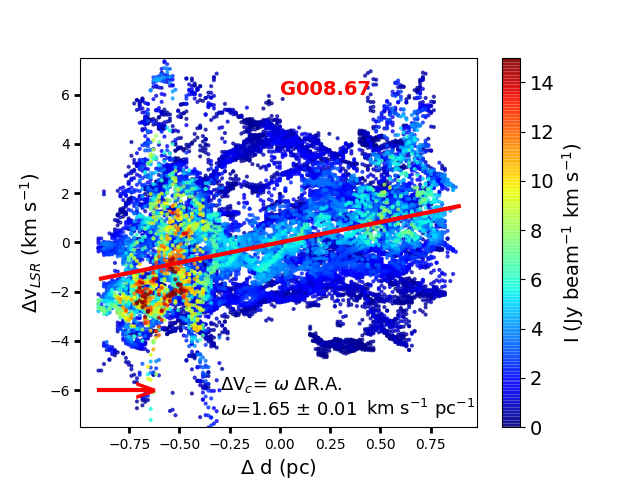}
    \caption{{\color{black}Left figure is the integrated intensity map (moment 0) of the G008.67 region obtained from 2.5$''$ circular beam. Here, the orange rectangular region denotes the area where
we examine the position-velocity (PV) diagram in two perpendicular directions (towards the green and red arrows). Middle figure is the intensity weighted PV diagram towards the direction of the green arrow mentioned in the left figure. Right figure is the intensity weighted PV diagram towards the direction of the red arrow mentioned in the left figure. Color bars in the Middle and Right figures denotes the integrated intensity of the components. The fitted velocity gradient is 1.95 $\pm$ 0.01 km s$^{-1}$ pc$^{-1}$ which is shown in red solid color.}}
    \label{fig:fig14}
\end{figure*}

\begin{figure*}[!hbt]
    \centering
    \includegraphics[width=2.3in, height=2.0in]{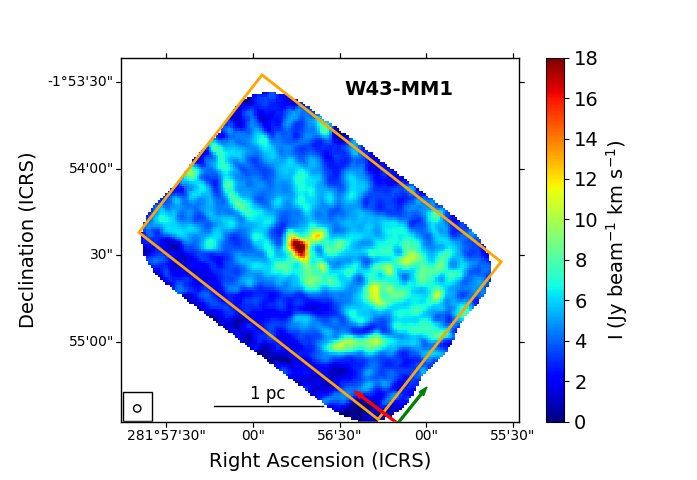}
    \includegraphics[width=2.3in, height=2.0in]{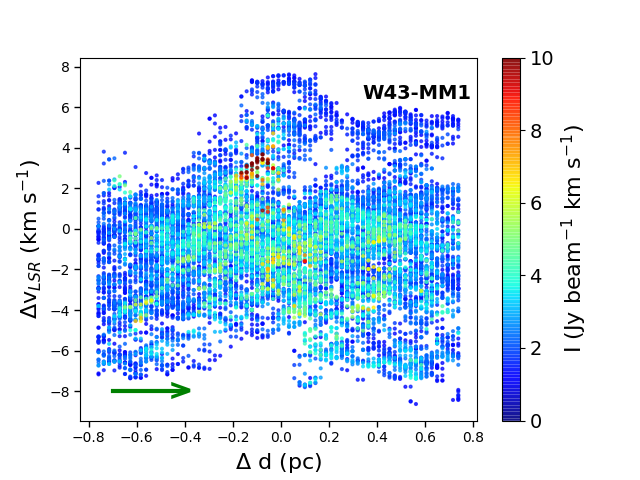} \includegraphics[width=2.3in, height=2.0in]{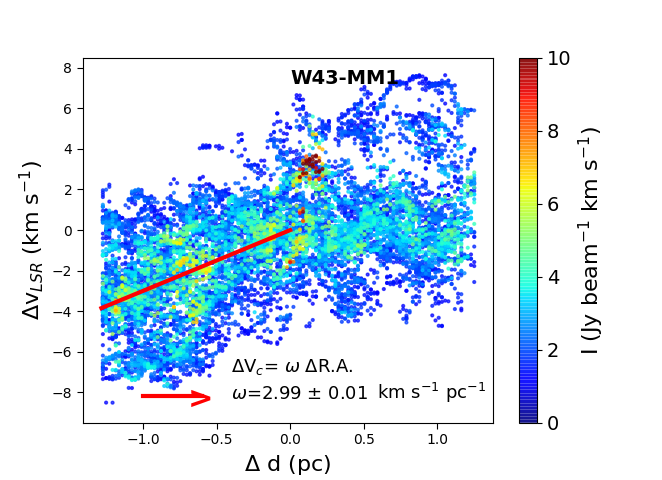}
    \caption{{\color{black}Left figure is the integrated intensity (moment 0) map of W43-MM1 region obtained from 2.5$''$ circular beam. Here, the orange rectangular region denotes the area where
we examine the position-velocity (PV) diagram in two perpendicular directions (towards the green and red arrows). Middle figure is the intensity
weighted PV diagram towards the direction of the green arrow mentioned in the left figure. Right figure is the intensity weighted PV diagram
towards the direction of the red arrow mentioned in the left figure. Color bars in the Middle and Right figures denote the integrated intensity of the components. The fitted velocity gradient is 2.99 $\pm$ 0.01 km s$^{-1}$ pc$^{-1}$ which is shown in red solid color.}}
    \label{fig:fig15}
\end{figure*}

\newpage

\section{Examining the effect of spatial smoothing on Sonic Mach number ($M_{\text{s}}$) distribution in G337.92 protocluster}\label{Appendix7}

For calculating the sonic Mach number ($M_{\text{s}}$) distribution in these 15 protoclusters, we required to smooth the C$^{18}$O (2$-$1) line cubes into the same dust image cubes which are at a slightly coarser angular resolution at 2.5$''$ \citep{2024A&A...687A.217D}. Consequently, we examine whether beam smoothing has a significant impact on the sonic Mach number ($M_{\text{s}}$) analysis. We tested this using one protocluster G337.92 whose original beam size was 0.70$''$ $\times$ 0.55$''$. First, we fit the pixel-wise spectra using the \texttt{Gausspy+} module and calculate the effective velocity dispersion ($\sigma_{\text{eff}}$) using the Eqn.\ref{eqn_2}. We show the spatial distribution and histogram plot of $\sigma_{\text{eff}}$ in the upper left and right panels of Fig. \ref{fig:fig16}. From the right panel, we notice that the median and the mean values are 0.96 and 1.11 respectively. The distribution also extends up to $\sim$ 5. Then, we smooth the original C$^{18}$O (2$-$1) line image cube with a circular beam of 2.5$''$ and fit the pixel-wise spectra with the same \texttt{Gausspy+} module. In the lower left and right panels of Fig. \ref{fig:fig16}, we show the spatial distribution and the histogram plots of the sonic mach number ($M_{\text{s}}$) distribution at the 2.5$''$ circular beam. From the histogram plot, we notice that the median and mean values are 0.93 and 1.11 respectively. Here also the distribution extends up to $\sim$ 5. From the comparison of the histogram plots in two beam sizes, we notice that due to the beam smoothing, the values of $\sigma_{\text{eff}}$ are not significantly affected, which indicates that there has been an insignificant effect on the sonic Mach number ($M_{\text{s}}$) analysis. Maybe, significantly large beam smoothing ($>>$ 2.5$''$) will have a noticeable impact on the distribution of $\sigma_{\text{eff}}$. Therefore, in our study, we first smooth all the regions with 2.5$''$ beam and study the sonic mach number ($M_{\text{s}}$) distribution.

\begin{figure*}
	\centering 
	\includegraphics[width=3.6in,height=2.6in,angle=0]{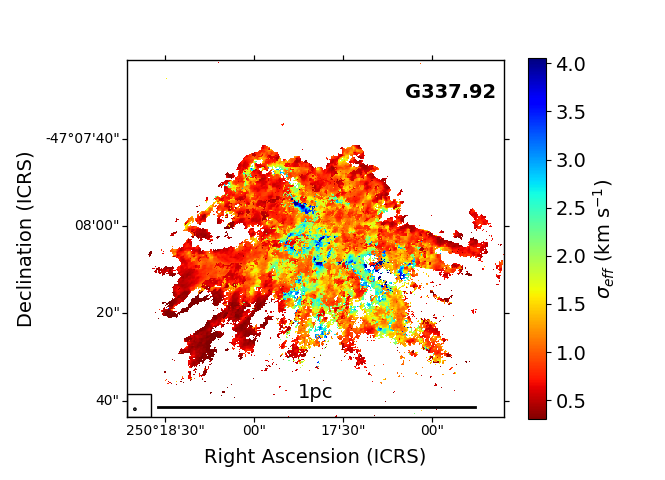}\includegraphics[width=3.6in,height=2.6in,angle=0]{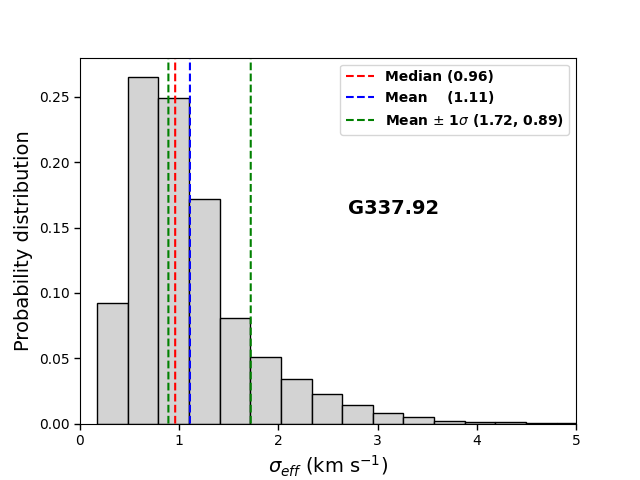}\\
 \includegraphics[width=3.6in,height=2.6in,angle=0]{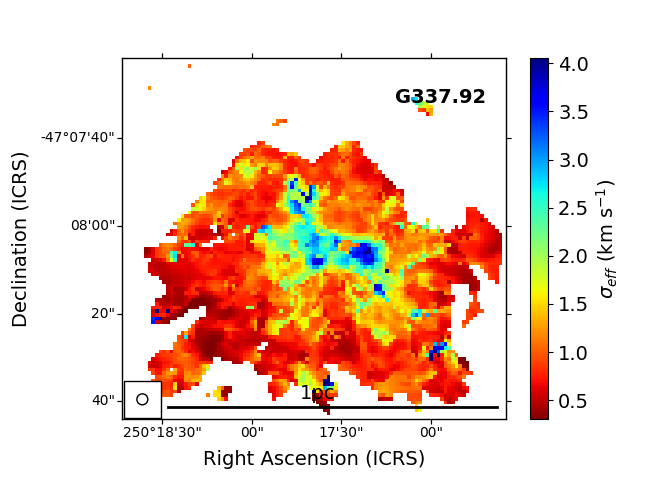}\includegraphics[width=3.6in,height=2.6in,angle=0]{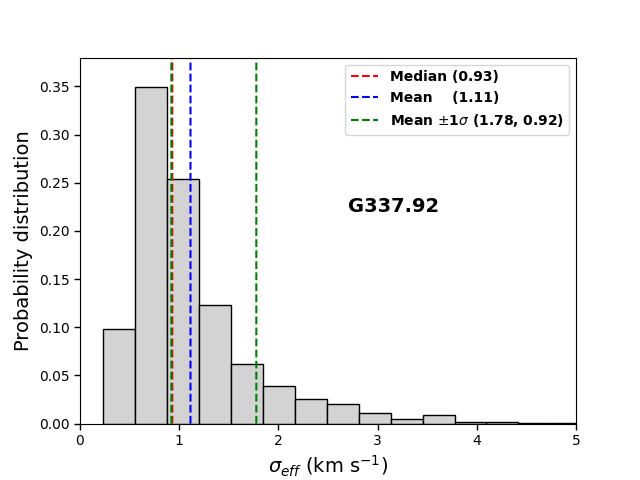}
	\caption{Upper left and right figures are the spatial distribution and the histogram plot of the effective velocity dispersion ($\sigma_{\text{eff}}$) in G337.92 region when the beam size is 0.70$''$ $\times$ 0.55$''$. Lower left and right figures are the spatial distribution and the histogram plot of the effective velocity dispersion ($\sigma_{\text{eff}}$) in G337.92 region when the beam size is 2.5$''$ $\times$ 2.5$''$.}
	\label{fig:fig16}
\end{figure*}

\section{Distribution of Sonic Mach number ($M_{\text{s}}$) in the protoclusters}\label{A:sonic}
We show the spatial distribution and histogram plot of $M_{\text{s}}$ for the protoclusters in Fig.\ref{AF:sonic}.

\begin{figure*}
    \centering
\includegraphics[width=0.33\linewidth]{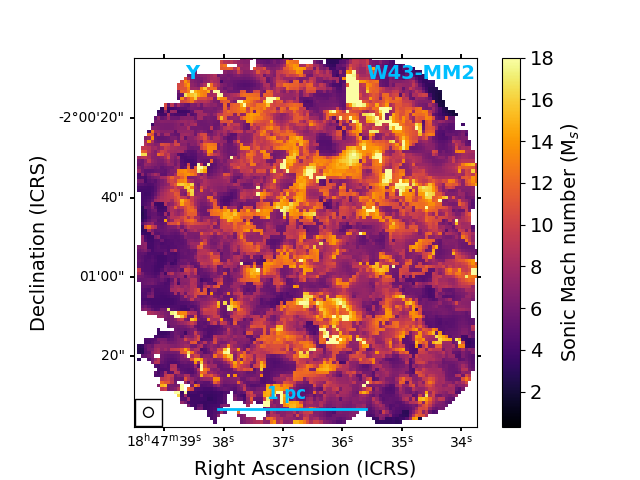}
\includegraphics[width=0.33\linewidth]{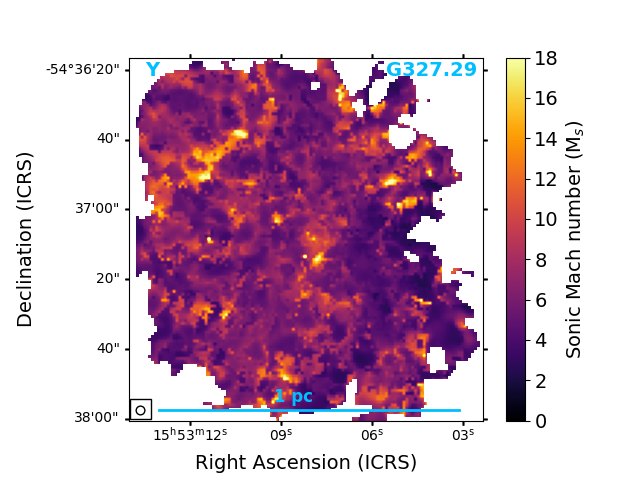}
\includegraphics[width=0.33\linewidth]{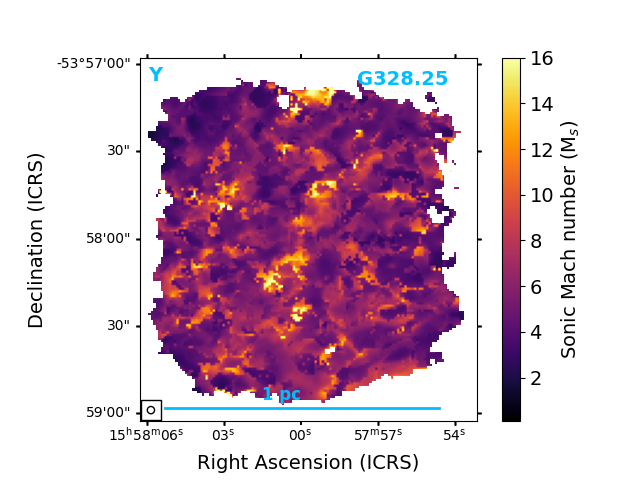}\\
\includegraphics[width=2.3in,height=1.75in,angle=0]{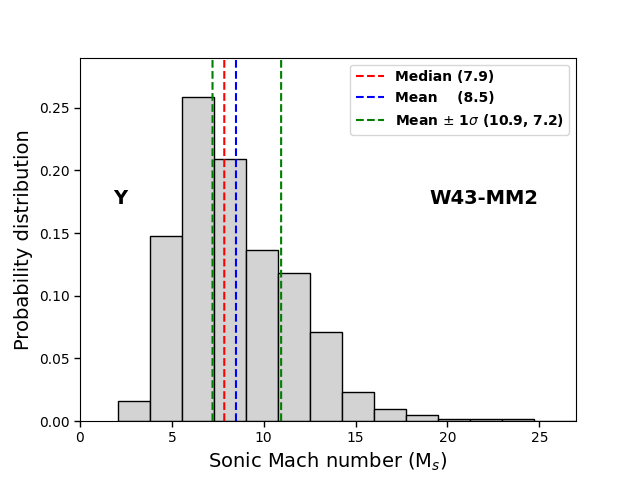}
\includegraphics[width=2.28in,height=1.75in,angle=0]{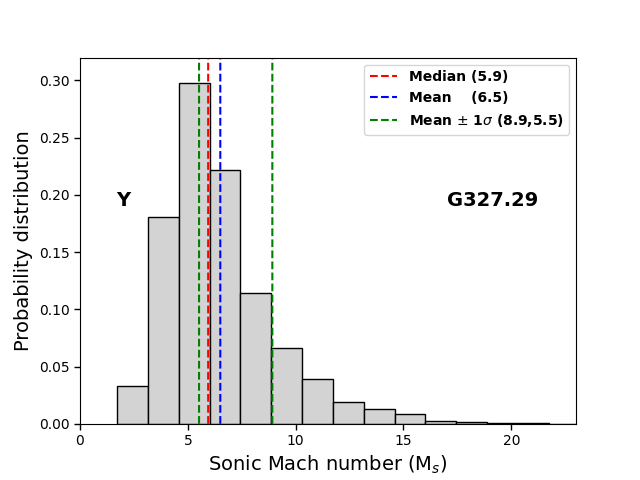}
\includegraphics[width=2.3in,height=1.75in,angle=0]{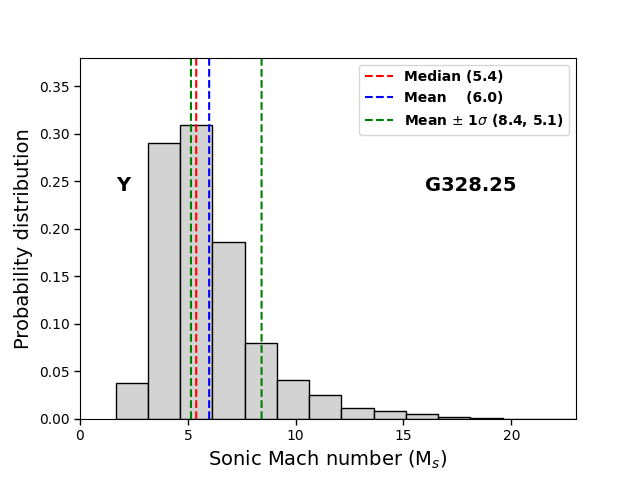}\\
 \includegraphics[width=0.33\linewidth]{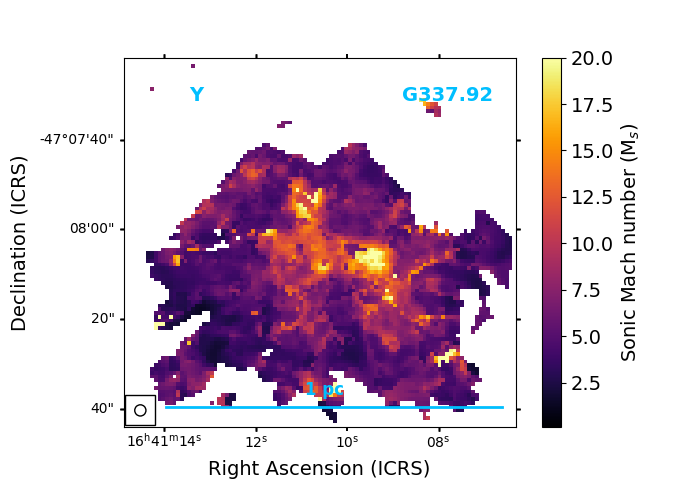}
 \includegraphics[width=0.33\linewidth]{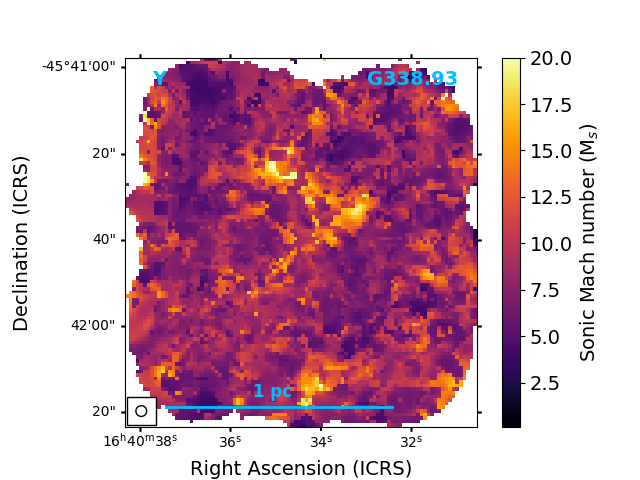}
 \includegraphics[width=0.33\linewidth]{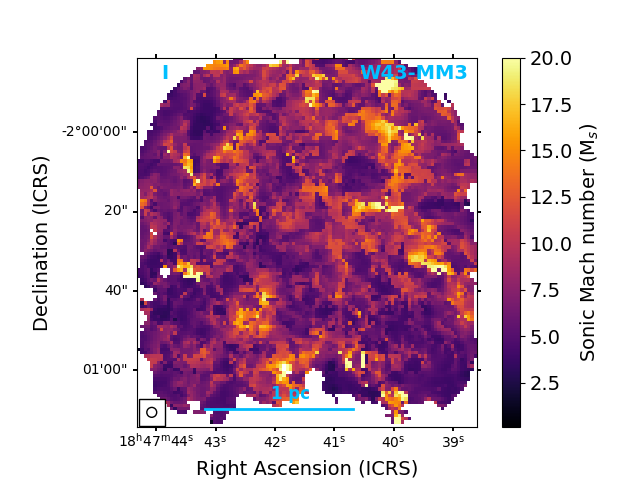}\\
  \includegraphics[width=2.3in,height=1.75in,angle=0]{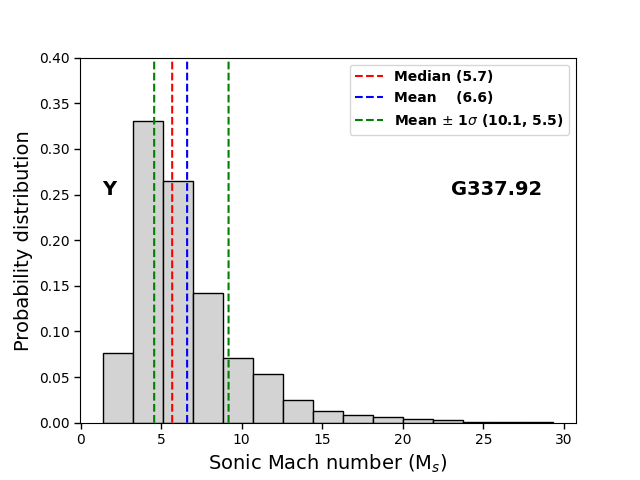}
 \includegraphics[width=2.3in,height=1.75in,angle=0]{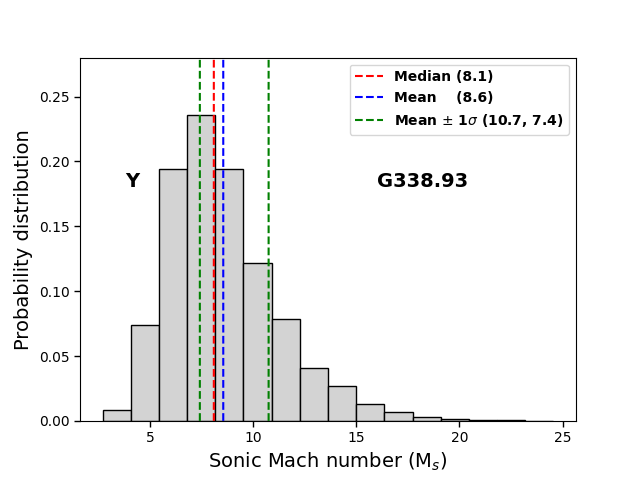}
\includegraphics[width=2.3in,height=1.75in,angle=0]{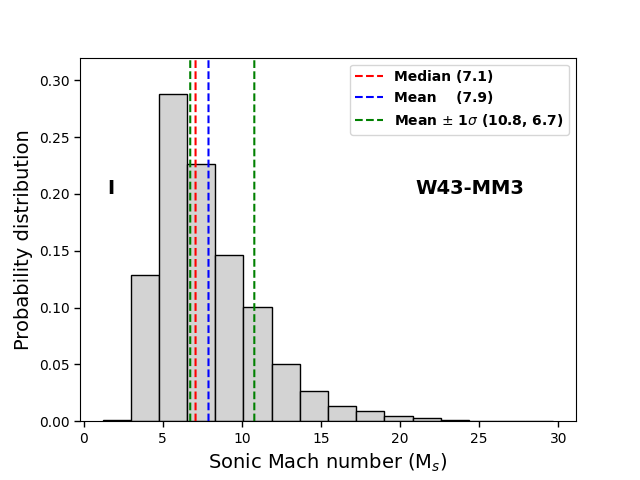}

\caption{For each protocluster, we display the map of the sonic Mach number (color map) and the histogram plot of its distribution below. Symbols Y  and I in figures indicate the young and intermediate stage protoclusters (see Section \ref{section_0}.}
\label{AF:sonic}
\end{figure*}

\begin{figure*}
    \centering
\includegraphics[width=0.33\linewidth]{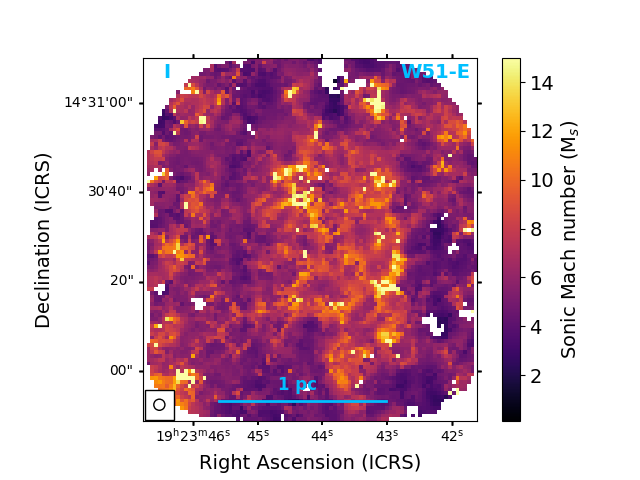}
 \includegraphics[width=0.33\linewidth]{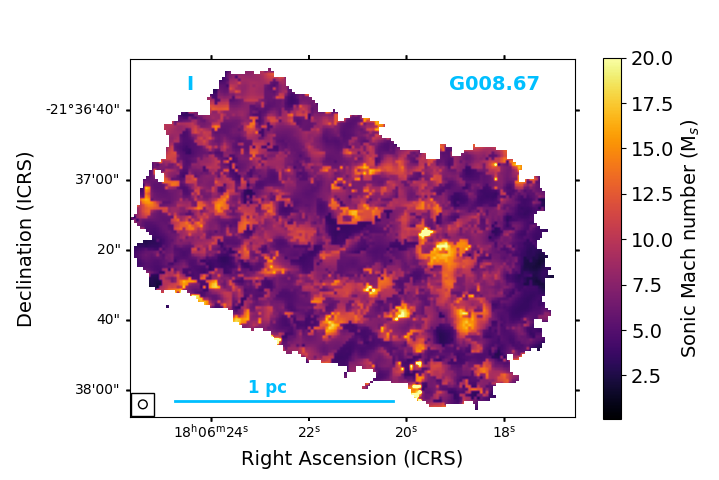}
 \includegraphics[width=0.33\linewidth]{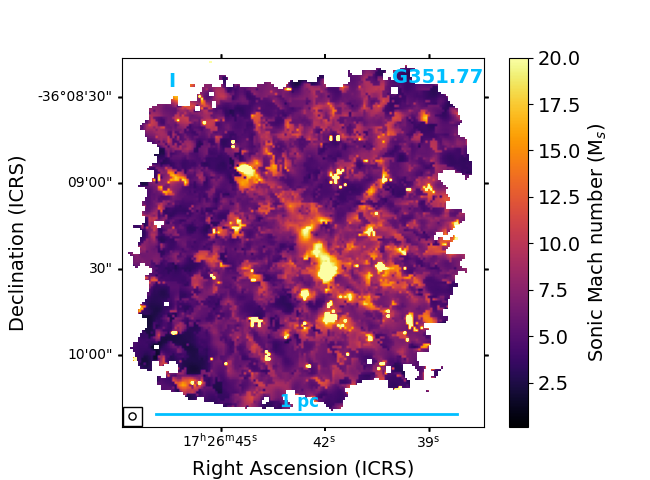}\\
\includegraphics[width=2.3in,height=1.75in,angle=0]{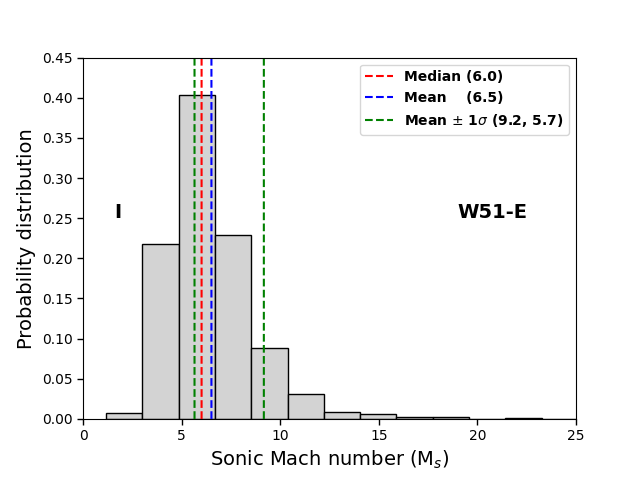}
 \includegraphics[width=2.3in,height=1.75in,angle=0]{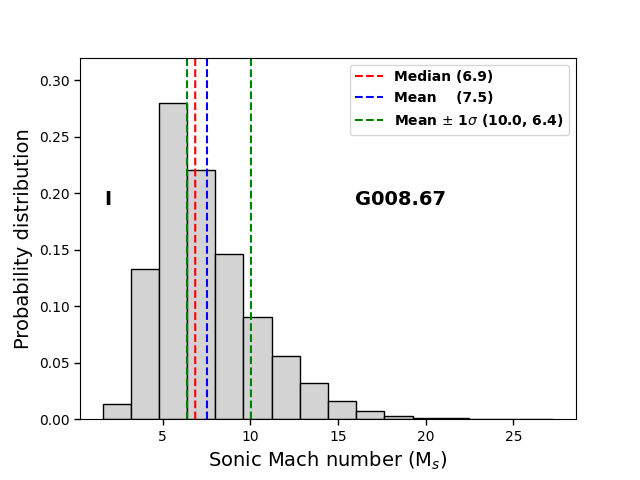}
 \includegraphics[width=2.3in,height=1.75in,angle=0]{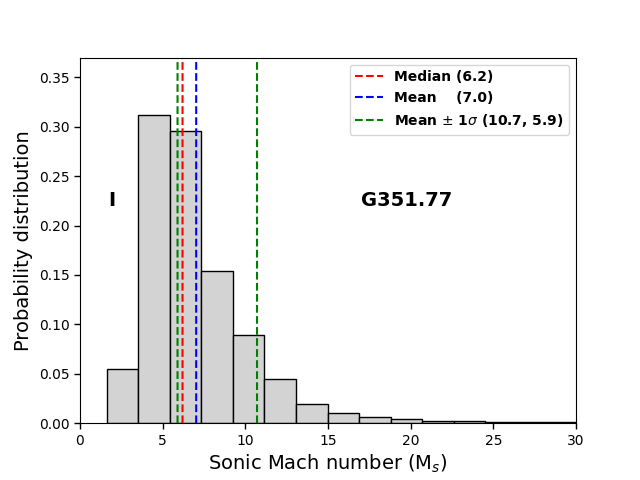}\\
\includegraphics[width=0.33\linewidth]{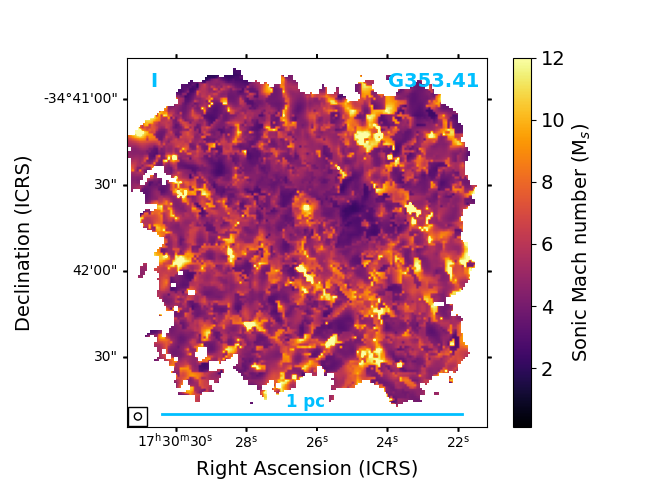}
 \includegraphics[width=0.33\linewidth]{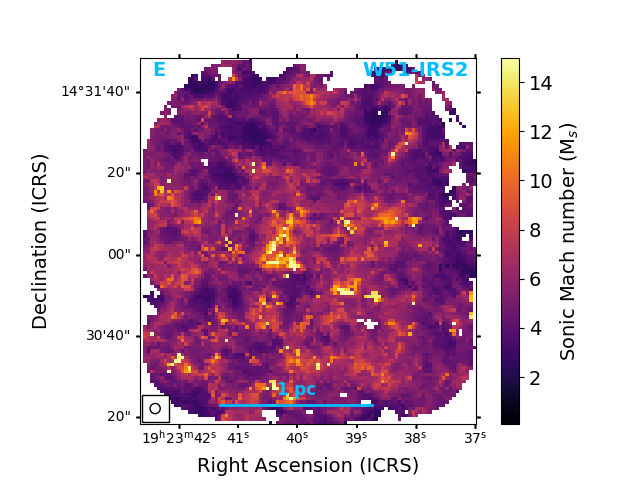}\\
 \includegraphics[width=0.33\linewidth]{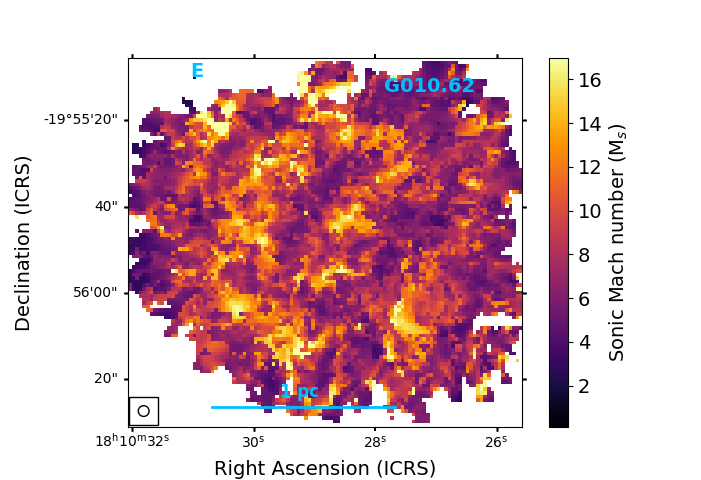}\\
  \includegraphics[width=2.3in,height=1.75in,angle=0]{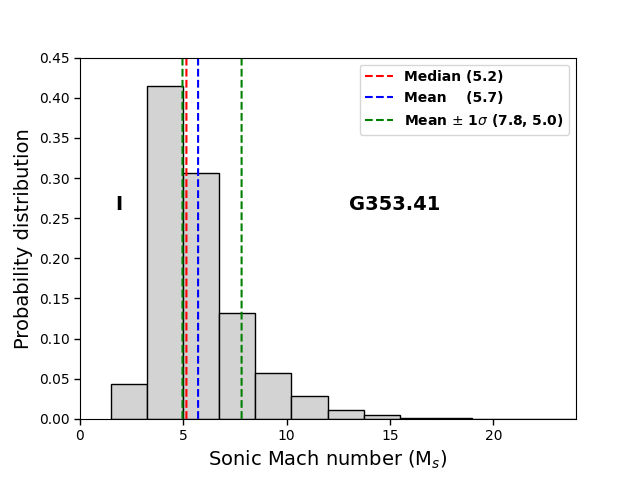}
 \includegraphics[width=2.3in,height=1.75in,angle=0]{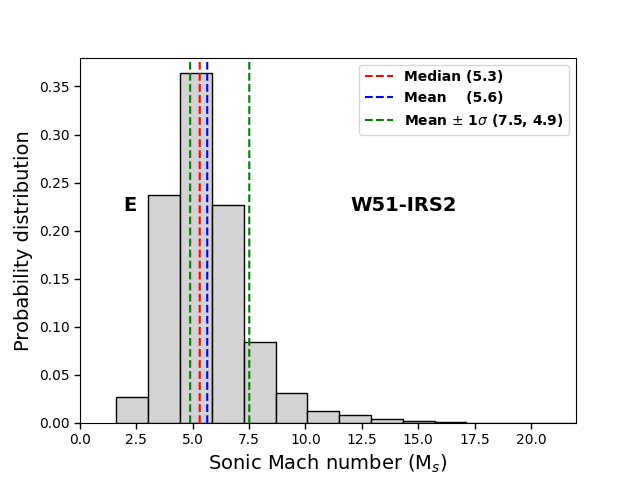}
\includegraphics[width=2.3in,height=1.75in,angle=0]{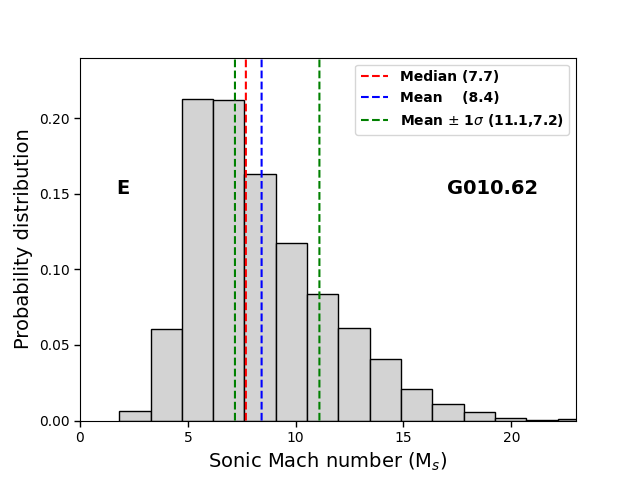} 
    \caption{Following of Fig.~\ref{AF:sonic}. . Symbols I and  E in figures indicate the intermediate and evolved protoclusters (see Section \ref{section_0}.}
    
\end{figure*}

\begin{figure*}
    \centering
\includegraphics[width=0.33\linewidth]{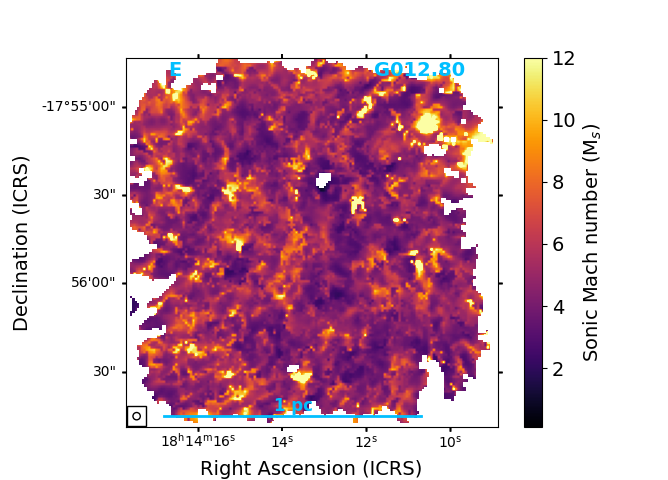}
 \includegraphics[width=0.33\linewidth]{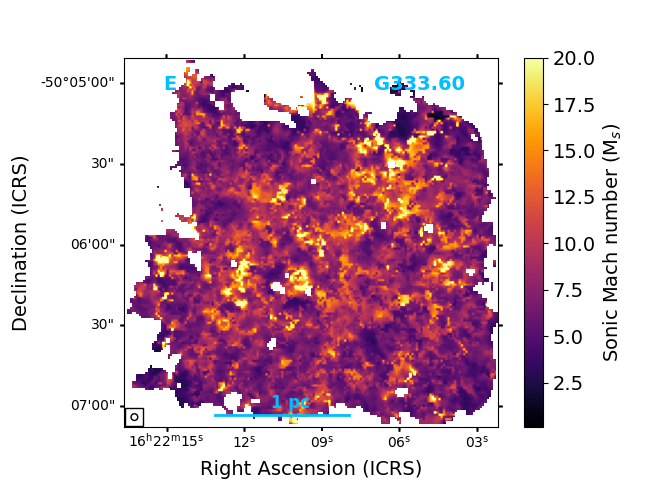}\\
 \includegraphics[width=2.3in,height=1.75in,angle=0]{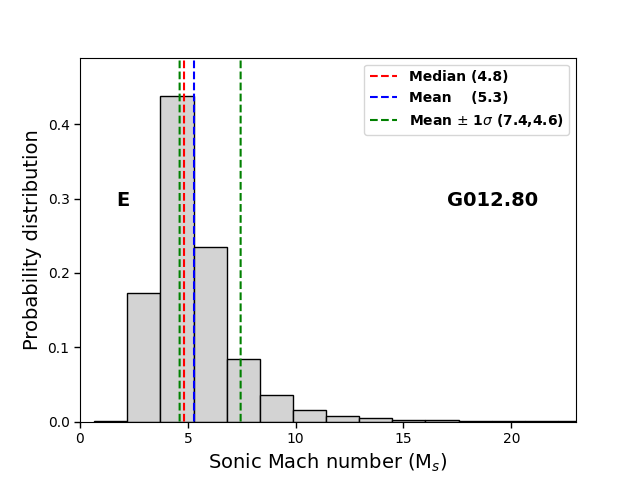}
 \includegraphics[width=2.3in,height=1.75in,angle=0]{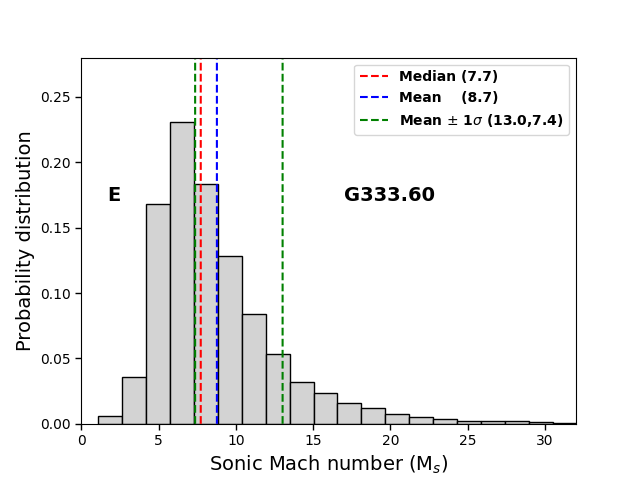}

    \caption{Sonic Mach (continued). Symbol E in figures indicates the evolved protoclusters (see Section \ref{section_0}.}
   
\end{figure*}

\section{Correlation between sonic Mach number ($M_{\text{s}}$) and hydrogen column density [$N(\text{H}_2)$] for the protoclusters} \label{A:correlation_column_density}

We show the correlation between sonic Mach number ($M_{\text{s}}$) and hydrogen column density [$N(\text{H}_2)$] for protoclusters in Fig. \ref{fig:fig27}.

\begin{figure*}
	\centering 
 \includegraphics[width=2.35in,height=1.88in,angle=0]{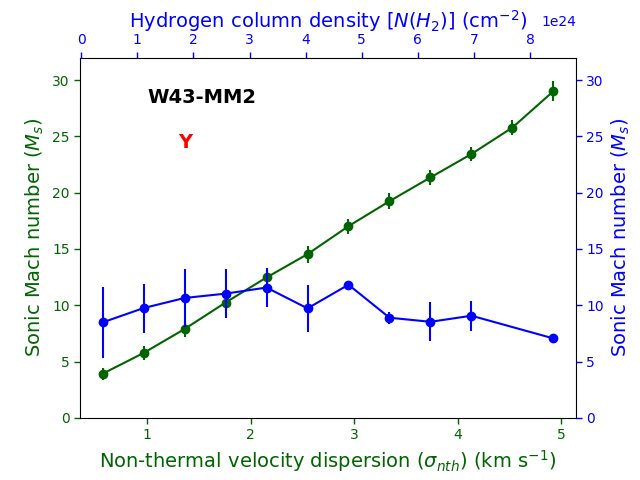} \includegraphics[width=2.35in,height=1.88in,angle=0]{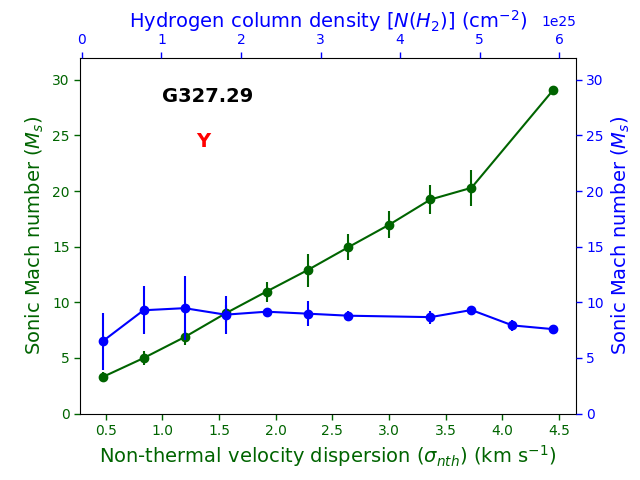}\\
    \includegraphics[width=2.35in,height=1.88in,angle=0]{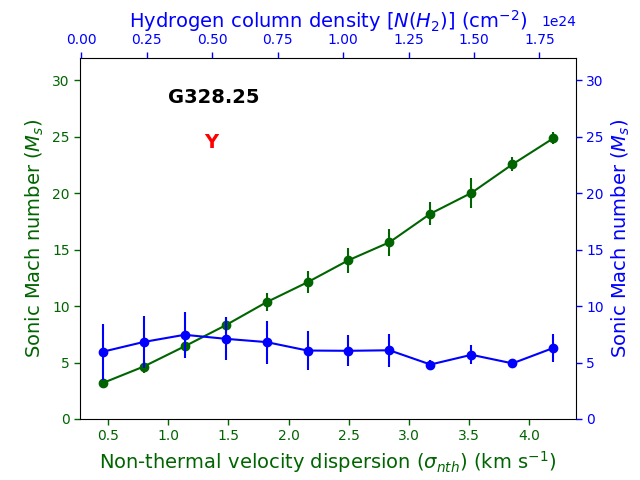}\includegraphics[width=2.35in,height=1.88in,angle=0]{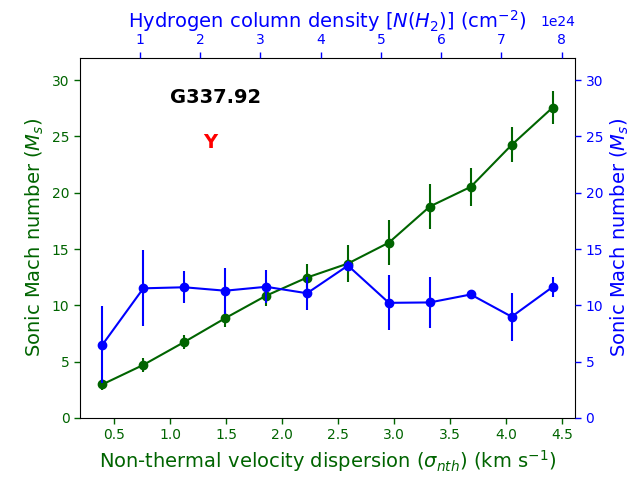} \includegraphics[width=2.35in,height=1.88in,angle=0]{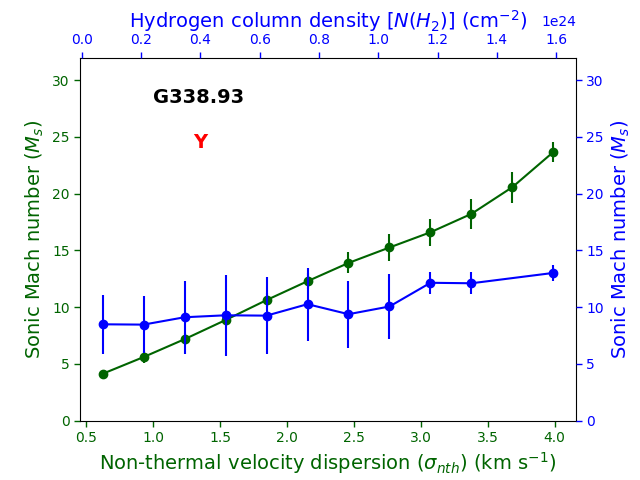}\\
    \includegraphics[width=2.35in,height=1.88in,angle=0]{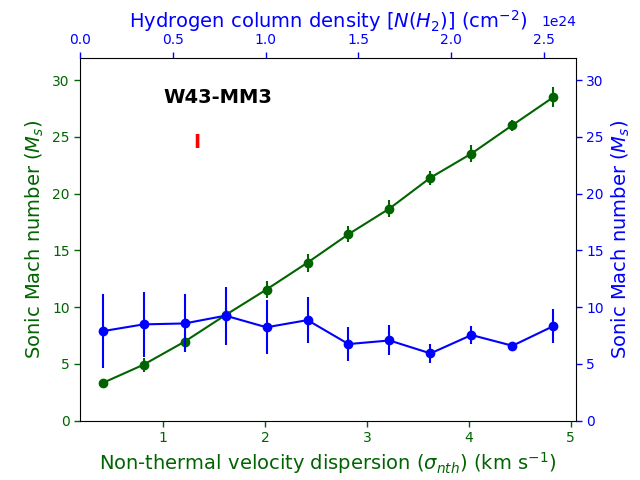}\includegraphics[width=2.35in,height=1.88in,angle=0]{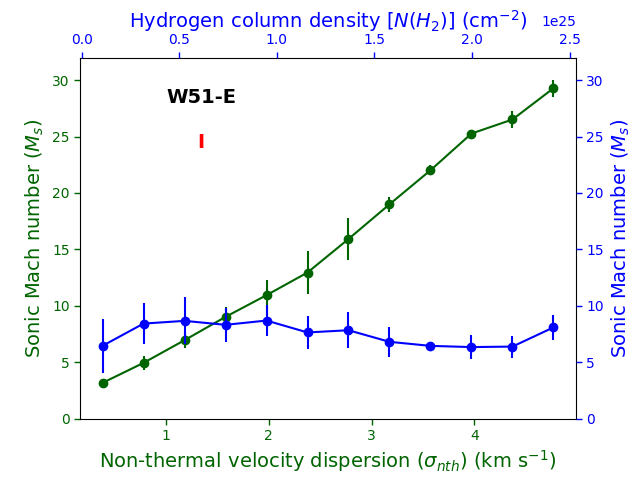} \includegraphics[width=2.35in,height=1.88in,angle=0]{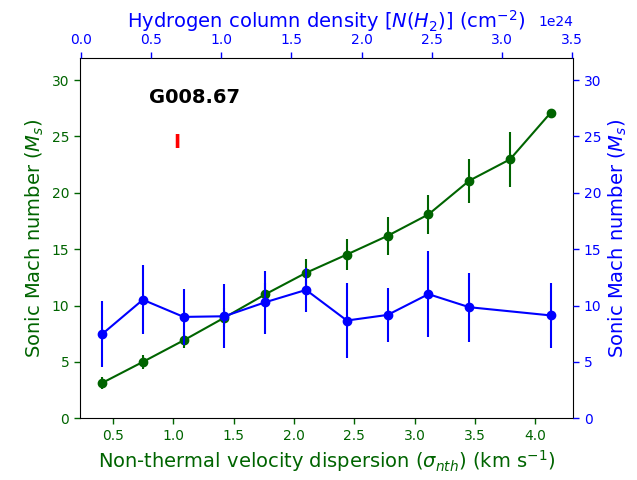}\\
    \includegraphics[width=2.35in,height=1.88in,angle=0]{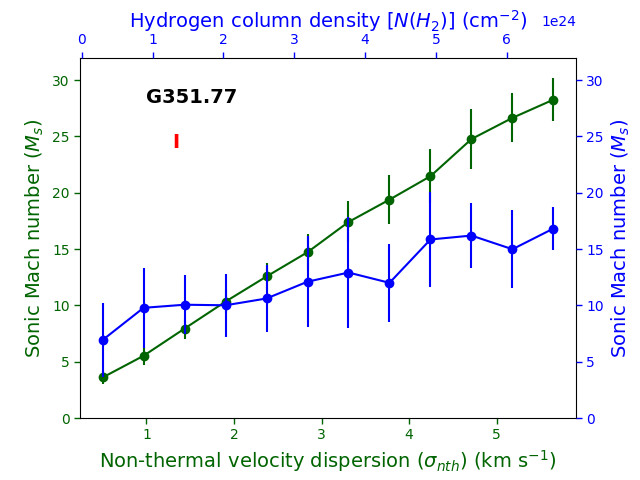}\includegraphics[width=2.35in,height=1.88in,angle=0]{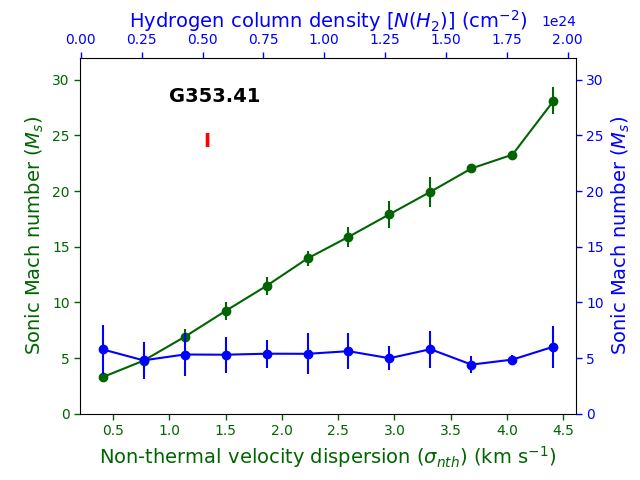} \includegraphics[width=2.35in,height=1.88in,angle=0]{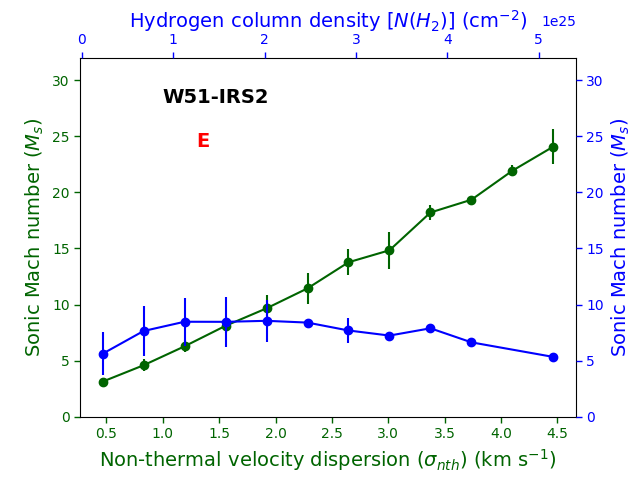}\\
    \includegraphics[width=2.35in,height=1.88in,angle=0]{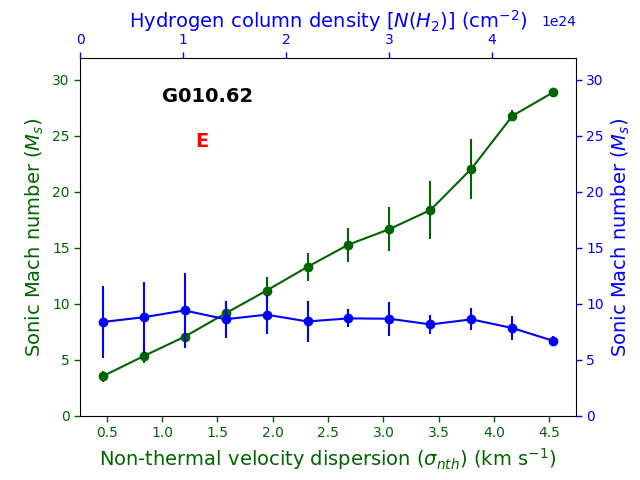}\includegraphics[width=2.35in,height=1.88in,angle=0]{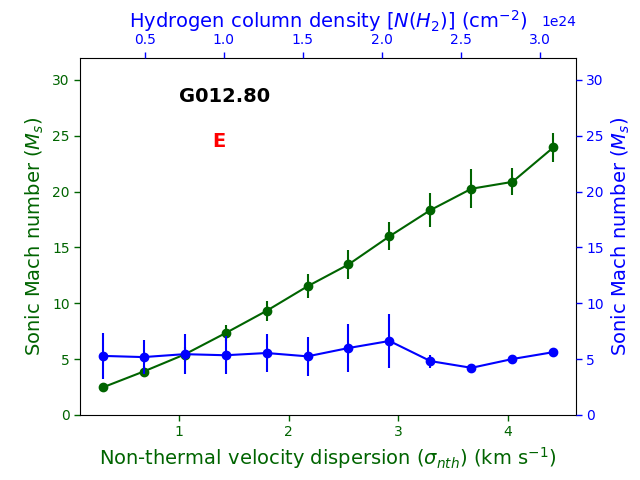} \includegraphics[width=2.35in,height=1.88in,angle=0]{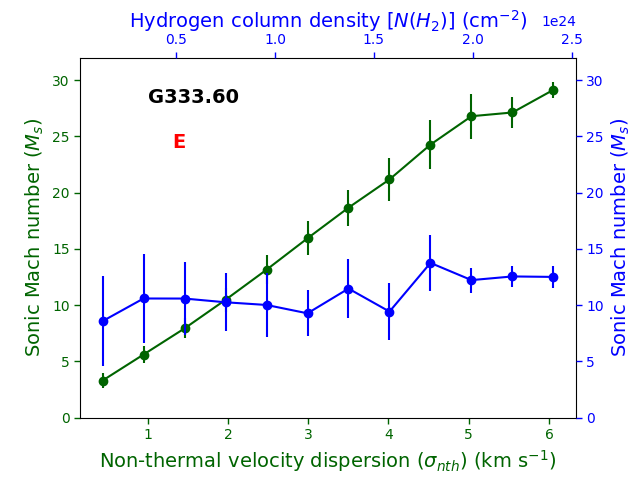}\\
	\caption{Correlation between (a) sonic Mach number ($M_{\text{s}}$) and non-thermal velocity dispersion ($\sigma_{\text{nth}}$) (green color) and (b) sonic Mach number ($M_{\text{s}}$) and hydrogen column density [$N({\text{H}_{2}})$] (blue color) for protoclusters. Symbols Y, I, and E in the figures indicate the young, intermediate and evolved protoclusters (see Section \ref{section_0}).}
	\label{fig:fig27}
\end{figure*}

{\color{black}

\section{Examining the effect of an intensity cut (\texttt{min\_value} = 2.5$\sigma_{\text{rms}}$) on the velocity dispersion ($\sigma_{\text{tot}}$) of the wider lines in the dendogram analysis}\label{A:structure_intensity_cut}

We checked whether the parameter \texttt{min\_value} in the Astrodendro module reduced the velocity dispersion of the wider components. While extracting the structures from the PPV cube, we set \texttt{min\_value} to 2.5$\sigma_{\text{rms}}$ for all the protoclusters. On the one hand, it is necessary to avoid taking noise in the analysis; on the other hand, this criterion may truncate a significant fraction of the large wing for the broad components. Currently, there is no effective way to resolve this issue. However, we have checked for four protoclusters whether this effect is severe or not. The four protoclusters are W43-MM1, G333.60, G338.93, and W51-E. In Fig. \ref{fig:fig28}., we first plot the histogram of the FWHM of the C$^{18}$O (2$-$1) line components that we obtain from the \texttt{Gausspy+} module. From the figures, it shows that the histogram plots of the FWHM for the number of components exhibit a log-normal type profile. The FWHMs extend up to $\sim$ 16 km s$^{-1}$ in all four cases. We set the cut-off point for FWHM at 5 km s$^{-1}$, beyond which we consider the components to have wings-like structures. Although this cut-off point is somewhat arbitrary, there is no way to set any absolute value. We then plot the histogram of the components whose FWHM is greater than 5 km s$^{-1}$ into two categories. One for which the peak intensity of the components is greater than 3 $\times$ \texttt{ min\_value} and the other for which the peak intensity of the components is below 3 $\times$ \texttt{min\_value}. These histograms are plotted for four protoclusters in blue and red colors on the right panels of Fig. \ref{fig:fig28}. From these figures, it indicates that between 6 and 10 km s$^{-1}$, components of which $I_{\text{peak}}$ is $>$ 3 $\times$ \texttt{min\_value} are dominated compared to the number of components for which  $I_{\text{peak}}$ is $<$ 3 $\times$\texttt{ min\_value}. However, beyond $\sim$ 10 km s$^{-1}$, the number of components in both categories becomes more or less equal. Consequently, due to the cutoff of \texttt{ min\_value} = 2.5$\sigma_{\text{rms}}$, the velocity dispersion of half of the larger components whose velocity dispersion is greater than 10 km s$^{-1}$ may reduce significantly. This can be considered as a caveat for the Astrodendro module, in addition to the contamination issue of the structures. 

}

\begin{figure*}
	\centering 
	 \includegraphics[width=3.0in,height=2.2in,angle=0]{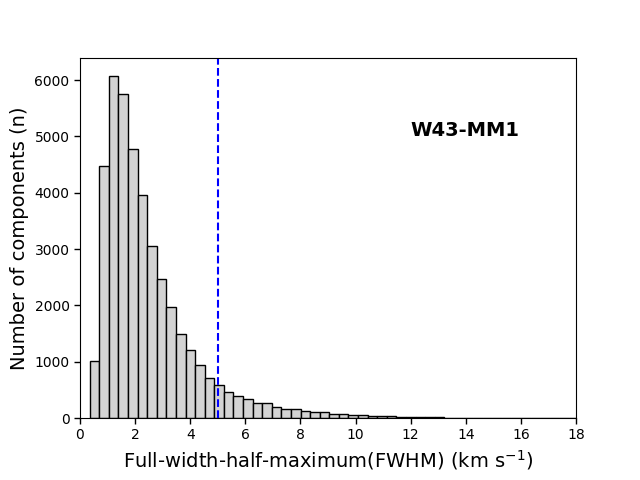} \includegraphics[width=3.0in,height=2.2in,angle=0]{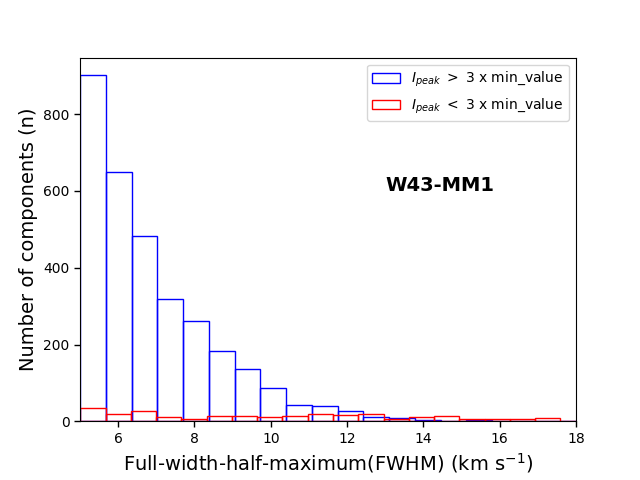}\\
  	 \includegraphics[width=3.0in,height=2.2in,angle=0]{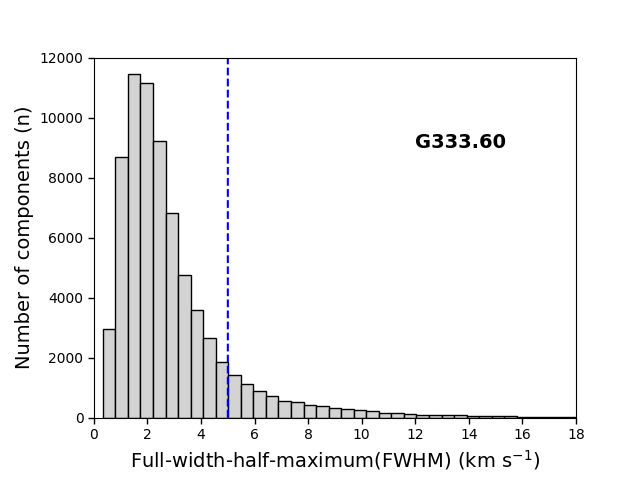} \includegraphics[width=3.0in,height=2.2in,angle=0]{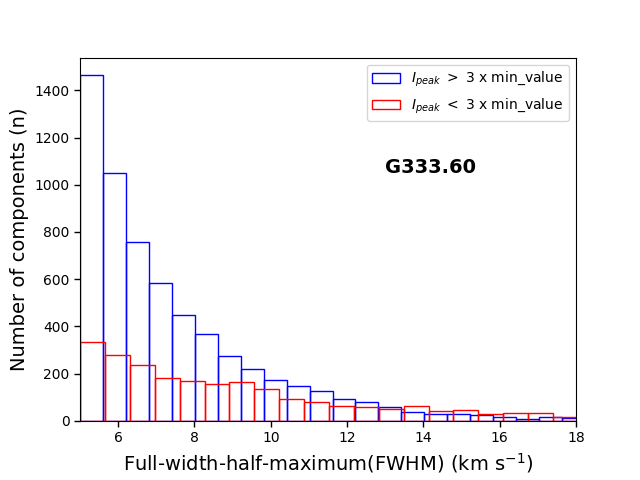}\\
     	 \includegraphics[width=3.0in,height=2.2in,angle=0]{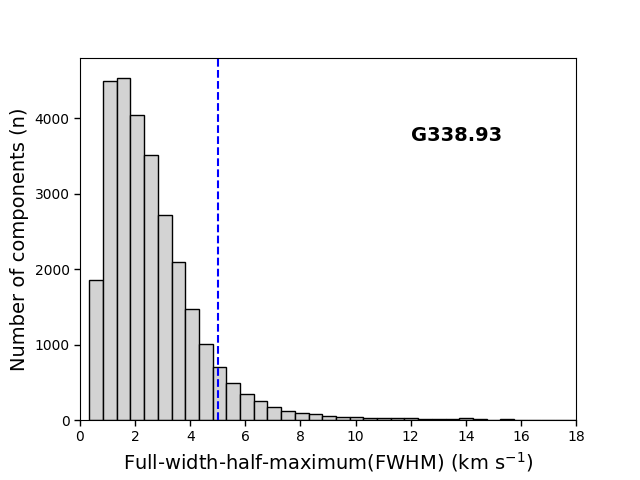} \includegraphics[width=3.0in,height=2.2in,angle=0]{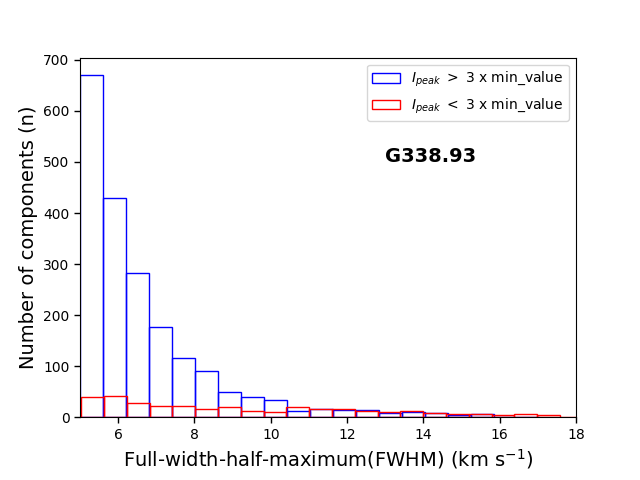}\\
         	 \includegraphics[width=3.0in,height=2.2in,angle=0]{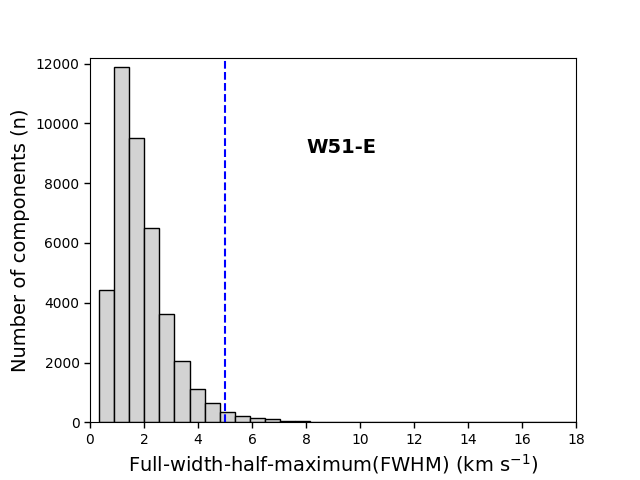} \includegraphics[width=3.0in,height=2.2in,angle=0]{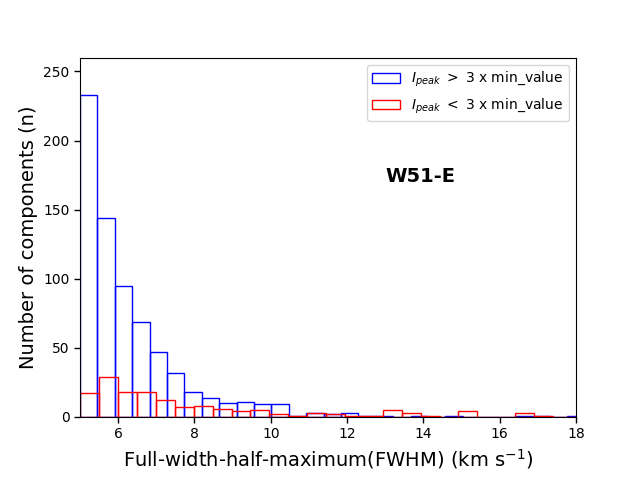}\\
	\caption{{\color{black}Left panel in four rows show the histogram plots of the full-width-half-maximum (FWHM) of the C$^{18}$O (2$-$1) line components for the W43-MM1, G333.60, G338.93 and W51-E protoclusters respectively. Here the blue dashed line denotes the FWHM value of 5.0 km s$^{-1}$, above which we consider the components have large wing like structures. Right panel in four rows show the histogram plots of the components for the four protoclusters whose FWHM are above 5.0 km s$^{-1}$ and intensity peaks ($I_{\text{peak}}$) are above 3$\times$ \texttt{min\_value} (blue color) and below 3$\times$ \texttt{min\_value} (red color).}} 	
    \label{fig:fig28}
\end{figure*}

\end{appendix}

\vspace{20 mm}


\end{document}